%% file: main.tex
\documentclass[11pt, twoside]{article}

\input{Preamble/preamble_packages} 

\input{Preamble/bibliography_code} 

\title{

How Small is Big Enough? Open Labeled Datasets and the Development of Deep Learning
}

\author[1]{Daniel Souza\thanks{Corresponding author. Email address: danielfernando.desouza@polimi.it}}
\author[2,3,4]{Aldo Geuna}
\author[5]{Jeff Rodríguez}
\affil[1]{{\small Department of Management, Economics and Industrial Engineering, Polytechnic University of Milan, Via Raffaele Lambruschini, 4/B, 20156, Milan, Italy}}
\affil[2]{{\small Department of Cultures, Politics and Society, University of Turin, Lungo Dora Siena, 100A, 10153, Turin, Italy}}
\affil[3]{{\small Collegio Carlo Alberto, Piazza Vincenzo Arbarello, 2, 10122, Turin, Italy}}
\affil[4]{{\small Programme Innovation, Equity \& The Future of Prosperity, Canadian Institute for Advance Research (CIFAR), MaRS Centre, West Tower 661 University Ave., Suite 505, Toronto, ON M5G 1M1, Canada}}
\affil[5]{{\small OECD, 2 rue André Pascal, 75775, Paris, France}}

\date{\today}

\begin{document}

\raggedbottom

\maketitle

\begin{center}
\begin{abstract}

\small

\noindent
We investigate the emergence of Deep Learning as a technoscientific field, emphasizing the role of open labeled datasets. Through qualitative and quantitative analyses, we evaluate the role of datasets like CIFAR-10 in advancing computer vision and object recognition, which are central to the Deep Learning revolution. Our findings highlight CIFAR-10's crucial role and enduring influence on the field, as well as its importance in teaching ML techniques. Results also indicate that dataset characteristics such as size, number of instances, and number of categories, were key factors. Econometric analysis confirms that CIFAR-10, a small-but-sufficiently-large open dataset, played a significant and lasting role in technological advancements and had a major function in the development of the early scientific literature as shown by citation metrics.

\blfootnote{

We extend our gratitude to the interview participants for sharing their insights on the development of Deep Learning. Special thanks go to Stefano Bianchini, Stijn Kelchtermans, Michele Pezzoni, and attendees of the 6th BETA Workshop, OIS Research Conference 2023, Munich Summer Institute Workshop 2023, REGIS Summer School 2023, Atlanta Conference on Science and Innovation Policy 2023, DRUID 2023, Eu-SPRI Conference 2023, 2nd International Conference on Science of Science and Innovation, and the International Joseph A. Schumpeter Society Conference 2024 for their valuable suggestions. We also appreciate the comments from seminar participants at the University of Manchester, Université Côte d'Azur, and Università Cattolica del Sacro Cuore. Special thanks to Fazliddin Shermatov for his outstanding research assistance.

We gratefully acknowledge the funding provided by the Canadian Institute for Advanced Research (CIFAR), Innovation Equity and the Future of Prosperity Program Catalyst Grant, as well as the financial support from the GRINS project, financed by PNRR (Piano Nazionale di Ripresa e Resilienza, Missione 4 (Infrastruttura e ricerca), Componente 2 (Dalla Ricerca all’Impresa), Investimento 1.3 (Partnership Estese), Tematica 9 (Sostenibilità economica e finanziaria di sistemi e territori)).}
\end{abstract}
\end{center}

{\small

\vspace{0.25cm}

}
\onehalfspacing


\section{Introduction}
\label{sec:intro}
Artificial Intelligence (AI) technologies promise to revolutionize the knowledge production process. At the core of one of the most important approaches to the AI revolution are machine learning (ML) algorithms: computer programs that improve performance as they are exposed to an increasing amount of data. An example of disruptive technology based on ML is AlphaFold – an AI algorithm developed by Google’s offshoot DeepMind first released in 2018, which solved one of the most challenging problems in the field of biology: the prediction of protein’s structures based on amino-acid sequences \parencite{jumperHighlyAccurateProtein2021, callawayItWillChange2020}.  A more recent example is ChatGPT, a Large Language Model (LLM) developed by OpenAI. It is based on the GPT (Generative Pre-training Transformer) architecture and is trained to generate human-like text. ChatGPT and other LLMs available in the early 2020s have been identified as having impact in diverse areas that go from medicine \parencite{jeblickChatGPTMakesMedicine2022} to journalism \parencite{pavlikCollaboratingChatGPTConsidering2023} and their impact on science is heavily discussed \parencite{stokel-walker_what_2023}.

These breakthroughs and many others are underpinned by developments in Deep Learning (DL), a subset of ML models that relies on neural networks and requires vast amounts of data to be trained \parencite{lecunDeepLearning2015}. Due to the extremely promising results in wide areas of application, DL has been regarded as a new method of invention and potentially a general-purpose technology in which the next industrial revolution maybe based \parencite{craftsArtificialIntelligenceGeneralpurpose2021}. Although a growing literature has studied the impact of DL on the knowledge production process \parencite{bianchiniArtificialIntelligenceScience2022, klingerDeepLearningDeep2021}, little attention has been given to its inception and to the specific role played by Open Labelled Datasets (OLDs).

In this paper we analyze the emergence of DL as a technoscientific field, that is, a domain in the middle of scientific enquiry and technical problem-solving \parencite{kastenhoferCommunityIdentityContemporary2021}. More specifically, we examine how OLDs have contributed to the growth and consolidation of DL, focusing on their distinct characteristics. Within this perspective, we regard OLDs as technological artifacts that allow the development of the field. We draw on the literature discussing the emergence of new scientific disciplines to provide a picture of the development of DL as the dominant approach in ML \& AI, and the role of OLDs in that process. We perform an analysis of the technological and scientific use of OLDs that includes both qualitative and quantitative elements. We devote particular attention to the role played by CIFAR-10, the most used dataset in the ML literature indexed at the \emph{Papers with Code} website\footnote{See Section \ref{subsec:data} for details.}. We carried out a set of semi-structured interviews with relevant actors and we implemented a survey of academics and ML practitioners who have used CIFAR-10 in their work; on the basis of the qualitative evidence we modeled the use OLDs in technological and scientific development proxied by patent (technology) and scholarly (science) citations in the period 2000-2022. 

Compute, data and algorithmic advances are the needed ingredients of the DL revolution \parencite{KochPeterson2024, Sevilla2022}. In early 2010s increased computing power availability (see the arrival of 2D and 3D GPUs) was in line with the doubling  approximately every 6 months of computing requirements  by new DL algorithms running on OLDs \parencite{Sevilla2022}. The main tenet of this paper is that once the bottleneck of computer power was not any longer a major problem, the potential of neural network approaches to AI - theoretically developed over the last fifty years of the 20th century - could be realized and further advanced through the use of OLDs. Given that AI as a field shifted towards an evaluation system based on \emph{benchmarking} - quantification of progress based on predictive accuracy on example datasets \parencite{KochPeterson2024}, OLDs became fundamental to develop better algorithms/architectures.  Models (algorithms and architectures) were developed to solve specific tasks using specific OLDs; they would not exist without the dataset, as the specific OLD allowed the development of more refined and accurate models. OLDs that required less computing power, such as CIFAR-10, a small-but-sufficiently-large dataset, enabled the testing and refinement of new model architectures like AlexNet, which succeeded in solving tasks using huge and complex datasets that were previously unattainable with the same computational resources. OLDs should be considered as the necessary testing tool that had to be developed to allow progress in the DL modelling. 



The qualitative evidence we put together support the view that OLDs, and CIFAR-10 in particular, were fundamental for the technological and scientific developments which lead to the DL revolution and still shape the trajectory of the field. We trace the creation of the CIFAR-10 to the CIFAR NCAP Summer School in 2008, where the labelling of the dataset was conducted mostly by graduate students over the supervision of Geoffrey Hinton, a prominent scholar in the field, and two of his students, Alex Krizhevsky and Vinod Nair. We also learned through our interviews that CIFAR-10 became a benchmark due to its technical specifications, namely the nature of the images, their size, the number of samples and categories.  The survey confirms the insights of the interviews and highlights that CIFAR-10 is used extensively in the training of computer scientists working with ML. Many researchers not only teach courses using CIFAR-10, but also were themselves exposed to the dataset while following graduate programs. This finding highlight teaching as an important channel through which CIFAR-10 impacted the field of DL. 

By examining data from 28,393 conference proceedings and journal publications in the ML literature that utilized OLDs to train models between 2010 and 2022, we assess the technological and scientific relevance of these papers based on their citations in patents and academic literature. Our econometric analysis confirms the significant role of CIFAR-10 in the technological and scientific development of DL. Specifically, we find that papers using CIFAR-10 — a small but sufficiently large dataset — had a substantial early impact on the scientific literature, as evidenced by high academic citation counts, and continue to be relevant today, as shown by their higher patent citation counts. This indicates that the technical characteristics that initially contributed to the dataset's success continue to drive research and technological advancements in DL, particularly in computer vision and image recognition. We compared the CIFAR-10 and ImageNet datasets, demonstrating that CIFAR-10 has been and continues to be significant for technological developments, while ImageNet keep on playing a prominent role in scientific developments within the DL literature.

The rest of the paper proceeds as follows. In the next section, we present the conceptual framework used followed by the historical and institutional background of DL research and OLDs in Section 3. Section 4 describes the empirical methodology, data collection, the construction of the sample and presents descriptive statistics. Section 5 reports and discusses the results of the analysis. Section 6 concludes the paper.

\section{Conceptual framework}
\label{sec:concept_frame}

Since Kunh’s \emph{The structure of scientific revolutions} \parencite{kuhnStructureScientificRevolutions1970}, the sociology of science - and more recently the economics of science - has been interested in studying the conditions of emergence of new disciplines or subdisciplines within the scientific endeavor. The most important idea presented by Kuhn is how scientific knowledge does not always grow in a stable and incremental fashion, but it can also go through short periods of big changes, in which new paradigms emerge and consolidate.

In this paper we explore in particular the question of how OLDs contributed to the process of making DL into the dominant paradigm within AI \parencite{kerstingMachineLearningArtificial2018, chahDeepRabbitHole2019, schmidhuberDeepLearningNeural2015}, after being dismissed for a long time in favor of symbolic AI \parencite{waldropWhatAreLimits2019, weberRiseDatadrivenAI2021}. To do so, we mostly rely the theoretical contributions of \textcite{frickelGeneralTheoryScientific2005}.  They argued that there are parallelisms between social movements and what they call Scientific/Intellectual Movements (SIMs). Just as social movements, SIMs involve the pursue of common projects and objectives by a group of people that must rely upon repertoires of collective action to face the resistance from others in the scientific or intellectual community. Since SIMs resemble social movements that emerge to challenge some previous paradigm and therefore inevitably face some level of resistance, they also must deal with the problems of collective action: "The emergence of new social forms in science and academe invariably requires some level of spatial, temporal, and social coordination." \parencite{frickelGeneralTheoryScientific2005}. 

Following \textcite{kuhnStructureScientificRevolutions1970}, we also consider that SIMs emerge at times of scientific crisis, when research anomalies linked to old paradigms have accumulated beyond a tolerable threshold. However, the contempt towards the dominant paradigm is only a prerequisite and never enough to generate a SIM. For an intellectual movement of that sort to be successful, the leaders must articulate a a distinctive research program program. Doing so requires certain structural conditions, especially the access to resources, such as employments for the members of the SIM, access to laboratories, academic positions that allow to publish their results, and organizational resources that allow the members of the SIM to come together and create  \emph{epistemic cultures}, and discuss repertoires of thought and action that allow them to advance their intellectual agenda.

After the initial conditions are given, SIMs also have the need (like social movements) to recruit new members, to do so a locus of exchange and discussion where novel research is presented to old members and potential new recruits become a major condition for the success of the movement. This scenarios of micromobilization can take the form of seminars, conferences, PhD positions, or summer schools.  SIM must find ways to validate itself both internally, building a narrative of its history and identity, and externally, against opponents  \parencite{frickelGeneralTheoryScientific2005}.

In the case in hand, in the 80s and 90s, there was a group of scientists, in different universities around the United States, Europe and Canada who were not satisfied with the direction of the research programs in AI, based mostly on symbolic systems. Among them, Geoffrey Hinton, a University of Toronto professor, who was convinced that DL "had to be the future of AI" \parencite{goldman10YearsLater2022}. He and some of his colleagues – particularly Yan LeCun and Yoshua Bengio – were at the forefront of the DL revolution. \textcite{waldropWhatAreLimits2019} describes what happens during this contentious period of the 80s and early 90s: 

\begin{quote}
\small
Today’s deep-learning revolution has its roots in the "brain wars" of the 1980s when advocates of two different approaches to AI were talking right past each other. On one side was an approach—now called "good old-fashioned AI"—that had dominated the field since the 1950s. Also known as symbolic AI, it used mathematical symbols to represent objects and the relationship between objects. […] But by the 1980s, it was also becoming clear that symbolic AI was impressively bad at dealing with the fluidity of symbols, concepts, and reasoning in real life. In response to these shortcomings, rebel researchers began advocating for artificial neural networks, or connectionist AI, the precursors of today’s deep-learning systems \parencite[p. 1075]{waldropWhatAreLimits2019}.
\end{quote}

The Canadian Institute of Advanced Research (CIFAR), a research funding organization based in Toronto that finances basic research with a high-risk, high-reward philosophy was, since its foundation in the 1980s, consistently interested in the advancement of AI and was at the forefront of the upsurge of ML technologies  \parencite{chahDeepRabbitHole2019}. CIFAR became the institutional setting on which those "rebel researchers" were able to join forces and form their own epistemic culture. CIFAR provided access to symbolic resources in the form of positions - like fellowships - for some of them, but also material resources in the form of funding (not in a significant amount) for conferences, meetings and summer schools, all of them part of the micromobilization scenarios needed to recruit new members, discuss novel ideas, and in general advance their agenda.

\subsection{Open Science}
\label{subsec:open_sci}

In a series of works in the early 2000s, Paul A. David elaborated on the concept of \emph{open science} contrasting it with the increasing reliance on Intellectual Property Rights (IPRs) in the production of science \parencite{davidEconomicLogicOpen2003, davidCanOpenScience2004, davidDigitalTechnologyBoomerang2005}. Open science  in its original conception, takes a descriptive sense, referring to a new paradigm born:

\begin{quote}
\small
with Renaissance mathematics, the cultural ethos and social organization of western European scientific activities during the late sixteenth and seventeenth centuries […] --departing from the previously dominant regime of secrecy in the pursuit of ‘Nature’s secrets’ \parencite[p. 15]{fadfc20e-20f4-3bc9-8e9f-d067328c5712}.
\end{quote}

This new paradigm shaped the organization of the scientific endeavor in the West, including the imperatives of public disclosure of discoveries, and the methods that lead to those discoveries. This openness was supported by a public (open) system of Universities and research communities, and a series of norms, including communalism, universalism, desinterestedness, originality and skepticism \parencite{mertonSociologyScienceTheoretical1973}, that created a reward system based on collegiate reputation that was achieved by validated claims to priority in discovery or invention \parencite{davidEconomicLogicOpen2003}.

Since its inception, the concept positions itself as opposed to a "closed" science based on IPRs like patents and copyrigts, that jeopardize the traditional ethos of open science. Scholars like Dasgupta \& David \parencite{parthaNewEconomicsScience1994, davidCanOpenScience2004} have warned about the social and economic problems that might arise from the enclosure of scientific knowledge within the framework of IPRs. Among others potential hazards, they mention a suboptimal level of production of basic science, which have the greatest spillovers; and scientists getting more and more engaged in duplicitous work, unable to access a big part of the stock of codified knowledge in the form of patents created by a culture of "intellectual capitalism" \parencite{davidCanOpenScience2004}.

More recently, the advent of the digital technologies, and in particular the internet, has given rise to a slightly different conceptualization of open science that lies "between the age-old tradition of openness in science and the tools of information and communications technologies (ICTs) that have reshaped the scientific enterprise" \parencite{oecdMakingOpenScience2015}. In this conception, open science is (loosely) defined as "efforts by researchers, governments, research funding agencies or the scientific community itself to make the primary outputs of publicly funded research results – publications and the research data – publicly accessible in digital format with no or minimal restriction as a means for accelerating research; these efforts are in the interest of enhancing transparency and collaboration, and fostering innovation" \parencite{oecdMakingOpenScience2015}. In this conception, open science is part of an ‘open ecosystem’ that encompasses open access journals, open data, open software, open collaboration, open peer review, among others.

One element of the open science ecosystem is particularly relevant for this work is open data. The European Commission provides a definition, stating that "Open Data is data that is made available by (public) organisations, businesses and individuals for anyone to access, use and share" \parencite{europeancommissionAIOpenData2018}. Access to data can have many advantages or purposes. Data (for example from public records) can be used for original research; for reproducing and validating (or not) existing knowledge; or to explore new research avenues.

Certainly data has become relevant for many areas of scientific inquiry; but for DL in particular, data is a \textit{conditio sine qua non} for its very existence, since the neural networks on which it is built rely on the availability of large amounts of data. Open data, built collaboratively, clearly labeled and free to access on the Internet was key to the emergence and eventual dominance of DL within AI  \parencite{martensImpactDataAccess2018}.

\subsection{The GPU Revolution}
\label{subsec:gpu_rev}

To achieve the status of a dominant paradigm within the machine learning (ML) literature, deep learning (DL) had to overcome a series of systemic bottlenecks that impeded its development. Although the theoretical basis for AI, based on ML algorithms and convolutional neural networks, was established in the 1980s, the first significant bottleneck from the 1990s onwards was the availability of large amounts of training data necessary to "feed" the DL models.

Besides the availability of training data, the development of DL depended also on the increase of computing power. Because of the enormous amounts of data to be processed and the increasing complexity of the algorithms used to analyze that data, computing capacity became a bottleneck for the development of DL until the second decade of the twenty-first century. Despite the excitement with neural networks in the 80s and 90s "computers were not powerful enough to allow this approach to work on anything but small, almost toy-sized problems" \parencite{deanDeepLearningRevolution2020}.

The paradigm of general-purpose computing on GPU cards, originally used for gaming, "because of GPU cards’ high floating point performance relative to CPUs, started to allow neural networks to show interesting results on difficult problems of real consequence." \parencite{deanDeepLearningRevolution2020}. In particular, from mid 90s the performance of GPUs increase significantly with 2D and 3D acceleration on the same unit. The coming on the market of Nvidia GeForce 256 in 1999 is usually considered the turning point of the industry. The consequence of those technological advances was that "computers finally started to become powerful enough to train large neural networks on realistic, real-world problems" \parencite{deanDeepLearningRevolution2020}. 

By 2009 when CIFAR-10 was launched, the technological conditions for its use and exploitation were mature. CIFAR-10 became a dataset that could be manipulated on personal computers (see Table \ref{tab:comp_requirement} below for a comparison of computing requirements of mostly used OLDs), and used as a toy-dataset to train and improve algorithms that could later be used on more complex datasets such as ImageNet. 

\subsection{AI as a Technoscience}
\label{subsec:ai_technosci}

Different from other intellectual movements, AI in general and DL in particular can be better understood as a technoscience, located in the intersection between traditional scientific research and technological applications \parencite{raimbaultEmergenceTechnoscientificFields2021}. Different from pure sciences, the quest of technoscience is not only motivated by a search for new knowledge in an abstract way, but to the solution of practical problems. In fact, "technoscience is ‘face to face’ with the things. It is less interested in what they are or what regular behaviors they are naturally disposed to exhibit, and more interested in what they can become or what they might offer" \parencite{bensaude-vincentMattersInterestObjects2011}. Usually, in the policy arena, technosciences are often referred to as Pasteur Sciences \parencite{stokes_pasteurs_2011}.

In the case of DL, the practical applications go from medical image analysis, language translation, object detection for autonomous vehicles, content filter, and many others \parencite[\textit{Cfr.}][]{bengioDeepLearningAI2021}. It is no surprise then that technosciences develop strong links with industry, as shown by the fact that most of the academics that ignited the DL revolution ended up joining the industry (Geoffrey Hinton in Google; Yan LeCun in Facebook, and Yoshua Bengio in his own venture, Element AI).

The theoretical implications of considering AI as a technoscience and DL as a paradigm within it, mean a deviation from a traditional analysis of a scientific discipline. For example, even if traditional criteria, like the priority of discovery \parencite{mertonPrioritiesScientificDiscovery1957} still apply, it does in a different way. More than publications in academic journals, breakthroughs are shown through \emph{competitions}, in which the new techniques (in this case the algorithms) are tested against a certain benchmark to validate the real-world performance of the \emph{discovery}. In other words, the understanding of causal mechanisms in the aim of proving or disproving a certain theoretical perspective become secondary, while practical (technological) results are of the utmost importance.

This is very clear in the case of DL. \textcite{bengioDeepLearningAI2021} highlight that "DL scored a dramatic victory in the 2012 ImageNet competition, almost halving the error rate for recognizing a thousand different classes of object in natural images". Other authors like \textcite{lecunDeepLearning2015} and \textcite{schmidhuberDeepLearningNeural2015}, also underscore the performance of the algorithms in those competitions as the most important milestones in the paradigm shift. 2012 became to be know as the year of the "DL revolution" The practical performance becomes then more relevant than the actual understanding of the mechanism that drive those results. In fact, "once a DL system has been trained, it’s not always clear how it’s making its decisions" \parencite{waldropWhatAreLimits2019}. Articles in scientific journals play a role in the development of this new technoscience but other forms of knowledge diffusion and creation of reputation such as conference presentations, conference proceedings and patents are of similar or higher importance \parencite{franceschet_role_2010, meyer_viewpointresearch_2009, fortnow_viewpointtime_2009}. For example, in the full sample of 37,242 articles identified in this paper as composing the relevant literature in DL for computer vision and image recognition around 55\% were conference proceedings. Science and technology are interlinked and publications and proceedings are cited more frequently and faster in patents protecting downstream technological development.  To try to capture developments in the field we must therefore use both patents and publications because the latter would only provide a limited representation of the evolution of the science. 


\section{Institutional background}
\label{sec:inst_background}

\subsection{Winning the \emph{brain wars}: The emergence of DL as a dominant paradigm within AI}
\label{subsec:emergence_dl}

DL is a subfield of ML that is inspired by the structure and function of the brain’s neural networks. It involves training artificial neural networks, which are composed of layers of interconnected nodes or "neurons" to learn from large amounts of data. These networks can be used to perform a wide variety of tasks, such as image and speech recognition, natural language processing, and decision making. DL is often used in combination with other techniques, such as reinforcement learning, to solve complex problems \parencite{lecunDeepLearning2015}.

DL is a subset of ML, which in turn is defined as "concerned with the question of how to construct computer programs that automatically improve with experience" \parencite{mitchellMachineLearning1997}. ML, on the other hand, is one of the most important approaches of AI.
AI, ML and DL are interrelated but differ from each other. \textcite{chahDeepRabbitHole2019} makes the following distinction: 

\begin{quote}
\small
Although the three terms —AI, ML and DL— are intricately linked, nuanced differences in their specific definitions can make the difference between whether the term is used precisely or whether the actual operations on the ground are obfuscated. The dynamic definition of AI affects what state-of-the-art advances are considered as AI for a particular time and place. ML is primarily concerned with training machines to learn from data, following closely the original definition by Arthur Samuel in 1959. To implement the ever-changing state-of-the-art techniques that exhibit AI capabilities, DL is one of the most popular sets of ML techniques in use today \textcite[p. 3]{chahDeepRabbitHole2019}.
\end{quote}

The concept of DL was coined in 2006 by Geoffrey Hinton and his colleagues \parencite{lecunDeepLearning2015, chahDeepRabbitHole2019}. However, the concepts on which this technology is based, started to develop long before with the work on artificial neural networks in the 1940s \parencite{schmidhuberDeepLearningNeural2015, chahDeepRabbitHole2019} that evolved in the 1980s into the covolutional neural network \parencite{fukushima_neocognitron_1980}.

Despite being a growing field, DL was at the peripheries of AI for many decades. According to Yan LeCun, one of the main proponents and intellectual architects of the DL revolution "In the late 1990s, neural nets and backpropagation were largely forsaken by the machine-learning community and ignored by the computer-vision and speech-recognition communities" \parencite{lecunDeepLearning2015}. By 2006, a paper by \textcite{hintonFastLearningAlgorithm2006} reignited the interest, by showing the possibility of using DL to achieve state-of-the-art results (1.25 percent error rate) in recognizing handwritten digits. In 2012, a DL algorithm developed by \textcite{krizhevskyImageNetClassificationDeep2017} won the ImageNet classification competition, one of the most challenging image recognition databases at the time. AlexNet, the winning algorithm, became a milestone that positioned DL as the dominant paradigm within ML and consequently within AI.

AlexNet was developed by Alex Krizhevsky in collaboration with Ilya Sutskever and Geoffrey Hinton. Both Krizhevsky and Hinton were also behind the creation of CIFAR-10, which became the basis for the development of the AlexNet algorithm (Fergus, 2022, Interview No. 2; Bengio, 2022, Interview No. 5).


\subsection{The development of Open Labeled Datasets and CIFAR-10}
\label{subsec:develop_olds}

CIFAR support rational aimed to finance risky basic research with networking type of money and provided the institutional space for alternative research approaches. In 2004, CIFAR supported a diverse group of unorthodox scientists led by Geoffrey Hinton into pursuing an ambitious program in AI called Neural Computation and Adaptive Perception Program (NCAP) (Silverman, 2022, Interview No. 7; \cite{brownellHowArtificialIntelligence2016a}).

On his account of the beginning of the NCAP program, Prof. Silverman highlights how this group of people were full of new ideas, but did not have the institutional spaces to present and discuss them: 

\begin{quote}
\small
But when I’m trying to create a scenario, where, as they spoke, […] that they sort of, had come together as a group, informally because they didn’t have anybody to talk to in their own departments. There they were, they had their own disciplines, they made it into, they had faculty appointments, they were achieving in their own departments, but basically, their interests had taken them in a much broader, different way [to] understand how the brain processes information, not really staying in a single lane, if you know what I mean. And so that, that, that was that resonated with me." He continues saying that they thought "We’re smart, and nobody wants us. Because we’re trying to work on this really tough problem. (Silverman, 2022, Interview No. 7).
\end{quote}

CIFAR became then the institutional space that provided with basic resources for that group of people to come together and start exploring their common interests. Geoffrey Hinton was joined in his research effort by Yoshua Bengio, and Yann LeCun. Their work became a seminal piece in the paradigm shift that saw DL become the dominant approach in AI. "Their work together led to a number of advances, including a breakthrough AI technique called DL, which is now integral to computer vision, speech recognition, natural language processing, and robotics" \parencite{farrowTuringAwardHonours2019}. Because of this work, they received the A.M. Turing Award, considered as the "Nobel Prize of Computing".\vspace{0.5cm}

\noindent\textbf{Open Labeled Datasets}.
Before 2009, the two main datasets used for computer vision and object recognition tasks were CALTECH-101 and MNIST (Modified National Institute of Standards and Technology database). CALTECH-101  is a dataset that contains pictures of objects belonging to 101 categories. It contains about 40 to 800 images per category, with most categories containing about 50 images. It was collected in September 2003 by Fei-Fei Li, Marco Andreetto, and Marc'Aurelio Ranzato \parencite{li_andreeto_ranzato_perona_2022}. MNIST was one of the first annotated datasets used in ML models and consists of large collection of images of handwritten digits taken from the CENSUS bureau, it contains 60,000 black and white training images and 10,000 testing images, it was released by AT\&T Bell Labs in 1998 \parencite{Goltsev2004}. The CIFAR team worked mostly with MNIST. 

In 2006, Rob Fergus, Antonio Torralba and William T. Freeman released the "80 million tiny images", a new dataset that could overcome some of the limitations of the existing ones. They automatically collected low-resolution images from different search engines (Altavista, Ask, Flickr, Cydral, Google, Picsearch and Webshot) and loosely labeled with one of the 53,464 non-abstract nouns in English, as listed in the Wordnet lexical database \parencite{torralba80MillionTiny2008}. However, in the original paper in which he introduces the CIFAR databases, \textcite{krizhevsky2009learning} mentions that he is trying to solve 

\begin{quote}
\small
A […] problematic aspect of the tiny images dataset is that there are no reliable class labels which makes it hard to use for object recognition experiments. We created two sets of reliable labels. […]. Using these labels, we show that object recognition is significantly improved by pre-training a layer of features on a large set of unlabeled tiny images. \parencite[p. 1]{krizhevsky2009learning} 
\end{quote}

In 2008, Geoffrey Hinton, along with two of his students, Vinod Nair and Alex Krizhevsky, had the idea to manually label a sub-set of the "80 million tiny images" (Fergus, 2022, Interview No. 2), to address one of the problems they encounter when using this large dataset for unsupervised training. To label the images, they took advantage of the NCAP Summer School that took place in August 2008. The students that participated occupied some of the time labeling the images according to a protocol written by Alex Krizhevsky and Rob Fergus (Fergus, 2022, Interview No. 2). Rob Fergus account of the process is that 

\begin{quote}
\small
the data [From the 80 million tiny images] need[ed] to be manually cleaned in order to make a sort of good supervised training dataset. And Geoff wants you to do this. And so he organized the CIFAR summer school, he got all the summer school students sitting down. So how did it work? So I think Alex Krizhevsky and I wrote a labelling routine to actually, you know, have labelling interface where all the students would sit down, and we will go through the images, cleaning them up, and we decided that, Geoff decided he was going to pick, you know, these 10 Super categories, and then each one of which had subcategories that form the CIFAR 100. (Fergus, 2022, Interview No. 2).
\end{quote}

In that way, the CIFAR team created the CIFAR-10 dataset, which consists of 60,000 32x32 color images in 10 classes, with 6,000 images per class; and the CIFAR-100 is just like the CIFAR-10, except it has 100 classes containing 600 images each.  The datasets overcame some of the problems encountered in older open datasets, while keeping an architecture similar to that of MNIST. These two datasets were subsequently used to train computer vision algorithms through a procedure called \emph{supervised learning}. \textcite{lecunDeepLearning2015} explain supervised learning as follows: 

\begin{quote}
\small
Imagine that we want to build a system that can classify images as containing, say, a house, a car, a person or a pet. We first collect a large data set of images of houses, cars, people and pets, each labeled with its category. During training, the machine is shown an image and produces an output in the form of a vector of scores, one for each category. We want the desired category to have the highest score of all categories, but this is unlikely to happen before training. We compute an objective function that measures the error (or distance) between the output scores and the desired pattern of scores. The machine then modifies its internal adjustable parameters to reduce this error. These adjustable parameters, often called weights, are real numbers that can be seen as ‘knobs’ that define the input–output function of the machine. In a typical deep-learning system, there may be hundreds of millions of these adjustable weights, and hundreds of millions of labeled examples with which to train the machine \parencite[p. 436]{lecunDeepLearning2015}.
\end{quote}

It is clear from this definition that supervised learning requires vast amounts of data and very well labeled, so that the machine can be trained with this clean set of images and thus learn how to recognize objects of the same classes. By 2009, there were not datasets the combined a large number of images, a rigorous labelling process, and well constructed categories that made it easy to manipulate. CIFAR-10 and CIFAR-100 became quickly benchmarks for new algorithms of computer vision using DL. However, it was CIFAR-10 that had the most impact. According to experts’ accounts [Fergus, 2022, Interview No. 2] beyond the reliability, the size of this dataset and its simplicity were key for its success. In fact, it was light enough so that it was easy to manipulate and work with, specially to train algorithms that require a lot of computer power, but had enough data to properly train a neural network. As a corollary, Fergus (2022, Interview No. 2) added that the fact that "every student can run CIFAR-10 on their laptop" make a difference in terms of usage.

CIFAR databases were made available for free on the web from the University of Toronto, very much in line with the Open Data paradigm mentioned above. Along with the other characteristics, the easy availability became one the distinctive characteristics and main advantages of the CIFAR datasets.

During the same period, other research teams were developing similar image databases. One notable example is ImageNet, created by Fei-Fei Li (currently at Stanford University, formerly at the University of Illinois Urbana-Champaign) and Christiane Fellbaum (Princeton University), and introduced in 2009. ImageNet includes a large collection of labeled object images and rapidly became a benchmark for state-of-the-art computer vision algorithms. The dataset was annotated using Amazon Mechanical Turk (MTurk)\footnote{MTurk is a crowdsourcing platform provided by Amazon Web Services that connects businesses and researchers with a global pool of remote workers. It is designed to handle tasks that are difficult for machines but relatively easy for humans, such as labeling datasets.}. For a more detailed description of ImageNet and other similar datasets, see Section \ref{subsec:data} and Table \ref{tab:datasets_descriptives}.

In the following sections, we will evaluate whether OLDs, particularly CIFAR-10, have influenced the development and evolution of DL as a technoscience.  Specifically, we will explore whether the unique characteristics of these open datasets contributed to the development of the foundational Convolutional Neural Network (CNN) architectures that sparked the DL revolution.

\section{Methods and data}
\label{sec:method_data}

\subsection{Method}
\label{subsec:method}

For the empirical analysis we use a mixed methods approach (see Appendix \ref{app:qualitative_methodology} for details). The initial step was to conduct semi-structured interviews with relevant actors, including prominent academics working on the field of AI and DL, as well as CIFAR personnel linked directly or indirectly to the creation of CIFAR-10. Two kinds of interviews were conducted: general interviews with academics working on AI, not necessarily related to CIFAR datasets, with the aim of getting an understanding of the field and some general features that practitioners might look for in a training dataset; and more specific interviews with strategic individuals that were directly or indirectly related to the development of the CIFAR datasets. In total we conducted 7 interviews, out of which 2 were with field experts not linked to CIFAR; and 5 with persons linked to CIFAR.

Second, we surveyed researchers and practitioners who referred to CIFAR datasets in their articles. The survey aims to validate the information obtained from the interviews and develop a broader assessment of the impact of the CIFAR-10 on DL and Computer Vision. The sample population includes corresponding author from the subset of papers referencing CIFAR datasets out of 6,060 paper we were able to retrieved 3,033 valid emails (see Qualitative Methodology Appendix \ref{app:qualitative_methodology} for the response analysis and survey questions). We were able to collect 295 answers to the survey, which corresponds to a response rate of 9.4\%. The response analysis indicates that our sample is representative for most variables available for the population (total and valid email).

Finally, we concentrated on publications in the ML literature that utilized OLDs for model training between 2010 and 2022, following the release of the CIFAR and ImageNet datasets. We conducted an econometric analysis aimed at examining the relationship between the use of specific OLDs and receiving citations from patent and scientific publications. Our method involved comparing publications that referenced CIFAR-10 and ImageNet with those that did not but used one of the other similar labeled datasets, while controlling for various confounding factors. Specifically, we estimate regressions of the following models:

\begin{equation}
\label{eqn:main_cifar}
\begin{split}
\mathbb{E}[Citations_{jst}] = & \exp(\beta_1 \mbox{CIFAR-10 (only)}_{jst} + \beta_2 \mbox{CIFAR-10 (others)}_{jst} + \\
& + \beta_3 X_{jst}  + \alpha_j + \delta_s + \gamma_t + \varepsilon_{jst} )
\end{split}
\end{equation}

\begin{equation}
\label{eqn:main_cifar_imagenet}
\begin{split}
\mathbb{E}[Citations_{jst}] = & \exp(\beta_1 \mbox{CIFAR-10 (only)}_{jst} + \beta_2 \mbox{CIFAR-10 (others)}_{jst}  + \beta_3 \mbox{ImageNet}_{jst} + \\
& + \beta_4 X_{jst}  + \alpha_j + \delta_s + \gamma_t + \varepsilon_{jst} )
\end{split}
\end{equation}

For each focal paper published of type $j$ (scholarly journal or conference proceeding), in a scientific area $s$ and year $t$, we measure its outcome using different metrics that capture their technological and scientific citation impact. Our main explanatory variables are binary indicator s that assumes value 1 if the paper mentioned only CIFAR-10, CIFAR-10 and other datasets or ImageNet. Our main dependent variables which we use as measure of technological and scientific relevancy are the total number of patents citing the articles and the total number of scientific citations.

To ensure a fair comparison of our articles, we develop an empirical design that allows us to compare similar ML/DL articles that differ only in their use of the CIFAR-10 dataset for model training. Thus, we incorporate a set of control variables, $X_{jst}$, which describe various characteristics of the focal papers and are related to citation impact. These controls include the number of authors, the number of references, the presence of international collaboration, and the share of authors affiliated with  companies, as these factors may influence both the use of CIFAR-10/ImageNet and citation impact.

Additionally, we include as control variables observable characteristics of the labeled datasets used in the papers, such as the number of OLDs mentioned, the number of modalities (i.e., different types of data beyond images, such as text and audio), and the number of ML prediction tasks performed with these datasets\footnote{Further considerations regarding tasks are provided in Section \ref{subsec:data}. As the field advances, datasets are increasingly applied to new tasks. Nonetheless, we use the number of unique tasks identified for each dataset up to July 2023 as a proxy for the breadth of application of a specific OLD in these fields.}. These dataset characteristics serve as proxies for the types of DL models being developed and refined, helping us partially control for factors that could simultaneously affect both citation counts and the use of the primary OLDs under investigation.

We then estimate a fixed-effects Poisson model, including the independent and control variables discussed above and a set of fixed effects which includes type of publication $\alpha_j$, scientific fields $\delta_s$ and calendar year $\gamma_t$ to control for time-invariant features that may also explain citation impact. In the robustness checks we use Negative Binomial model and different variable operationalization and sample definitions to test the consistence with. We use the same setting for both technology and scientific citations impacts.

We do a split sample analysis focusing on two periods 2010-2014 and 2015-2022. We identify 2014 a year of structural change in the motivations to use CIFAR-10 because it was the year where state-of-the-art DL models consistently surpassed human-level accuracy in image classification tasks\footnote{The human error rate on CIFAR-10 is estimated to be around 6\%. Working papers originally published in the end of 2014 achieved an error rate of around 3-4\% \parencite{graham_fractional_2015}.}. Thus, from 2015 onward CIFAR-10 was essentially a "solved problem": prediction accuracy and error rates of trained models were comparable to humans-levels. We perform the split sample analysis for both model \ref{eqn:main_cifar} and \ref{eqn:main_cifar_imagenet}: while the former allow us to estimate how many citations papers using CIFAR-10 are expected to receive compared to papers that do not (conditional on observable paper characteristics), the latter model enables us to assess the expected citations for papers using CIFAR-10 or ImageNet in comparison to others. By comparing the coefficients of the CIFAR-10 and ImageNet variables, we can evaluate how papers utilizing datasets of different dimensions and complexity perform in terms of citations.

\subsection{Data}
\label{subsec:data}

We constructed a novel and unique dataset that includes both detailed bibliometric information on publications and patents in the ML literature on image recognition and object classification and the OLDs used by them to train DL models. Our data collection process involves identifying OLDs similar to CIFAR-10, the most used dataset in the ML literature indexed at the \emph{Papers with Code} website\footnote{The Papers with Code platform offers access to all its contents under the CC BY-SA licence, which can be downloaded from the website paperswithcode.com/about. In the context of this study, we obtained the data to perform our analysis on July 17, 2023.}. Papers with Code is a platform launched by ML practitioners in July 2018 to share open resources associated with AI development, with a focus on ML \parencite{MartnezPlumed2021}. Presently, Papers with Code comprises approximately 135 thousand research papers, encompassing over 11 thousand benchmarks that address 5,000 distinct tasks. 

First, we identified all the tasks mentioned in Papers with Code that involved models trained with CIFAR-10. These tasks refer to different types of predictions or inferences made using models trained on specific data. For example, image datasets can be used to train ML models to solve tasks such as image classification, object detection, and anomaly detection. See Table \ref{tab:task_list} in Appendix \ref{app:data_construction} for a detailed list of these tasks related to CIFAR-10. Our analysis identified 46 tasks related to CIFAR-10 that have been utilized in developing ML models. We then used these tasks to identify other datasets used to train models that handle at least one task similar to those involving CIFAR-10. For information on the most frequently used datasets, see Table \ref{tab:datasets_descriptives} in Appendix \ref{app:data_construction}.

The Papers with Code platform aggregates ML research papers that are openly accessible and accompanied by source code, mostly sourced from the open access online repositories like \emph{arXiv}. To obtain better bibliographic and citation coverage, we collected scientific publications using a list of annotated datasets of interest on \emph{Scopus}, Elsevier's citation database. Specifically, we used the Scopus database\footnote{We utilized pybliometrics, a Python package for accessing the Scopus API. See \textcite{rose_pybliometrics_2019} for further details on the package.} to find any publication that mention a datasets in our list in their titles, abstracts, or keywords. As annotated datasets frequently have variations or subsets tailored to specific tasks, we searched for the full names, shortened names, and variants of each dataset.This approach allowed us to identify 37,242 Scopus indexed scientific publications that mention a total of 264 unique labeled datasets\footnote{In the remainder of this work, we will consider publications that \emph{mention a dataset in the title, abstract, or keywords} and \emph{use a dataset to train ML models} as equivalent. We acknowledge that this approach has limitations, as authors might not always mention the dataset used to train their models in these sections, or may mention only some of them. We explored using backward citations to introductory papers, but this method proved less precise and biased. Despite these limitations, we believe that referencing datasets in the title, abstract, or keywords provides a strong indication of the dataset's importance in the publication and is the most reliable way to identify relevant papers in this literature.}. 

\begin{figure}[ht]
\centering
\caption{Distribution of Publications by Subject Area}
	\includegraphics[width=\linewidth]{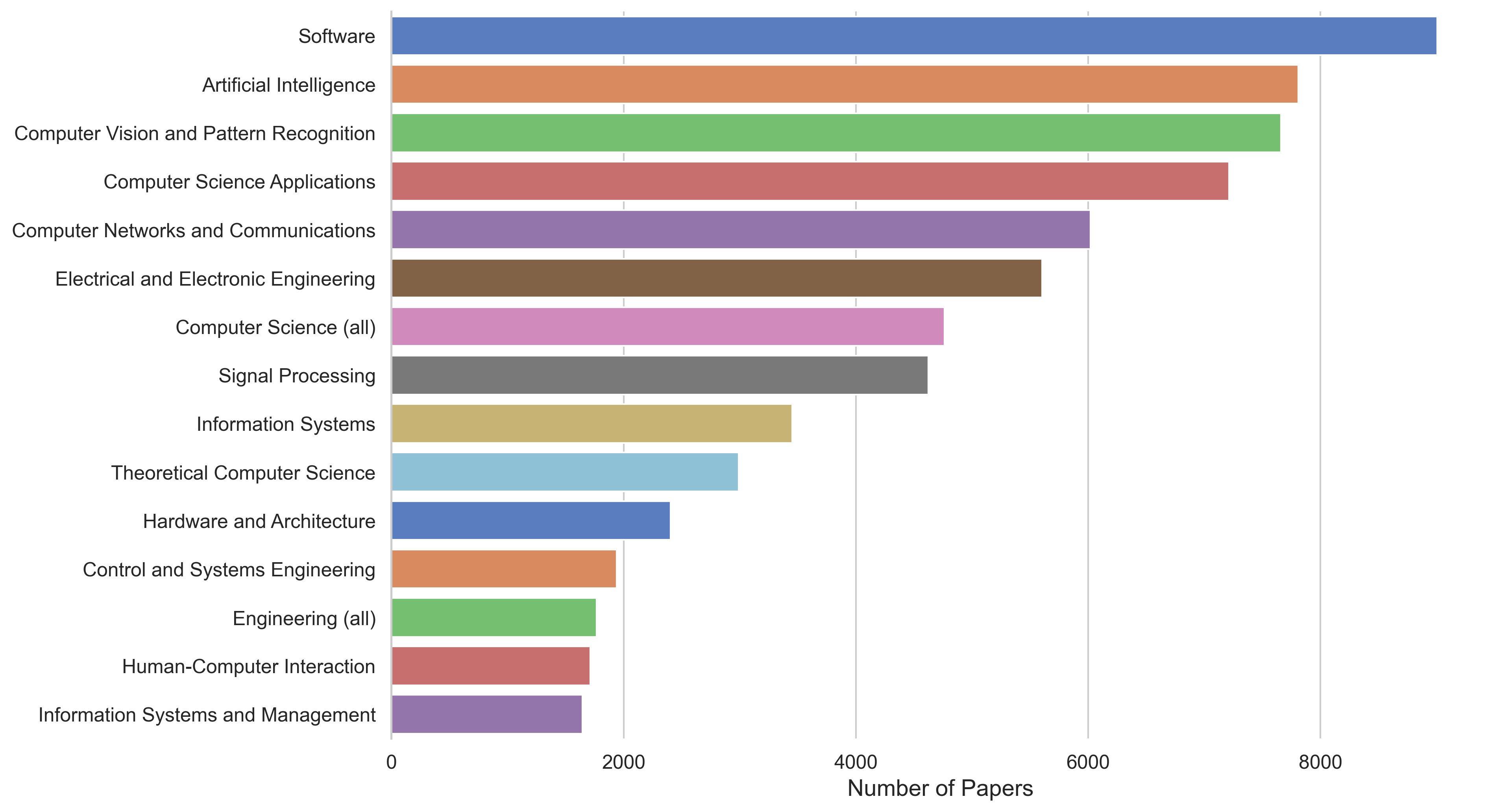}
	
	\label{fig:subject_areas}
        
        \floatfoot{\emph{Notes}: This figure displays the distribution of All Science Journal Classification (ASJC) codes assigned by Scopus to each paper based on its journal, conference, or other publication venue. Note that papers may be assigned multiple ASJC codes.}
\end{figure}

Using Scopus' All Science Journal Classification (ASJC) codes, we identified the scientific fields of the publications that frequently use the datasets in our study, as shown in Figure \ref{fig:subject_areas}. Unsurprisingly, most of the papers in our sample are published in journals from the fields "Software", "Artificial Intelligence", "Computer Vision and Pattern Recognition" and "Computer Science Applications". Interestingly, there is also significant representation from "Electrical and Electronic Engineering", "Hardware and Architecture", and "Control and Systems Engineering", suggesting that these datasets have technological applications beyond strictly computer science disciplines.

\begin{figure}[ht]
\centering
\caption{The Rise of Annotated Image Datasets}
	\includegraphics[width=\linewidth]{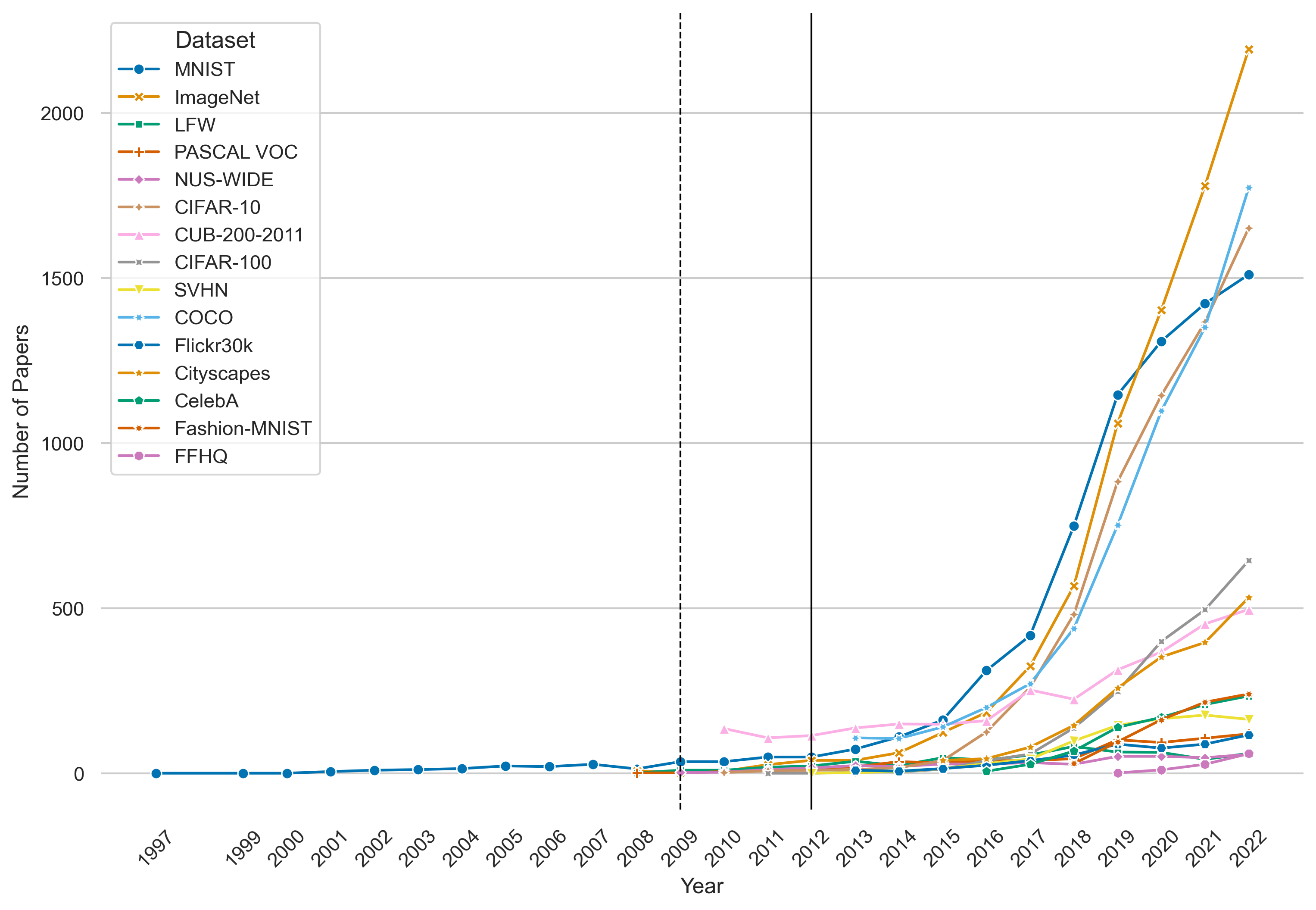}
	
	\label{fig:yearly_pubs}

	\floatfoot{\emph{Notes}: This figure illustrates the annual growth in the number of publications referencing the 15 most commonly used annotated image datasets. The vertical dashed line denotes 2009, the year ImageNet and CIFAR-10 were introduced, while the solid horizontal line marks 2012, the year of the DL revolution.}
\end{figure}

To understand the technological developments linked to papers using OLDs, we complement the Papers With Code and Scopus data with patent-publication citation links from the \emph{Reliance on Science} dataset \parencite{marx_reliance_2020, marx_reliance_2022}. We gather all the front-page and in-text citations of patents granted worldwide that reference scientific papers in our sample. Since our focus is on technological developments rather than intellectual property concerns, we aggregate these patents into patent families using data from the EPO-PATSTAT database (Autumn 2023 version). We identified 31,170 patents families citing 14,435 papers from our sample of journal articles and conference proceedings, either on the front-page, in-text, or both\footnote{We have decided to include in-text citations as well, because we believe that non-patent literature cited only on the front page would not adequately cover less conventional scientific publications, such as conference proceedings, data introductory papers, and other sources likely to be found throughout the full patent text.}.

Considering the relative distribution of the datasets used by publications in our sample, Figure \ref{fig:yearly_pubs} illustrates the number of yearly publications citing the fifteen most common datasets in computer vision and image recognition literature using ML, accounting for 81.67\% of papers. The significant increase, particularly after 2012 — the year of the DL revolution — was primarily driven by papers citing ImageNet, MNIST, COCO, and CIFAR-10\footnote{For more information on these datasets, visit their official sites: \href{https://image-net.org/}{ImageNet}, \href{https://yann.lecun.com/exdb/mnist/}{MNIST}, \href{https://cocodataset.org/}{COCO}, and \href{https://www.cs.toronto.edu/~kriz/cifar.html}{CIFAR-10}.}, which account for 62.71\% of the total sample. CIFAR-10 and ImageNet represent 33.23\% of the publications in our sample and show similar trends: both were introduced in 2009 by young scholars who believed in the potential of labeled datasets to advance DL. These datasets consist of natural images and have been used for various tasks such as image classification, object recognition, and image generation\footnote{According to our analysis using the Papers with Code platform, CIFAR-10 has been utilized in 46 unique tasks and ImageNet in 64 tasks, with 24 tasks overlapping between the two (52.17\% of CIFAR-10 tasks). See Table \ref{tab:task_list} for details.}. They are the first two most used databases in Papers with Code. The main difference between them is that CIFAR-10 is much smaller, with 60,000 images and 10 categories, compared to ImageNet's 14,197,122 images and over 20,000 categories. Additionally, ImageNet was central to the ImageNet Large Scale Visual Recognition Challenge (ILSVRC) from 2010 to 2017, which incentivized the development of ML models using this dataset.

Regarding the other two datasets, MNIST is a dataset of handwritten digits introduced in 1998. It comprises 60,000 training examples and 10,000 test examples, with digits that have been size-normalized and centered in fixed-size images. COCO, introduced in 2014 by a Microsoft group, contains images of complex everyday scenes with common objects in their natural context. It includes 91 object categories, 82 of which have more than 5,000 labeled instances, totaling 2,500,000 labeled instances in 328,000 images. Unlike ImageNet, COCO has fewer categories but more instances per category, and it is used for tasks such as detection, segmentation, and captioning.

\input{Tables/datasets_computational_requirements}


Table \ref{tab:comp_requirement} provides an overview of the computing capacity required for the four most commonly used datasets.\textbf{ }Using the best supercomputer in 2024 and a typical research laptop as benchmarks, and referencing state-of-the-art algorithms that outperform humans on CIFAR-10 and MNIST, it is clear that MNIST is now too simple for complex tasks, while COCO remains too challenging to solve fully. Therefore, we believe ImageNet is the best candidate for comparison with CIFAR-10. It is important to note that CIFAR-10 is both sufficiently complex and manageable in size. For instance, training a model to achieve human-level accuracy of 94\% on CIFAR-10 would take an average research laptop about 10 seconds \parencite{jordan_94_2024}.

To analyze the citation patterns described in Section \ref{subsec:method}, we consider only publications between 2010 and 2022, reducing our sample to 36,859 publications\footnote{Few papers used large labeled datasets before the DL revolution in 2012, thus we lose very few papers in this step}. We chose 2010 as the starting year because it marks the introduction of the two most popular OLDs to the ML community: CIFAR-10 and ImageNet. We further restrict our sample to conference proceedings and journal articles, as these are the types of publications where we expect to see ML models trained using OLDs, excluding review papers and data introduction papers. This restriction leaves us with a sample of 35,705.

Since we want to compare similar articles, we removed those lacking fundamental bibliometric information used as control variables\footnote{We have 28 publications missing author information, 1,761 missing references, 1,074 missing affiliation information, and 4 missing subject area information. We also exclude 2 papers that came from dataset that are described in Papers With Code, but do not have any paper indexed to it in the platform.}. Additionally, not all publications indexed by Scopus can be found in Reliance on Science and vice versa, due to duplicated IDs and other issues. Thus, to ensure reliable information about patent citations, we drop from our sample 4,944 conference proceedings and journal articles for which we cannot confirm the number of patent citations received. This leaves us with 28,416 papers and proceedings\footnote{We perform robustness checks using the sample without excluding those publications in Appendix \ref{app:robustness}.}.

Among these, we identified 23 papers (top 0.1\% in the citation distribution of the full sample and 0.08\% in the restricted sample) as outliers, which received citations orders of magnitude higher than the others\footnote{The most cited paper in our sample is "Deep Residual Learning for Image Recognition," published in the Proceedings of the IEEE/CVF Conference on Computer Vision and Pattern Recognition by a Microsoft group. This paper introduced the Residual Networks (ResNet) architecture, a key component in modern DL models (e.g., Transformers, AlphaGo Zero), and has received 95,139 citations. It was the most cited paper globally for five consecutive years, according to Google Scholar (source: \href{https://www.nature.com/nature-index/news/google-scholar-reveals-most-influential-papers-research-citations-twenty-twenty-one}{Nature Index}).}. These publications represent breakthroughs so significant that they are not comparable with the average paper in the field. To ensure the comparability, robustness, and stability of our estimations, we excluded these outliers from our sample.

\input{Tables/summary}

Our final data sample for the econometric analysis includes 28,393 journal articles and conference proceedings, as well as 252 labeled datasets. Table \ref{tab:summary_statistics} provides descriptive statistics for our regression sample. Approximately 15.4\% of the sample cited the CIFAR-10 dataset in the ten years following its release, while 19.9\% cited the ImageNet dataset. However, only 4.1\% cited CIFAR-10 exclusively, indicating that CIFAR-10 is often used alongside other datasets. On average, each paper cites only 1.3 datasets, with the third quartile being 1 dataset. The average number of authors per paper is 4.32, and 24.5\% of the papers include at least one international collaboration (i.e., authors from multiple countries). Private companies are represented as well, with 4\% of authors affiliated with companies (9.4\% of the papers have one company affiliated author). Focal papers have an average of 36 backward citations and 16 forward citations, with considerable variation. The average number of different modalities used is 1.25, indicating that most papers focus solely on images. Finally, the number of unique tasks overlapping with CIFAR-10 tasks across all datasets used in the focal papers is approximately 16, representing 34.5\% of similar tasks. In summary, this sample consists of papers using the most common labeled datasets in computer vision, with tasks closely related to those central to the DL revolution.

\section{Findings}
\label{sec:findings}

In this section we explore the role played by OLDs and in particular of CIFAR-10 on the development of DL using our three sources of information: the qualitative interviews, the survey and the bibliometric data. We use different approaches to triangulate and pinpoint how CIFAR-10 contributed to making DL a dominant paradigm within AI, as well as the factors that explain the widespread use of the CIFAR-10 in industrial and academic settings. 


\subsection{Interviews analysis}
\label{subsec:interviews}

\textbf{Bridging the gap - Dataset characteristics}. From the interviews that we conducted, the first element that prominent practitioners mentioned, was how CIFAR-10 went a step ahead of the MNIST but was more manageable than ImageNet, creating a sort of bridge between those two moments of the development of DL.
Yoshua Bengio (2022, Interview No. 05) mentions how the team at CIFAR had achieved some success with MNIST but "we didn’t have datasets of comparable size for natural images". This was confirmed by Rob Fergus (2022, Interview 02), and Yan LeCun (2022, Interview No. 4) both of whom mentioned that there was a "gap" that CIFAR-10 helped to fill.

An important element that made CIFAR-10 a bridge is that it used a small number of categories, like MNIST, but also natural images, like ImageNet. 

\begin{quote}
\small
It was much harder than the 10 digits [of MNIST], it was much, much harder. So it was useful, but the size was the same 60,000 training examples. So that mean, we could use the same kind of architectures. (Bengio, 2022, interview No. 05).
\end{quote}

That also helps explain why it was CIFAR-10 (and not CIFAR-100) the one that had the most impact: 

\begin{quote}
\small
Yes, CIFAR 10 was the one that really had a big impact. For one it was exactly the same format that MNST, 10 categories. When people started working with CIFAR 100, it was much harder. So there are 100 categories, yeah, but you have the same amount of data so that the accuracy is much worse. So CIFAR 100 has been used, but as far as I know, not nearly as much as CIFAR 10. (Bengio, 2022, Interview No. 05).
\end{quote}


\noindent\textbf{Testing architectures and scaling up}. The second element that emerges from the interviews it that CIFAR-10 was simple enough to test and iterate different algorithms and architectures, without requiring prohibitive amounts of computer power. Those architectures could then be used in more challenging datasets. Yoshua Bengio, Yan LeCun and Rob Fergus insisted, in very similar terms, on the potential of CIFAR-10 for trying different architectures and iterate experiments:

\begin{quote}
\small
So in a way, what I’m trying to say is working with CIFAR-10 we discovered architecture tricks, if you want a methodology, for training deeper networks that Alex was then able to apply to ImageNet. So, yeah, CIFAR-10 was kind of instrumental on the path to the modern revolution of computer vision with DL. (Bengio, 2022, Interview No. 05).
\end{quote}

\begin{quote}
\small
So I think, when you’re trying to develop a new method, you’ve got to be able to iterate experiments quickly. And ImageNet is still, […] a bit too big to do that with. And it turns out that the performance on CIFAR-10 generalizes quite well to other data sets like ImageNet. So, you can prototype on CIFAR-10. And then, you know, get some promising stuff, and then move over to something a bit bigger and harder. (Fergus, 2022, Interview No. 02).
\end{quote}

This characteristic became instrumental in the development of AlexNet, which marked the turning point in the DL revolution:

\begin{quote}
\small
The AlexNet paper, I’m not sure would have happened, had it not been for CIFAR-10. Because otherwise, it would have been very difficult for them [Alex and Ilya] to go directly to the ImageNet dataset, which was quite new at the time, and definitely quite big at the time, too, and challenging to use. (Fergus, 2022, Interview No. 02)
\end{quote}

\noindent\textbf{Pedagogical potential}. The third element that according to the interviewees help explain the success and persistence of CIFAR-10 as relevant tool for DL, is its pedagogic value. Since working on it does not require onerous computational capabilities, it can be easily used for teaching purposes.
Bengio notes that:

\begin{quote}
\small
My students started to use it pretty soon, like, we were hungry for that. And we were aware of it even before it was released, because Geoff [Hinton] was talking about it. And you know, we were in close communication with Geoff [Hinton]. Bengio (2022, Interview No. 05) 
\end{quote}

In the same line of reasoning, Fergus stated: 

\begin{quote}
\small
Once you’ve got a lot more people interested in DL, it was a great sort of introductory data set. I mean, small enough, you can do it in, you know, if you’re teaching a class, you can use it, because every student can run CIFAR-10 on their laptop, more or less. Fergus (2022, Interview No. 02)
\end{quote}

\subsection{Survey analysis}
\label{subsec:survey}

\textbf{Survey and Respondents Description}. The survey has received 295 complete responses, with a total response rate of 9.4\% at the time of closure. The vast majority of respondents (228) hold a Doctorate degree (PhD), and most of the respondents are employed in academia. 20\% of respondents work in industry or in a combination of industry and academia, when we look at the article affiliation we find a much lower share, about 11\%, confirming the importance of mobility of researchers from academia to industry. \vspace{0.5cm} 

\begin{figure}[ht]

\centering
\caption{Survey Results - CIFAR-10 Datasets Impact on DL \& Computer Vision}
	\includegraphics[width=\linewidth]{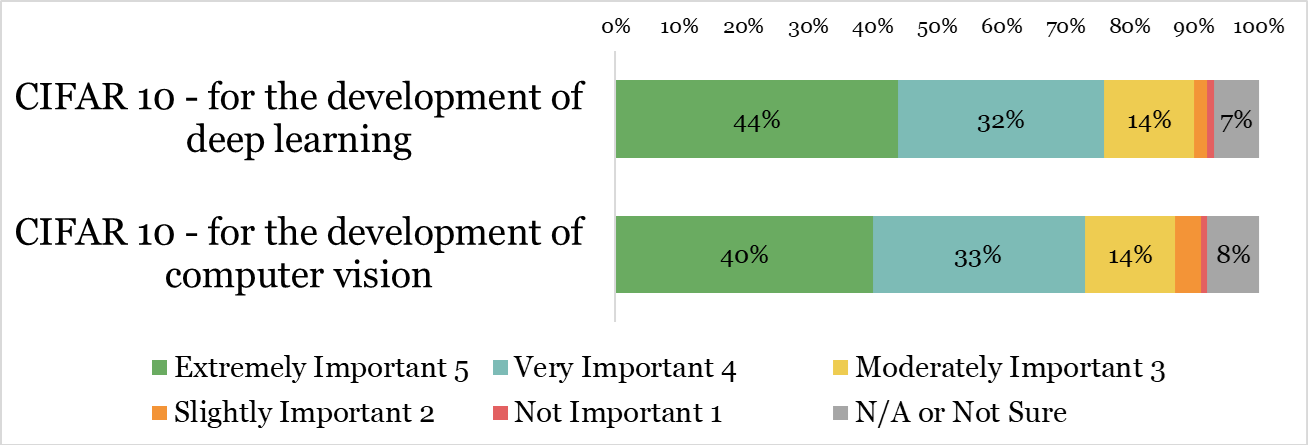}
 
    \label{fig:survey_cifar_impact}
    
	\floatfoot{\emph{Notes}:  This figure shows the distribution of answers for the Impact question.}
\end{figure}

\noindent\textbf{The Importance of the Datasets}. Figure \ref{fig:survey_cifar_impact} reports responses on the importance of CIFAR-10 for the development of DL and Computer Vision were overwhelmingly positive. 76\% believe  CIFAR-10 was very or extremely important for the development of DL and 73\% for the development of Computer  Vision.  44\% considered CIFAR-10 as extremely important for the development of DL in general, not only for Computer Vision. Though CIFAR-10 included labeled images, it is considered important for the development of the general field of DL. \vspace{0.5cm}

\begin{figure}[ht]

\centering
\caption{Survey Results - Comparing CIFAR-10 with Similar Datasets}
	\includegraphics[width=\linewidth]{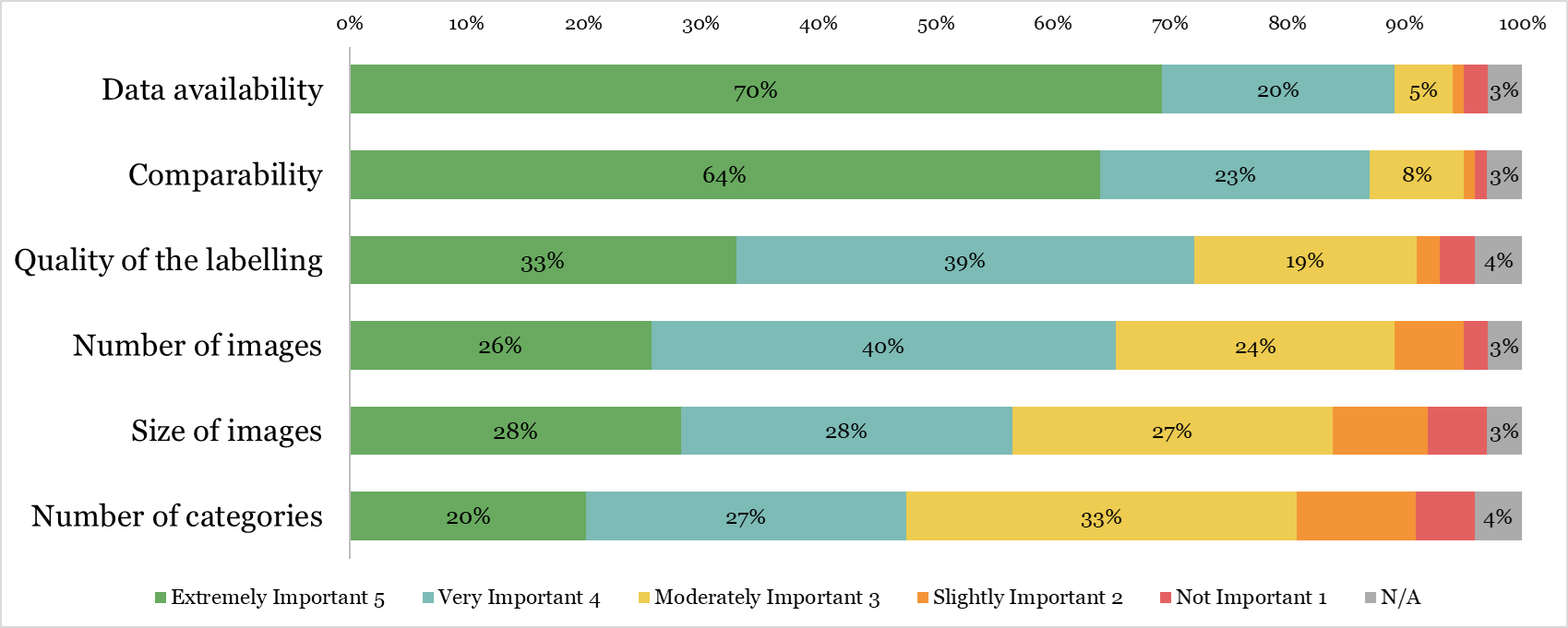}
	
	\label{fig:survey_cifar_comparison}
 
	\floatfoot{\emph{Notes}:  This figure shows the overall distribution of response for the question comparing CIFAR-10 to other datasets.}
\end{figure}

\noindent\textbf{Use compare to other OLDs}.  We asked the respondents to rate the reasons why they choose CIFAR-10 compared to similar datasets in the public domain.  Based on our interviews we included the quality of labelling, comparability as a benchmark, number of categories and images, image size, and data availability. Respondents rated each section on a Likert scale ranging from 1 (not important) to 5 (extremely important). Figure \ref{fig:survey_cifar_comparison} present the results of this questions. Around 90\% of respondents rated availability and comparability as very or extremely important. Quality of labelling and number of images were also considered important in explaining the choice of CIFAR-10 by 72\% and 66\% of the respondents.\vspace{0.5cm}

\begin{figure}[ht]
\centering
\caption{Survey Results - Integration of CIFAR-10 in Teaching environment}
	\includegraphics[width=\linewidth]{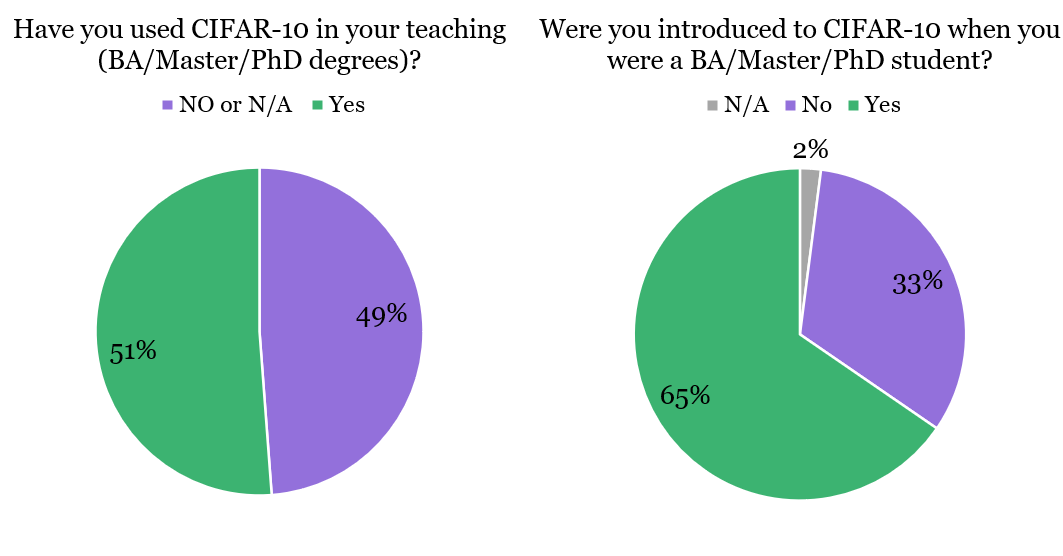}
	
	\label{fig:survey_teaching_learning}

	\floatfoot{\emph{Notes}:  The graphs illustrate the responses of participants regarding their usage of CIFAR-10 in teaching, as well as their introduction to CIFAR-10 based on their academic background.}
\end{figure}

\noindent\textbf{Pedagogical use of the datasets}. Figure \ref{fig:survey_teaching_learning} reports an interesting dimension of the survey: the pedagogical use of CIFAR-10. A significant number of respondents - 193 (65\%) - reported that they were introduced to the dataset during their studies (at the Bachelor, Master, or PhD level), and most respondents in academia routinely use it in their teaching programs. Furthermore, the responses to the open-ended question highlight the importance of CIFAR-10 as a pedagogical tool. \vspace{0.5cm}

The last question of the questionnaire was open, we asked to describe why they thought that CIFAR-10 was important for the development of DL or CV. Out of the 295 complete questionnaires analysed, we have got 182 quite detailed answers with a lot of interesting insights. To analyse them we have used the premium version of ChatGPT asking the algorithm to "identify the 5 main themes in the list of answers". 
 
\begin{itemize}
    \item \textbf{1. Benchmarking and Comparison}: CIFAR-10 is frequently cited as a standard benchmark for evaluating and comparing the performance of various algorithms and models. It provides a common platform for fair comparisons and validation, which is essential for developing and testing new methods in DL and computer vision.
    \item \textbf{2. Accessibility and Ease of Use}: The dataset is noted for its accessibility and ease of use. It is readily available, simple to download, and manageable in terms of size and computational requirements. This makes it an ideal choice for both beginners and researchers without access to extensive computational resources.
    \item \textbf{3. Educational Value and Prototyping}: CIFAR-10 serves as an excellent educational tool for new learners and students. Its simplicity and comprehensibility makes it a good starting point for understanding and experimenting with DL concepts. Additionally, it is suitable for rapid prototyping and initial testing of new ideas before scaling up to more complex datasets.
    \item \textbf{4. Quality and Characteristics of the Dataset}: The dataset is appreciated for its well-labeled, high-quality images. It offers a balanced number of categories and samples, which are sufficiently challenging for various image classification tasks. Its small image size and the diversity of the data allow for efficient experimentation and training..
    \item \textbf{5. Historical and Continued Relevance}: CIFAR-10 has historical significance in the field of computer vision and DL, having been used in many foundational studies and developments. Despite advancements in technology and the availability of larger datasets, it remains relevant due to its widespread use and the wealth of existing research that has utilized it as a benchmark.
\end{itemize}

\begin{figure}[!ht]
\centering
\caption{Word cloud of main terms used in the open-ended question in the survey}
	\includegraphics[width=\linewidth]{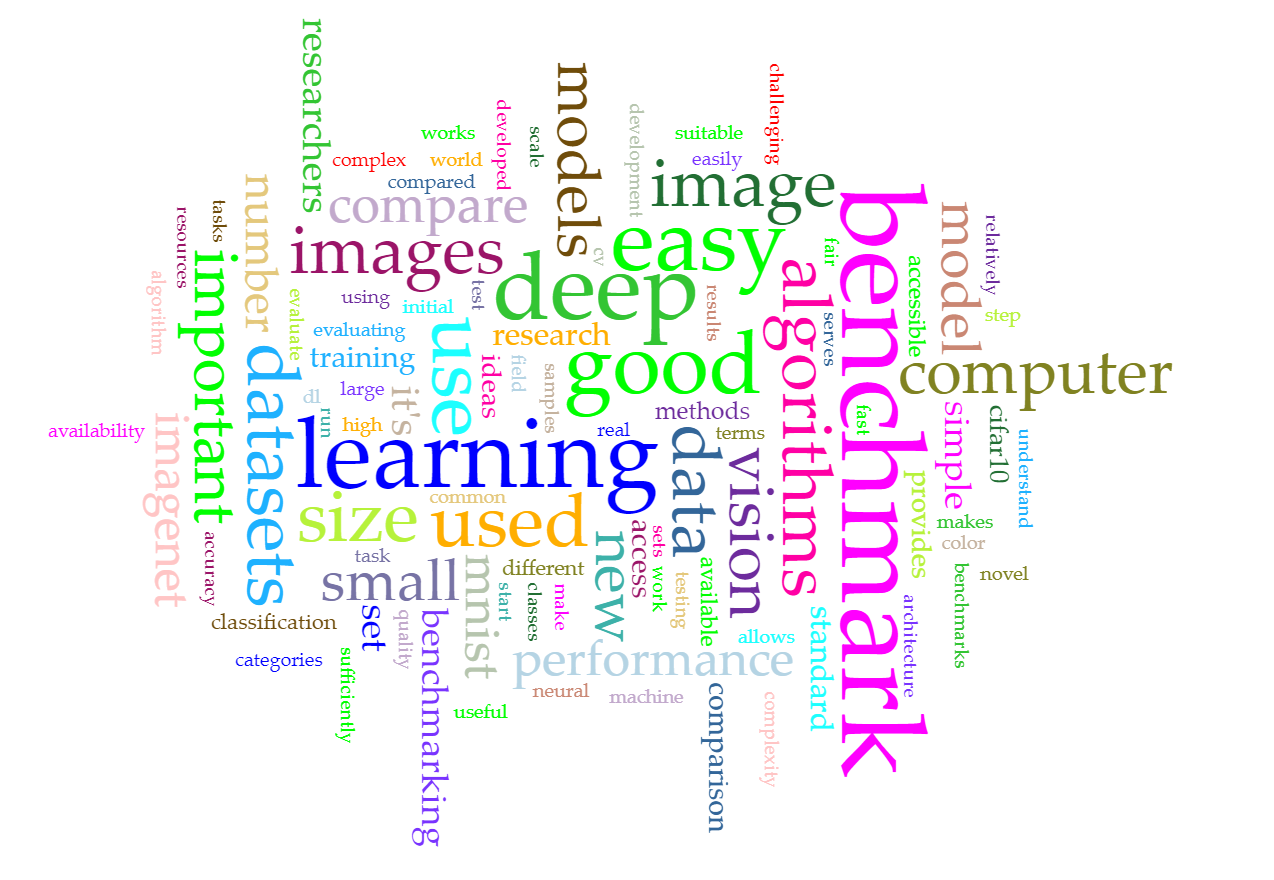}
	
	\label{fig:open_q_word_cloud}
    \floatfoot{\emph{Notes}:  This figure shows the most common terms used by respondents in the open-ended question on why CIFAR-10 was important for the development of DL or CV.}
\end{figure}

We also created a word cloud of the most common terms used in the answers to the open question\footnote{The word cloud was generated using Voyant Tools, an online open-source text analysis software available at https://voyant-tools.org/.}. We excluded some frequently used terms like "CIFAR" and "database", to get a more accurate idea of the reasons respondents assign importance to CIFAR-10. Figure \ref{fig:open_q_word_cloud} shows that "benchmarking" and "learning" are the most used terms, with 49 and 41 times respectively. These results are consistent with the analysis made through ChatGPT.

The evidence from both interviews and survey is consistent in highlighting that the specific characteristics (size, complexity, generalization) of CIFAR-10 made it the technological tool needed to develop and test convolutional neural  network  algorithms that gave rise to DL revolution. We also find consistent evidence that the accessibility, versatility, use as benchmark and pedagogical use of CIFAR-10 supported its continuous use and relevance even if much more complex and targeted OLDs became available in the ten years after its release. 

\subsection{Econometric analysis}
\label{subsec:econometrics}

\textbf{Results}. In Table \ref{tab:main_results_patents} we present the results of estimation equations \ref{eqn:main_cifar} and \ref{eqn:main_cifar_imagenet} using as outcome variable the number of patent citations received by a paper using OLDs.  Column 1 shows that papers mentioning only CIFAR-10  in the title, abstract or keywords received, on average, $e^{0.418}-1= 51.89\%$ more patent citations than papers that do not mention it. Considering only the first half of the decade after the creation of CIFAR-10 (2010 to 2014), as shown in column 2, papers mentioning CIFAR-10 accrued, on average, nearly double the citations ($e^{0.692}-1= 99.77\%$) compared to those using other datasets. In the later period (2015 to 2022), as shown in column 3, papers using only CIFAR-10 continued to receive on average a higher number of citations than other papers, though the effect is less significant in terms of magnitude ($e^{0.366}-1= 44.20\%$) and statistical significance. Papers using CIFAR-10 and other datasets receive, on average, the same number of citations as those not using CIFAR-10 across all periods. In columns 4-6 we estimate regressions using specification \ref{eqn:main_cifar_imagenet} and find similar results for papers using only CIFAR-10 and those using CIFAR-10 along with other datasets. Papers using ImageNet received, on average, 24.86\% more patent citations over the entire period ($e^{0.222}-1= 24.86\%$), primarily driven by papers published before 2015 (column 5, $e^{0.959}-1= 160.91\%$). CIFAR-10-only papers are expected to receive, on average, more citations ($e^{1.090}-1= 197.43\%$ in column 5) than papers using ImageNet, although the effect magnitude is comparable. Interestingly, more recent papers using ImageNet receive (not statistically significant for those published between 2015-2022, column 6), on average, a similar number of patent citations as papers using datasets other than CIFAR-10. 

\input{Tables/main_results_patents}


Table \ref{tab:main_results_science} shows the results of Poisson regressions for scientific citations. Columns 1-3 presents results for specification \ref{eqn:main_cifar}. As we can observe in column 1, on average papers that mention only CIFAR-10 or CIFAR-10 and other datasets between 2010 and 2022 receive less citations than paper that do not cite it, but this difference is not statistically significant at conventional confidence levels.  However, if we taken into consideration only the first half of the 2010 decade (2010 to 2014) as in column 2, we find that papers using only CIFAR-10 or CIFAR-10 and others accrued on average 64.38\% and 181.51\% more citations than articles using other datasets.  When considering the period post-2014 (2015 to 2022) in column 3, we see that papers using CIFAR-10, both alone and in combination with other datasets, received fewer citations compared to the average citations received by other papers. The results in column 3 are significant in both magnitude ($e^{-0.346}-1= -29.25\%$) and statistical terms for papers using CIFAR-10 along with other datasets, suggesting that the scientific citation impact of CIFAR-10 was primarily concentrated in its early years.

\input{Tables/main_results_science}


In columns 4-6 of Table \ref{tab:main_results_science}, we consider model specification \ref{eqn:main_cifar_imagenet}, where we compare papers citing either CIFAR-10 or ImageNet with those using other similar datasets. The results in column 4 indicate that, overall, papers mentioning CIFAR-10 do not differ significantly in terms of scientific citations compared to those mentioning other datasets. In contrast, papers mentioning ImageNet receive significantly more citations on average ($e^{0.386}-1 = 47.11\%$). However, the difference in citations between papers mentioning CIFAR-10 and those not mentioning it is positive and significant in the first period (column 5) but becomes statistically insignificant in the last period (column 6). Meanwhile, papers using ImageNet consistently receive more citations on average in every period, though they received fewer citations than papers using CIFAR-10 in the first period. Specifically, during 2010-2014, papers mentioning CIFAR-10, either alone or with other datasets, received 101.58\% and 295.11\% more citations, respectively, compared to those using other datasets but not ImageNet. In comparison, papers using ImageNet received 81.14\% more citations than those not using CIFAR-10.

\vspace{0.5cm}

\noindent\textbf{Robustness Checks}.
In Appendix \ref{app:robustness}, we perform a series of robustness checks and sensitivity analysis using various estimation methods, different sample definitions, and alternative model specifications.

\vspace{0.5cm}

\noindent\emph{Alternative Statistical Models}. The Poisson model is not the only method for handling highly skewed count data\footnote{Another approach involves using OLS to estimate models with log-transformed dependent variables or employing the inverse hyperbolic sine transformation. We chose not to use these estimation methods because our dependent variables include many zeros, and results can be sensitive to the arbitrary addition of a constant to handle these zero observations.}.  Table \ref{tab:robustness_negbin} presents estimates from a negative binomial regression\footnote{Negative binomial regressions lack a fixed-effects estimator that is as consistent as the one in the Poisson fixed-effects model. To address this, we substitute fixed effects with categorical variables that control for the same factors as the fixed effects in the Poisson model.} based on specification \ref{eqn:main_cifar}. When scientific citation count is used as the dependent variable, the direction of the CIFAR-10 indicator variable coefficients remains consistent, although the significance levels differ. This alternative model does not alter our main finding: the influence of CIFAR-10 in scientific literature is predominantly concentrated in the earlier period. Results for patent citations are qualitatively similar.

\vspace{0.5cm}

\noindent\emph{Further Restricted Sample}. The publications in our initial sample are quite diverse, including both journal articles and conference proceedings. To improve comparability, we further refine our sample to include only conference proceedings. These proceedings are more likely to represent recent advancements in ML models using labeled datasets. We restrict this further to papers that utilize datasets covering at least 10\% of the tasks addressed by CIFAR-10 and are indexed in Papers With Code, aiming to minimize noise. Table \ref{tab:robustness_restricted_sample_patents} presents results for patent citations, while Table \ref{tab:robustness_restricted_sample_science} shows results for scientific citations. The findings are qualitatively similar to our main results and exhibit greater significance and magnitude, providing additional evidence of the influence of CIFAR-10 and ImageNet on technological and scientific advancements in DL. In patents citations is confirmed the stronger and continuous use of CIFAR-10 compared to ImageNet.

\vspace{0.5cm}

\noindent\emph{Enlarged Sample}. To ensure consistency in comparing patent and scientific citations, we initially removed a substantial number of observations where patent citations could not be accurately measured (13.86\% of papers with missing patent citation values), as well as various publication types such as reviews, book chapters, and data papers. We then re-estimated our main specifications using an enlarged sample that includes papers with missing patent citation counts and all publication types for both patent citations (Table \ref{tab:robustness_enlarged_sample_patent} and  scientific citations (Table \ref{tab:robustness_enlarged_sample_science}. The results are qualitatively consistent with our original findings.

\vspace{0.5cm}

\noindent\emph{Alternative Dataset Indicator Variable}. Since papers often benchmark new ML models against multiple datasets, isolating the citation impact of dataset size and complexity is challenging. To address this issue, we refined our indicator variable to distinguish between papers using only CIFAR-10 and those using CIFAR-10 in combination with other datasets. The variable for papers using only CIFAR-10 is more likely to reflect the effect of a small, yet sufficiently large, dataset, while the variable for papers using CIFAR-10 alongside other datasets also captures the influence of combining multiple datasets. To test the sensitivity of our analysis to this variable definition, we estimated models with an alternative specification where the independent variable is set to 1 for any publication mentioning CIFAR-10, regardless of whether it is used alone or with other datasets, and 0 otherwise. Table \ref{tab:robustness_alternative_indicator_patent} shows that while the results for patent citations are consistent in sign with the full sample and the 2015-2022 period, they are no longer statistically significant. This suggests that papers more significant in technological development predominantly use CIFAR-10 alone, supporting the idea that small-but-large-enough datasets play a unique role in advancing DL models in computer vision. Results in Table \ref{tab:robustness_alternative_indicator_science} are qualitatively similar and confirm our main findings.

\vspace{0.5cm}

\noindent\emph{Citation Lags}. DL has experienced rapid growth in recent years, especially following the 2012 revolution and the release of the ChatGPT chatbot at the end of 2022. This surge has arguably heightened interest in older publications in the field and altered citation patterns in ways that publication year fixed effects may not fully capture. To assess the sensitivity of our results to different citation specifications, we compare citation counts within a fixed number of years after publication. This approach helps account for the dynamic nature of citation trends. Given the recent nature of our sample, we focus on a 3-year citation window to avoid losing too many observations from more recent years. Tables \ref{tab:robustness_cit_3years_patent} and \ref{tab:robustness_cit_3years_science} present the results for patent and scientific citations within this 3-year window, respectively, demonstrating that the findings remain qualitatively consistent.\footnote{We also conducted the analysis using the citation count of patent families sourced from Elsevier's PlumX Analytics. This metric includes only front-page citations from the European Patent Office (EPO), World Intellectual Property Organization (WIPO), Intellectual Property Office of the United Kingdom (IPO), United States Patent and Trademark Office (USPTO), and Japan Patent Office (JPO). The results were qualitatively similar.}

\vspace{0.5cm}

Overall, our main results remain consistent, confirming the robustness of our findings across various control variables, fixed effects, sample definitions, and model specifications.

\section{Discussion}
\label{sec:discussion}

Through our interviews, we learned that CIFAR-10 became a benchmark due to its technical specifications, including the nature of the images, their size, and the number of samples and categories. The timing of its release was also crucial to its popularity, as no other similar OLD was available at the time. ImageNet, released in 2009 by a team of university researchers and associated with the ImageNet Large Scale Visual Recognition Challenge, was also significant but proved too large and complex. Even today, in 2024, solving ImageNet with the best model and the largest supercomputer would take more than three years.

The survey confirms the insights from the interviews and reveals an additional role that CIFAR-10 played in the diffusion of DL methods.  We present evidence that CIFAR-10 is extensively used in training computer scientists working with ML. Many researchers not only teach courses using CIFAR-10 but were also exposed to the datasets during their own graduate programs. This finding highlights teaching as a significant channel through which CIFAR-10 influences the field of DL.

The econometric analysis of the technological and scientific roles played by OLDs confirms that CIFAR-10 has had a significant influence on the development of DL compared to other OLDs, including its closest competitor, ImageNet. For science, we find that CIFAR-10's contribution was particularly important in the early years of DL development, with patent citations to CIFAR-10 remaining frequent in recent years. The role of ImageNet for the development of DL has been more prominent and continuous, likely due to its complexity, which allows for the testing and development of more advanced models. However, CIFAR-10 continued to outperform ImageNet (and all other OLDs) in technological citations even in recent years.

In terms of scientific complexity, CIFAR-10 was effectively "solved" by 2014, when state-of-the-art DL models achieved an error rate of around 3-4\%, surpassing human-level accuracy in image classification tasks. Its sufficient complexity and status as a benchmark make it particularly useful in applied industrial research, where the speed of research and cost controls are more important than new scientific achievements. This continued use and technological relevance can explain the frequency of patent citations in recent years.

Based on the qualitative and quantitative evidence collected, it can be argued that CIFAR-10's lower computational requirements, ease of use, and the availability of a trained workforce make it more suitable for technology-oriented developments, as reflected in patent activity. These developments, which focus less on pushing the scientific frontier, are likely to rely more on CIFAR-10 compared to ImageNet and other more recent, complex datasets. The latter's increased complexity and higher computational demands make it less accessible for such practical applications.

This study has some limitations. First, while we have tried to interview active researchers in computer vision during the DL revolution, we were unable to interview the creators of CIFAR-10, Geoffrey Hinton, Vinod Nair, and Alex Krizhevsky. Gaining further insight into their motivations could illuminate the choice of dataset characteristics and how these are related to the development of DL models they were working on. Another limitation is the difficulty in identifying the specific OLDs used in each paper. Despite experimenting with different approaches, pinpointing the datasets in ML papers remains challenging. Future studies could employ more precise extraction algorithms to identify the datasets used, leveraging the full text of papers. Additionally, this study is primarily descriptive, making it challenging to establish causal effects of dataset usage. We do not observe the full process of building and refining ML model architectures or which datasets were effectively used prior to publication. Forthcoming investigations could exploit exogenous shocks in the availability of OLDs to understand their impact on the development of the field.

\section{Conclusion}

This paper aims to shed light into the role played by OLDs in the development of DL. Understanding the fundamental building blocks of this emerging technoscience is crucial, as these foundational elements will likely impact socioeconomic development in the coming years. Current advancements continue to be influenced by early events.

We find that CIFAR-10, a small yet sufficiently complex, well-labeled, and easily accessible database, was fundamental for the developments leading to the DL revolution and continues to shape the field's trajectory. We identify CIFAR-10 as one of the most important technological artifact used to develop DL algorithms and architectures. We trace the creation of this dataset to the CIFAR NCAP Summer School in 2008, where graduate students, supervised by Geoffrey Hinton, a prominent scholar in the field, carried out the labeling of the datasets.
 
The evolution of AI in the early 2020s has been marked by significant investments by private companies in data collection and computing capacity to develop advanced large language models (LLMs) expected to profoundly impact society. A few large companies, which have been recruiting top DL scientists (similar to the career trajectory of the lead scientists behind CIFAR-10) and attracted a substantial share of new graduate and postgraduate DL researchers (as evidenced by the current debate on universities' challenges in retaining DL scientists and our own data on the share of researchers working for companies), have the capacity to shape both the scientific and technological trajectory of DL.

Previously, the field developed with an open science approach, where public and private actors adhered to the ethos of open science by sharing data and methods. However, this approach has changed significantly. We may be entering a new phase in DL development characterized by a more traditional separation between science and technology, consistent with \textcite{parthaNewEconomicsScience1994} characterization of traditional science. If this is the case, there is an urgent need for substantial investment in public science conducted at universities. The "small is beautiful" model exemplified by the CIFAR-10 database may no longer be viable. Nonetheless, the widespread diffusion of CIFAR-10 and its origins reflect a human capital imprint of "open science" ethics that could be leveraged to maintain competitive dynamics in the DL field.


\printbibliography

\clearpage


\begin{appendices}

\counterwithin{figure}{section}
\counterwithin{table}{section}
\renewcommand\thefigure{\thesection\arabic{figure}}
\renewcommand\thetable{\thesection\arabic{table}}


\section{Qualitative methodology}
\label{app:qualitative_methodology}

The qualitative empirical material for this article is derived from a series of interviews with DL experts involved in the DL revolution, managers and administrative staff at CIFAR associated with the DL funding program, and computer science PhD students. The survey was conducted among computer scientists and ML practitioners with scientific publications that mention CIFAR-10 in the title, abstract and keywords indexed by Scopus.

First, we outline our study design. Next, we detail the mechanics of our interviews, including how we used the interview guide and conducted the sessions. Then, we present the survey text and response analysis.

\vspace{0.5cm}

\noindent\textbf{Interviews}. For the qualitative section, we conducted a series of semi-structured interviews of two kinds: shorter conversations were held with academics working on AI, regardless of their direct involvement with CIFAR-10, to gain a broad understanding of the field and identify general features that practitioners might seek in a training dataset; and in-depth interviews  with key individuals who were directly or indirectly involved in the development of the CIFAR-10/CIFAR-100 datasets. Table \ref{tab:interviews_list} provides a comprehensive list of all the interviews conducted. Some of these interviews contributed to refining the research question, others provided empirical material for our conclusions, and some served both purposes.

\input{Tables/interviews_list}

The selection of interviewees was opportunistic, leveraging existing contacts. From these initial contacts, we employed a snowball sampling method to reach individuals outside our direct network, focusing on those recommended by interviewees and those with experience directly related to the creation of CIFAR-10 and CIFAR-100. Additionally, we conducted shorter interviews with individuals peripherally related to the topic, selected for their direct, personal knowledge of specific facts. This approach resulted in seven in-depth interviews that were transcribed, along with many off-the-record conversations.

We framed the interviews as conversations, with most conducted online and a few by phone, typically lasting between 15 minutes to an hour. Whenever permission was granted, we recorded the conversations, though participants could designate specific comments as off the record at any time. They were also given the option to review sections of the article in which they were mentioned before publication to ensure accuracy and agreement with how their comments were used. Interviewees had the choice to determine whether they wished to be identified by name or remain anonymous.

\vspace{0.5cm}

\noindent\emph{Interview Guide}. In-depth interviews were based on a guide reproduced below, which aws adapted in minor ways for each interview  to reflect the fact that not all interviewees would have the same information to impart.

\begin{center}
    \textbf{Final Interview Guide for In-Depth Interviews}
\end{center}

\noindent\emph{Research Question}:
What is the impact of CIFAR-10/CIFAR-100 on the development of Deep Learning? 

\vspace{0.5cm}

\noindent\emph{Interview goal}: Understand the role of CIFAR-10 and CIFAR-100 in the Deep Learning Revolution. We are also trying to better understand the chain of events that led to the development of these two datasets around 2008/2009, particularly the Summer School of August 2008.

\vspace{0.5cm}

\noindent\textbf{Questions}:

\begin{enumerate}
    \item What was the impact of CIFAR-10/CIFAR-100 databases and how would you measure it?
    \begin{enumerate}
        \item Did it help the development of neural network algorithms?
        \item Did it help the development of computer vision?
        \item Did it help the development of other research topics in artificial intelligence?
        \item Should we measure the impact on publications?
        \item Should we measure the impact on patents?
        \item Should we measure the impact on working papers?
        \item Should we measure the impact on conference proceedings papers?
        \item Should we measure the impact on media?
    \end{enumerate}

    \item Can you tell us about the history of the AI projects at CIFAR?

    \item How was the process of creating CIFAR-10/CIFAR-100?

    \item Do you remember the NCAP summer school of 2008? Was that the moment in which CIFAR-10/100 were born? Was the whole process of labelling finished during the summer school or did it require additional work?

    \item Who decided to give the name of CIFAR in CIFAR-10/100? Was it related to the funding of the project?
    
\end{enumerate}

\noindent\textbf{Wrap-up}
\begin{itemize}
    \item[-] Who else do you think I should engage on this in relation to their work with CIFAR-10/CIFAR-100?
    \item[-] Are you interested in seeing the results of this research?
    \item[-] Thank you, very grateful for your time and thoughts.
\end{itemize}

\vspace{0.5cm}

\noindent\emph{Transcription}. Most interviews were recorded and transcribed by the authors. When interviewees declined to be recorded, or when recording was impractical, shorthand notes were taken during the interview and subsequently expanded into detailed notes as soon as possible afterward.

\vspace{0.5cm}

\noindent\textbf{Survey}. The inputs from the interviews were used to produce a survey that was distributed to ML practitioners and academics.

The questionnaire consisted of 9 questions; 3 of the questions were related to the informant (education, place of work), and 4 directly to the evaluation of the CIFAR datasets. Figure \ref{fig:survey_text} shows the full battery of questions.

  \begin{figure}[!htb]
    \centering
    \includegraphics[width=\linewidth]{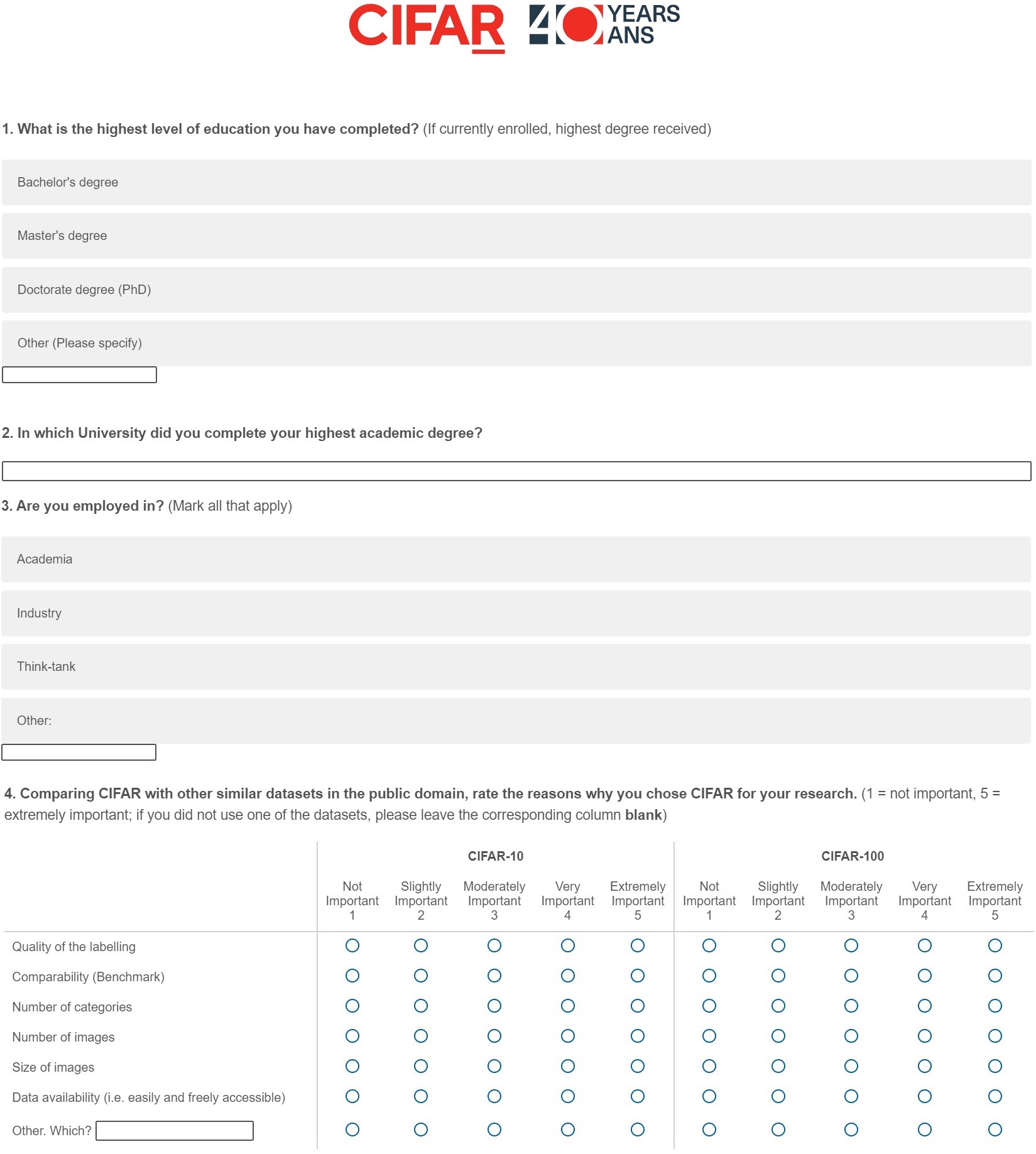}
    \caption{Survey Text}
    \label{fig:survey_text}
  \end{figure}

  \begin{figure}[!htb]
    \ContinuedFloat
    \captionsetup{list=off,format=cont}
    \centering
    \includegraphics[width=\linewidth]{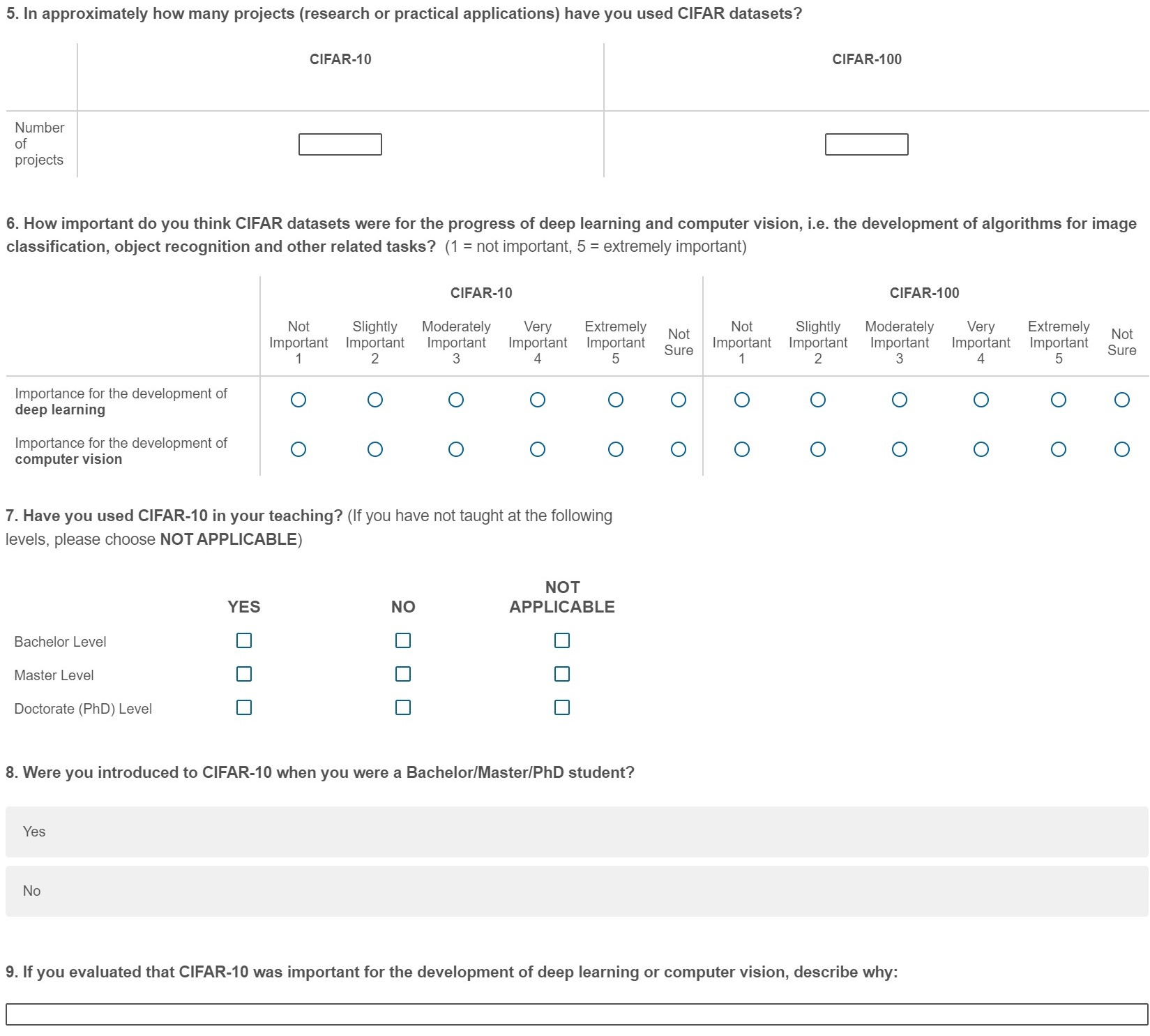}
    \caption{Survey Test Continued}
  \end{figure}
  \clearpage

To select the universe of possible respondents, we used the contact details of authors of papers extracted from Scopus that had used CIFAR-10 in their research. Out of the total of 6060 papers extracted, we were able to recover a valid email address of a corresponding author for 3033 papers. We sent a total of 4 requests to answer to the questionnaire to those authors in the period September 2022 to February 2023. 

The survey had a response rate of 9.7\%, with 392 authors starting the survey (13\%) and 295 completing it. The authors were from different geographical locations, with most affiliations in China and the US.
  \begin{figure}[!htb]
    \centering
    \includegraphics[width=\linewidth]{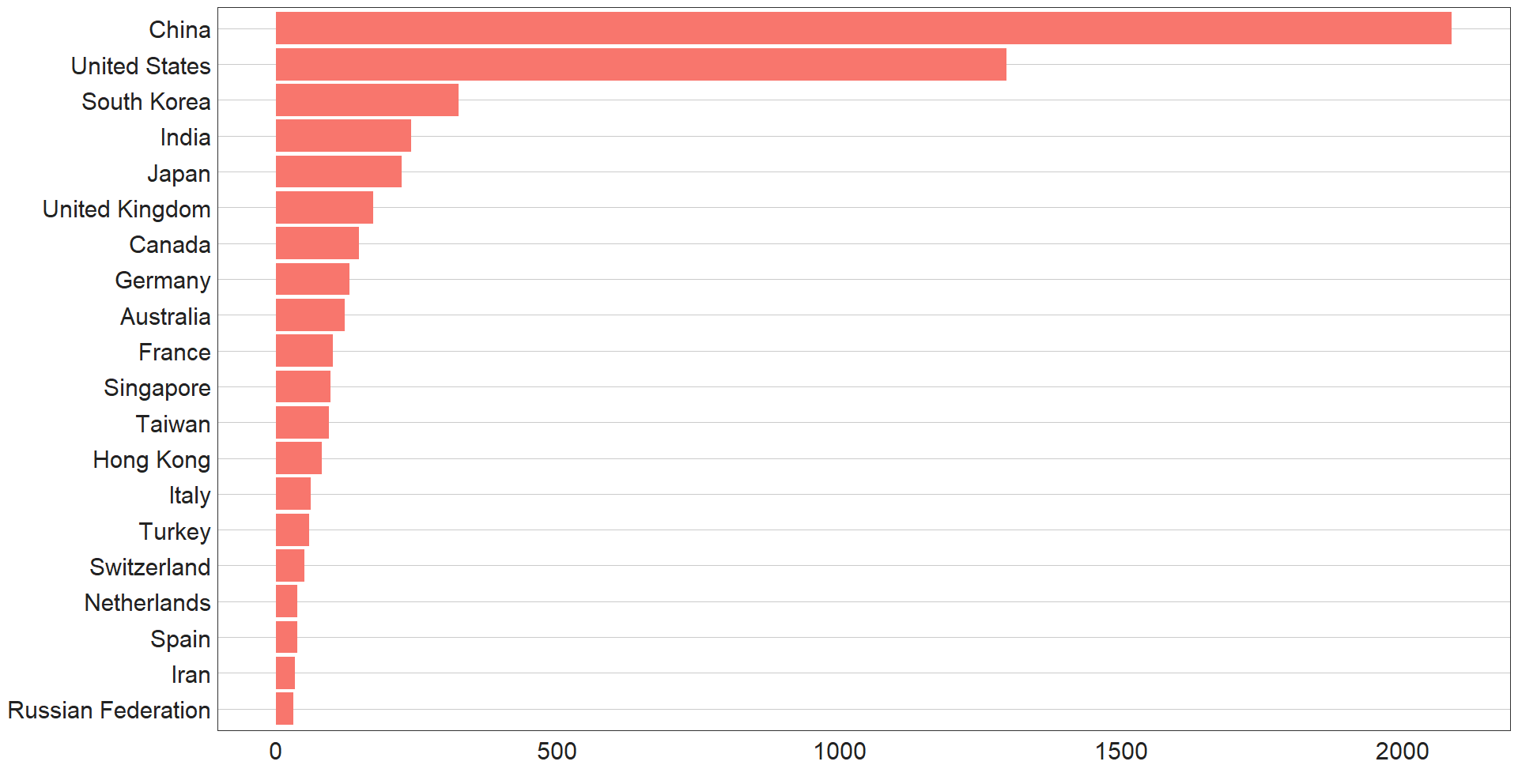}
    \caption{Distribution of CIFAR Papers among Top 20 Affiliations}
    \label{fig:countries_rate}
    	\floatfoot{Source: done by the authors}
\floatfoot
\footnotesize{\emph{Notes}:  The graph illustrates the fractional count of papers based on affiliations. The top 20 affiliations listed in the graph collectively account for 90\% of the the CIFAR papers.}
\end{figure}

\input{Tables/response_robustness}

Table \ref{tab:response_robustness_statistics} presents the summary statistics of our response analysis. The table includes  the Kolmogorov–Smirnov test (addressing the variance in the distribution) to compare the three sample considered: Total population, Population Survey Sent and Population Survey Answered.  We included in our response analysis the following variables: Year of publication, Number of authors, Citations count, Type publication (Journal versus others), International collaborations, Number of OLDs used, Use of ImageNet and Authors affiliated to a company. 

The year of publication was the only factor hypothesis rejected by all of the three tests. Respondents are associated to papers published more recently, however the difference is only in term of months, and is mainly due to no response from a few old papers. There are no significant differences between respondents and the population for which we had the email for the other seven variables we have considered.

When we compare respondents to the total population we see that journal articles are more frequent compare to other outlets, this is due to the fact that email addresses are difficult to be found in proceedings thus the email population was already biased in favour of journals. The respondent sample includes also papers with fewer citations compared to the total population; the bias was already present in the email population as the most highly cited articles are in category of proceedings.

\clearpage
\section{Dataset construction}
\label{app:data_construction}

In this Appendix, we report the procedure we followed to construct the dataset used in the econometric analysis.

We obtained data on labeled datasets' names, introduction dates, and associated tasks from Papers With Code on July 17, 2023. We identified 358 datasets, including their names, full names, and variants, which had at least one task overlapping with CIFAR-10 tasks. A list of CIFAR-10 and ImageNet tasks can be found in table \ref{tab:task_list}. Missing introduction dates were filled automatically by querying for the name of the introductory paper on Scopus or manually when the introductory paper was unavailable. Additionally, we collected data on the number of papers indexed in Papers With Code connected to each dataset.

We then queried the Scopus API using Pybliometrics with the following query structure for each dataset name:

\vspace{0.5cm}

\noindent\textbf{TITLE-ABS-KEY({"dataset"}) AND PUBYEAR AFT {intro-year}}.

\vspace{0.5cm}

This query identified papers published after the year of introduction, allowing for a two-year margin to account for discrepancies between the first online appearance and official publication dates. When the number of papers identified on Scopus using this query significantly exceeded\footnote{We considered double the number of papers indexed as the maximum threshold after preliminary tests.} the number of papers indexed on Papers With Code, we discarded the results. To refine the search for datasets with short or general names like "BSD," "Flowers," or "APRICOT," we appended "dataset" and "database" to the dataset names, ensuring the results were specific to machine learning papers. The query structure described above was executed on August 9, 2023. The following list of dataset names was queried using the outlined steps and yielded at least one publication indexed on Scopus:\vspace{0.5cm}

{\footnotesize
\noindent102 Category Flower Dataset, A Visible-infrared Paired Dataset for Low-light Vision, AFHQ, AFHQ Cat, AFHQV2, AI2 Diagrams, AI2D, APRICOT dataset, ARC-100, ARID dataset, ASIRRA, Abnormal Event Detection Dataset, AbstractReasoning, AdvNet, AmsterTime, AmsterTime: A Visual Place Recognition Benchmark Dataset for Severe Domain Shift, Animal Faces-HQ, Animal Species Image Recognition for Restricting Access, ArtDL, BAM!, BCI database, BCI dataset, BCN 20000, BCNB, BSD database, BSD dataset, BSDS300, BTAD, Bamboo dataset, BarkNet 1.0, Behance Artistic Media, Bentham dataset, Bentham project, Berkeley Segmentation Dataset, BigEarthNet, Boombox, BraTS 2016, BreakHis database, BreakHis dataset, Breast Cancer Histopathological Database, Breast Cancer Immunohistochemical Image Generation, CASIA-FASD, CCPD, CIFAR-10, CIFAR-10 Image Classification, CIFAR-10 image generation, CIFAR-100, CIFAR-100 vs CIFAR-10, CINIC-10, CIRCO, CIRR, CLEVR, CLEVR-Dialog, COCO, COCO 2014, COCO 2015, COCO 2017, COCO minival, COCO panoptic, COCO test-challenge, COCO test-dev, COCO+, COCO-Animals, COCO-CN, CORe50, COVID-19 Image Data Collection, COWC, CUB, CUB Birds, CUB-200-2011, CUB-LT, CURE-OR, CalTech 101 Silhouettes, Caltech-101, Caltech-256, Caltech-UCSD Birds-200-2011, Cars Overhead With Context, Cats and Dogs dataset, CelebA, CelebA-HQ, CelebA-Test, CelebAMask-HQ, CelebFaces Attributes Dataset, Challenging Unreal and Real Environments for Object Recognition, Chaoyang dataset, ChestX-ray8, Chinese City Parking Dataset, ChineseFoodNet, Cityscapes, Cityscapes test, Cityscapes val, Clothing1M, Cluttered Omniglot, Compose Image Retrieval on Real-life images, Compositional Language and Elementary Visual Reasoning, DF20, DF20 - Mini, DFUC2021, DTD dataset, Danish Fungi 2020, Deep PCB, Deep-Fashion, DeepFashion, DeepFashion2, DeepFish, DeepScores, DeepWeeds, Deepfashion2 validation, DensePose, DensePose-COCO, Describable Textures Dataset, DiagSet, Digits database, Digits dataset, Dry Bean Dataset, ELEVATER, EMNIST, EMNIST-Balanced, EMNIST-Digits, EMNIST-Letters, EgoHOS, EuroSAT, European Flood 2013 Dataset, Extended MNIST, Extended Yale B database, Extended Yale B dataset, Extended Yale-B, FER2013 database, FER2013 dataset, FFHQ, FGVC Aircraft, FGVC-Aircraft, FRGC database, FRGC dataset, Face Recognition Grand Challenge database, Face Recognition Grand Challenge dataset, FaceForensics++, Facial Expression Recognition 2013 Dataset, Fashion-Gen, Fashion-MNIST, Fishyscapes, Flickr database, Flickr dataset, Flickr-Faces-HQ, Flickr30k, FlickrLogos-32, Flowers database, Flowers dataset, Flowers-102, Food-101, Food-101N, Freiburg Groceries, Functional Map of the World, GOZ, GPR1200, Galaxy Zoo DECaLS, GasHisSDB, George Washington database, George Washington dataset, Google Landmarks, Google Landmarks Dataset v2, Grocery Store dataset, HR-ShanghaiTech, Hotels-50K, Hyper-Kvasir Dataset, IAM Handwriting, IAM database, IAM dataset, IAPR TC-12, IAPR TC-12 Benchmark, IARPA Janus Benchmark-B, IARPA Janus Benchmark-C, ICFG-PEDES, ICubWorld, IJB-B, IJB-C, ILSVRC 2015, ILSVRC 2016, INSTRE, IRMA database, IRMA dataset, ISBNet, Image Retrieval from Contextual Descriptions, ImageCoDe, ImageNet, ImageNet Detection, ImageNet-10, ImageNet-100, ImageNet-32, ImageNet-9, ImageNet-A, ImageNet-C, ImageNet-Caltech, ImageNet-LT, ImageNet-O, ImageNet-R, ImageNet-Sketch, ImageNet32, ImageNet64x64, Imagenette, In-Shop, InLoc, Incidents database, Incidents dataset, InstaCities1M, JFT-300M, JFT-3B, JHU CoSTAR Block Stacking Dataset, JHU-CROWD, JHU-CROWD++, Kannada-MNIST, Kitchen Scenes, Konzil, Kuzushiji-49, Kuzushiji-MNIST, Kvasir-Capsule, LFW database, LFW dataset, LHQ, LIDC-IDRI database, LIDC-IDRI dataset, LLVIP, LSUN, LSUN Bedroom, LaSCo, LabelMe, Labeled Faces in the Wild, Large-scale Scene UNderstanding Challenge, Lemons quality control dataset, Letter Recognition Data Set, Letter database, Letter dataset, Localized Narratives, Logo-2K+, MAMe, MIAD, MINC dataset, MLRSNet, MNIST, MNIST Large Scale dataset, MNIST-8M, MNIST-full, MNIST-test, MS-COCO, MSCOCO, MSRA Hand, MUAD, MVTEC ANOMALY DETECTION DATASET, MVTec AD, MVTec D2S, MVTecAD, Materials in Context Database, Melodic Design, Meta-Dataset, Microsoft Common Objects in Context, Million-AID, Moving MNIST, MuMiN, Multi-Modal CelebA-HQ, MultiMNIST, N-Caltech 101, NAS-Bench-201, NCT-CRC-HE-100K, NUS-WIDE, New Plant Diseases Dataset, Notre-Dame Cathedral Fire, NumtaDB, OFDIW, OMNIGLOT, ObjectNet, OmniBenchmark, Omniglot, OnFocus Detection In the Wild, Open Images V4, Open MIC, Open Museum Identification Challenge, Optical Recognition of Handwritten Digits, Oxford 102 Flower, Oxford 102 Flowers, Oxford Buildings, Oxford-IIIT Pet Dataset, Oxford105k, Oxford5k, PASCAL VOC 2007 database, PASCAL VOC 2007 dataset, PASCAL VOC 2011, PASCAL VOC 2011 test, PASCAL VOC 2012 database, PASCAL VOC 2012 dataset, PASCAL VOC 2012 test, PASCAL VOC 2012 val, PASCAL VOC database, PASCAL VOC dataset, PASCAL Visual Object Classes Challenge, PCam, PGM dataset, PKU-Reid, PROMISE12, Pano3D, PatchCamelyon, Patzig, Perceptual Similarity, PhotoChat, Places database, Places dataset, Places-LT, Places2, Places205, Places365, Places365-Standard, PlantVillage database, Procedurally Generated Matrices (PGM), Processed Twitter, QMNIST, Quick, Draw! Dataset, QuickDraw-Extended, RESISC45 database, RESISC45 dataset, RF100, RIT-18, RPC database, RPC dataset, RVL-CDIP, Recipe1M, Recipe1M+, Replica dataset, Retail Product Checkout, Ricordi, Riseholme-2021, Road Anomaly, Rotated MNIST, Rotating MNIST, SI-Score, STAIR Captions, STL-10, STN PLAD, STN Power Line Assets Dataset, SUN Attribute, SUN397, SVHN, SVLD, Saint Gall, Schiller dataset, Schwerin, Self-Taught Learning 10, Semi-Supervised iNaturalist, Semi-iNat, Sequential MNIST, Sewer-ML, ShanghaiTech, ShanghaiTech A, ShanghaiTech B, Shiller, ShoeV2, Silhouettes database, Silhouettes dataset, SketchHairSalon, SketchyScene, So2Sat LCZ42, Spot-the-diff, Stanford Cars, Stanford Dogs, Stanford Online Products, Street View House Numbers, StreetStyle, Structured3D, StyleGAN-Human, Stylized ImageNet, TMED, TUM-GAID, Tencent ML-Images, Thyroid Disease database, Thyroid Disease dataset, Thyroid database, Thyroid dataset, Tiny ImageNet, Tiny Images, Tiny-ImageNet, TransNAS-Bench-101, Tsinghua Dogs, Twitter100k, UBI-Fights, UCF-CC-50, UCSD Anomaly Detection Dataset, UCSD Ped2, UCSD-MIT Human Motion, UFPR-AMR, UMIST, UMist, UPIQ, USPS database, USPS dataset, Unified Photometric Image Quality, VOC12, VegFru, VehicleX, Verse dataset, Visual Madlibs, Visual Wake Words, VocalFolds, WHU-Hi, WIT dataset, Washington RGB-D, WebVision, WebVision-1000, Wikipedia-based Image Text, Wine Data Set, Wine database, Wine dataset, Wuhan UAV-borne hyperspectral image, YFCC100M, beanTech Anomaly Detection, cats vs dogs, ciFAIR-10, cifar10, cifar100, fMoW, fashion mnist, food101, iCartoonFace, iNat2021, iNaturalist, iNaturalist 2018, iNaturalist 2019, iSUN, imagenet-1k, mini-ImageNet-LT, smallNORB, tieredImageNet, xBD. }
\vspace{0.5cm}

From the original list of 358 unique labeled datasets, we managed to identify on Scopus 37,242 papers citing 264 unique labeled datasets. The discrepancy is due to some datasets not being identified precisely enough using the described steps, i.e. having too many results even after adding the words "dataset" and "database" in the query, or having no results at all. The labeled datasets we could not find were either not indexed by Scopus or did not mention the datasets in the title, abstract, or keywords. We then merged this information with Papers With Code's annotated datasets information to obtain the complete sample with all the necessary data.

\input{Tables/task_list}

\clearpage

\section{Additional descriptives}
\label{app:further_descriptives}

Table \ref{tab:datasets_descriptives} reports the main characteristics of 15 selected labeled datasets in our sample, including their names, supporting institutions,
introduction years, number of categories, and instance counts.

\input{Tables/datasets_descriptives}

Additionally, we constructed Table \ref{tab:comp_requirement} using estimates from \textcite{shermatov_national_2024} to provide computational requirements for running state-of-the-art (SOTA) models on the four most commonly used open labeled datasets (OLDs) in the literature: ImageNet, COCO, MNIST, and CIFAR-10. These estimates were derived from Epoch AI’s methods for assessing the training compute of deep learning systems, including operation counts, GPU time, and performing calculations based on SOTA data compiled by the \href{https://paperswithcode.com/sota}{Papers with Code} platform. More details can be found at \href{https://epochai.org/blog/estimating-training-compute}{https://epochai.org/blog/estimating-training-compute}.

These calculations are intended for illustrative and comparative purposes only. Computing power requirements can differ significantly across datasets and architectures. We provide rough estimates by comparing the compute demands of leading models with the compute capabilities of different hardware. We present estimates for hardwares with two levels of performance: supercomputers and research laptops. For supercomputers, we use the Frontier exascale machine, which delivers 1194 PFlop/s, as a benchmark. For research laptops, we reference average devices with \href{https://www.techpowerup.com/gpu-specs/geforce-rtx-4080.c3888}{NVIDIA GeForce RTX 4080 } or \href{https://www.techpowerup.com/gpu-specs/radeon-rx-7900-xtx.c3941}{AMD Radeon RX 7900 XTX GPUs}, which provide roughly 5x less flops/s. Actual flops allocated for deep learning tasks can vary greatly depending on the specific model and its configuration.

For example, the top CIFAR-10 model by Google Research Brain Team, ViT-H/14, requires a substantial amount of flops to achieve 99.5\% accuracy. A simpler model, “airbench”, requires 3.6 times fewer flops to achieve human-level accuracy of 94\% \parencite{jordan_94_2024}. On an average researcher laptop, training this model on CIFAR-10 would take approximately \textit{10 seconds to achieve 94\% accuracy}.

\clearpage

\section{Robustness checks and sensitivity analysis}
\label{app:robustness}

In this Appendix we report the robustness checks and sensitivity analysis we run and discussed in the Section \ref{subsec:econometrics} of this article.

\input{Tables/robustness_negbin}


\input{Tables/robustness_restricted_patent}


\input{Tables/robustness_restricted_science}


\input{Tables/robustness_enlarged_patent}


\input{Tables/robustness_enlarged_science}


\input{Tables/robustness_indicator_patents}


\input{Tables/robustness_indicator_science}


\input{Tables/robustness_3years_window_patents}


\input{Tables/robustness_3years_window_science}


\end{appendices}


\end{document}

%% file: Preamble/preamble_packages.tex
\usepackage{graphicx}
\usepackage[gen]{eurosym}
\usepackage[title]{appendix}
\usepackage[right = 1in, left = 1in, top= 1in, bottom= 1in]{geometry}
\usepackage{hyperref}
\hypersetup{colorlinks=true,
    linkcolor=blue,
citecolor=blue,
urlcolor=blue,
breaklinks}

\usepackage[american]{babel}
\usepackage{multirow}
\usepackage{booktabs}
\usepackage{array}
\usepackage[T1]{fontenc}
\usepackage{subfig}
\usepackage{amsmath}
\usepackage{amssymb}
\usepackage{authblk}
\usepackage[utf8]{inputenc}
\usepackage{bm}
\usepackage{tabularx}
\usepackage{adjustbox}
\usepackage[flushleft]{threeparttable}
\usepackage{amsfonts,caption,color,setspace,comment,footmisc,dcolumn}
\usepackage{longtable}
\usepackage{lipsum}
\usepackage[figuresright]{rotating}
\usepackage{makecell}
\usepackage{numprint}
\usepackage[capposition=top]{floatrow}
\usepackage{xcolor} 
\usepackage{geometry} 

\DeclareCaptionFormat{cont}{#1 (cont.)#2#3\par}

%% file: Preamble/bibliography_code.tex
\usepackage{xpatch} 
\usepackage[%
      style=authoryear,%
      maxcitenames=2,%
      maxbibnames=5,%
      backend=biber,		
      natbib=true,
      uniquename=init,
      giveninits=true,
      uniquename=false,
      url=true,
      uniquelist=false 
	]{biblatex}
\renewbibmacro{in:}{%
  \ifentrytype{article}{}{\printtext{\bibstring{in}\intitlepunct}}}
\usepackage[autostyle, english = american,maxlevel=3]{csquotes}
\MakeOuterQuote{"}

\DeclareFieldFormat{citehyperref}{%
  \DeclareFieldAlias{bibhyperref}{noformat}
  \bibhyperref{#1}}
\DeclareFieldFormat{textcitehyperref}{%
  \DeclareFieldAlias{bibhyperref}{noformat}
  \bibhyperref{%
    #1%
    \ifbool{cbx:parens}
      {\bibcloseparen\global\boolfalse{cbx:parens}}
      {}}}
\savebibmacro{cite}
\savebibmacro{textcite}
\renewbibmacro*{cite}{%
  \printtext[citehyperref]{%
    \restorebibmacro{cite}%
    \usebibmacro{cite}}}
\renewbibmacro*{textcite}{%
  \ifboolexpr{
    ( not test {\iffieldundef{prenote}} and
      test {\ifnumequal{\value{citecount}}{1}} )
    or
    ( not test {\iffieldundef{postnote}} and
      test {\ifnumequal{\value{citecount}}{\value{citetotal}}} )
  }
    {\DeclareFieldAlias{textcitehyperref}{noformat}}
    {}%
  \printtext[textcitehyperref]{%
    \restorebibmacro{textcite}%
    \usebibmacro{textcite}}}

\DeclareFieldFormat[online]{title}{#1} 
\DeclareFieldFormat[article]{doi}{} 
\DeclareFieldFormat[book]{doi}{} 
\DeclareFieldFormat[inproceedings]{doi}{} 
\DeclareFieldFormat[online]{url}{\href{https://#1}{#1}}

\AtEveryBibitem{%
  \clearfield{month}%
  \clearfield{day}%
  \clearfield{issn}%
  \clearfield{isbn}%
  \ifentrytype{article}{
    \clearfield{url}%
    \clearfield{urldate}%
    \clearfield{publisher}%
    \clearfield{series}%
    \clearfield{note}%
    \clearfield{doi} 
}{}
  \ifentrytype{book}{
    \clearfield{url}%
    \clearfield{urldate}%
    \clearfield{series}%
    \clearfield{note}%
    \clearfield{doi} 
}{}
  \ifentrytype{inproceedings}{
    \clearfield{url}%
    \clearfield{urldate}%
    \clearfield{series}%
    \clearfield{note}%
    \clearfield{publisher}%
    \clearfield{location}%
    \clearfield{doi} 
}{}
}

\xpatchbibmacro{volume+number+eid}{%
  \setunit*{\adddot}%
}{%
}{}{}
\DeclareFieldFormat[article]{number}{\mkbibparens{#1}}

\xpatchbibmacro{date+extrayear}{
  \printtext[parens]%
}{%
  \setunit{\addcomma\space}%
  \printtext%
}{}{}

\DeclareFieldFormat{pages}{#1}
\bibliography{references.bib}
\newcommand\blfootnote[1]{%
  \begingroup
  \renewcommand\thefootnote{}\footnote{#1}%
  \addtocounter{footnote}{-1}%
  \endgroup
}

%% file: Tables/datasets_computational_requirements.tex
\begin{sidewaystable}\centering 
  \caption{Computational Requirements} 
  \label{tab:comp_requirement} 
  \footnotesize
  \renewcommand{\arraystretch}{1.2}
  \begin{adjustbox}{max width=\textwidth}
\footnotesize\setlength{\tabcolsep}{18pt}
\begin{threeparttable}
\begin{tabular}
{p{1.5cm}p{2cm}p{0.75cm}p{3cm}p{1cm}p{1.75cm}p{1.5cm}p{1.5cm}p{1.75cm}p{1.75cm}}
   \tabularnewline \midrule \midrule
\textbf{Dataset} & \textbf{Description} & \textbf{Instances} & \textbf{Primary Task} & \textbf{Year Introduced} & \textbf{Creator Affiliation} & \textbf{Current Best Model Performance} & \textbf{Hardware Burden} & \textbf{Estimated Time on Super- computer} & \textbf{Estimated Time on Laptop}\\
\hline
COCO & Complex everyday scenes of common objects in their natural context & 2,500,000 & Object recognition & 2015 & Microsoft & Models today only reach error rates up to \textbf{38.7\%} & Target error rate of \textbf{10\% }requires estimated $10^{31}$ flops & $3.17 \times 10^5$ years & $6.34 \times 10^9$ years  \\
\hline
ImageNet & Labeled object image database & 14,197,122 & Object recognition, classification & 2009 & Princeton University & Best model \textit{OmniVec} reached error rate of \textbf{8\%} & Reaching human-level error rate of \textbf{5\%} requires $10^{26}$ flops   & $3.17$ years &  $6.34 \times 10^4$ years \\
\hline
CIFAR-10 Dataset & Many small, low-resolution, images of 10 classes of objects & 60,000 & Classification & 2009 & University of Toronto & Most models can reach \textbf{99\%+} accuracy & Reaching human-level error rate of \textbf{6\% }requires $10^{21}$ flops & $\sim 16$  minutes & 230 days \\
\hline
MNIST database & Database of handwritten digits & 70,000 & Classification & 1994 & AT\&T Bell Labs & Most models can reach \textbf{99\%+} accuracy  & To train a similar model $10^{12}$ flops  &  $10^{-6}$ seconds  &   $<10$ seconds\\
\midrule \midrule

\end{tabular}

\begin{tablenotes} 
\item \footnotesize{\textit{Notes}: This table shows the computational requirements to train State-Of-The-Art (SOTA) ML models using the 4 most common dataset in our sample. See methodological appendix for the construction of the table. Source: \textcite{shermatov_national_2024}.}

\textbf{}

\end{tablenotes}

\end{threeparttable}
\end{adjustbox}
\end{sidewaystable}

%% file: Tables/summary.tex
\begin{table}[!htbp] \centering 
  \caption{Summary Statistics} 
  \label{tab:summary_statistics} \footnotesize\setlength{\tabcolsep}{0.8pt}
  \begin{adjustbox}{max width=\linewidth}
\begin{threeparttable}
\begin{tabular}{@{\extracolsep{5pt}}lcccccccc} 
\\[-1.8ex]\hline 
\hline \\[-1.8ex] 

Statistic & \multicolumn{1}{c}{N} & \multicolumn{1}{c}{Mean} & \multicolumn{1}{c}{St. Dev.} & \multicolumn{1}{c}{Min} & \multicolumn{1}{c}{Pctl(25)} & \multicolumn{1}{c}{Median} & \multicolumn{1}{c}{Pctl(75)} & \multicolumn{1}{c}{Max} \\ 
\hline \\[-1.8ex] 

CIFAR-10 & 28,393 & 0.154 & 0.361 & 0 & 0 & 0 & 0 & 1 \\ 
CIFAR-10 (only) & 28,393 & 0.041 & 0.198 & 0 & 0 & 0 & 0 & 1 \\ 
ImageNet & 28,393 & 0.199 & 0.399 & 0 & 0 & 0 & 0 & 1 \\ 
Nb. Authors & 28,393 & 4.320 & 2.581 & 1 & 3 & 4 & 5 & 100 \\ 
Nb. References & 28,393 & 36.212 & 20.505 & 1 & 22 & 33 & 47 & 811 \\ 
International Collaboration & 28,393 & 0.245 & 0.430 & 0 & 0 & 0 & 0 & 1 \\ 
Share Company Affiliation & 28,393 & 0.040 & 0.152 & 0 & 0 & 0 & 0 & 1 \\ 
Nb. Patent Citations & 28,393 & 0.157 & 1.049 & 0 & 0 & 0 & 0 & 48 \\ 
Nb. Scientific Citations & 28,393 & 16.365 & 73.377 & 0 & 0 & 2 & 10 & 2,279 \\ 
Nb. Dataset & 28,393 & 1.300 & 0.657 & 1 & 1 & 1 & 1 & 7 \\ 
Nb. Modalities & 28,393 & 1.246 & 0.465 & 1 & 1 & 1 & 1 & 5 \\ 
Nb. Tasks & 28,393 & 46.647 & 29.430 & 1 & 25 & 54 & 67 & 183 \\ 
Nb. Tasks Similar CIFAR-10 & 28,393 & 15.876 & 16.082 & 1 & 2 & 5 & 24 & 46 \\ 
Share Tasks Similar CIFAR-10 & 28,393 & 0.345 & 0.350 & 0.022 & 0.043 & 0.109 & 0.522 & 1 \\ 

\hline \\[-1.8ex] 
\end{tabular}

\begin{tablenotes} 
\item \footnotesize{\textit{Notes}: Summary statistics for regression sample on publications mentioning annotated image datasets.}
\end{tablenotes} 
\end{threeparttable}
\end{adjustbox}
\end{table}

%% file: Tables/main_results_patents.tex
\begin{table}[!htpb] \centering 
  \caption{Labeled Datasets and Patent Citations} 
  \label{tab:main_results_patents} 
  \footnotesize
  \setlength{\tabcolsep}{9pt}
  \renewcommand{\arraystretch}{1.2}
  \begin{adjustbox}{max width=\linewidth}
\begin{threeparttable}

\begingroup
\centering

\begin{tabular}{lcccccc}
   \tabularnewline \midrule \midrule
    & \multicolumn{6}{c}{Patents Citations}\\
    \cline{2-7} 
 & \multicolumn{1}{c}{Full} & \multicolumn{1}{c}{2010-2014} & \multicolumn{1}{c}{2015-2022} & \multicolumn{1}{c}{Full} & \multicolumn{1}{c}{2010-2014} & \multicolumn{1}{c}{2015-2022} \\ 
   Model:                       & (1)           & (2)           & (3)           & (4)           & (5)           & (6)\\  
   \midrule 
   CIFAR-10 (only)               & 0.418$^{*}$   & 0.692$^{**}$  & 0.366$\textsuperscript{\textdagger}$   & 0.486$^{**}$  & 1.090$^{***}$ & 0.385$\textsuperscript{\textdagger}$\\   
                                 & (0.169)       & (0.211)       & (0.199)       & (0.179)       & (0.211)       & (0.207)\\   
   CIFAR-10 (others)             & -0.055        & 0.339         & -0.017        & 0.030         & 0.943         & 0.008\\   
                                 & (0.187)       & (0.423)       & (0.171)       & (0.202)       & (0.621)       & (0.181)\\   
   ImageNet                      &               &               &               & 0.222$^{*}$   & 0.959$^{**}$  & 0.067\\   
                                 &               &               &               & (0.105)       & (0.355)       & (0.105)\\   
   log(Nb. Authors)              & 0.480$^{***}$ & -0.103        & 0.681$^{***}$ & 0.474$^{***}$ & -0.091        & 0.680$^{***}$\\   
                                 & (0.099)       & (0.157)       & (0.109)       & (0.100)       & (0.153)       & (0.109)\\   
   log(Nb. References)           & 0.674$^{***}$ & 1.751$^{***}$ & 0.472$^{**}$  & 0.654$^{***}$ & 1.621$^{***}$ & 0.466$^{**}$\\   
                                 & (0.149)       & (0.160)       & (0.147)       & (0.144)       & (0.150)       & (0.146)\\   
   International Collab.         & 0.087         & 0.191         & 0.049         & 0.084         & 0.113         & 0.048\\   
                                 & (0.079)       & (0.241)       & (0.093)       & (0.079)       & (0.274)       & (0.093)\\   
   Share Company Affil.          & 1.147$^{***}$ & 2.518$^{***}$ & 0.964$^{***}$ & 1.117$^{***}$ & 2.199$^{***}$ & 0.955$^{***}$\\   
                                 & (0.198)       & (0.422)       & (0.145)       & (0.194)       & (0.433)       & (0.146)\\   
   Nb. Datasets                  & 0.014         & 0.162         & 0.044         & 0.007         & 0.182         & 0.042\\   
                                 & (0.090)       & (0.121)       & (0.081)       & (0.090)       & (0.136)       & (0.081)\\   
   Nb. Tasks                     & 0.009$^{***}$ & 0.010$^{**}$  & 0.007$^{***}$ & 0.007$^{***}$ & 0.002         & 0.006$^{**}$\\   
                                 & (0.002)       & (0.004)       & (0.002)       & (0.002)       & (0.004)       & (0.002)\\   
   Nb. Modalities                & 0.093         & -0.597        & 0.183$^{*}$   & 0.145$\textsuperscript{\textdagger}$   & -0.477        & 0.198$^{*}$\\   
                                 & (0.070)       & (0.392)       & (0.077)       & (0.078)       & (0.405)       & (0.085)\\   
   \midrule 
   Pub. Venue Type Fixed Effect  & YES           & YES           & YES           & YES           & YES           & YES\\  
   Subject Area Fixed Effect     & YES           & YES           & YES           & YES           & YES           & YES\\  
   Publication Year Fixed Effect & YES           & YES           & YES           & YES           & YES           & YES\\  
   \midrule 
   Observations                  & 27,905        & 1,620         & 26,220        & 27,905        & 1,620         & 26,220\\  
   Dependent variable mean       & 0.15951       & 0.54691       & 0.13596       & 0.15951       & 0.54691       & 0.13596\\  
   Pseudo R$^2$                  & 0.26575       & 0.25381       & 0.26234       & 0.26654       & 0.26795       & 0.26241\\  
   \midrule \midrule

\end{tabular}
\par\endgroup

\begin{tablenotes} 
\item \footnotesize{\textit{Notes}: This table reports estimates of regressions of the models described in equations \ref{eqn:main_cifar} and \ref{eqn:main_cifar_imagenet}. The dependent variable is the total number of patent families that cited the focal paper. The response variables are indicator variables that are equal to one if a paper mentions only CIFAR-10, CIFAR-10 among other datasets or ImageNet in the title, abstract or keywords. Columns (1) reports our baseline results of the estimates stemming from a Poisson regression.  Column (2) and (3) reports estimates of the same equation in a subset of the sample comprised of papers published from 2010 to 2014 and those published from 2015 to 2022, respectively. Columns (4 - 5) report estimates when adding a dataset indicator variable also for papers using ImageNet. Exponentiating the coefficients and differencing from one yields numbers interpretable as elasticities. All the specifications include publication venue type, publication year and scientific field fixed effects. Standard errors are clustered at the journal/conference level. Significance levels: \textdagger p<0.1; * p<0.05; ** p<0.01; *** p<0.001.}
\end{tablenotes} 
\end{threeparttable}
\end{adjustbox}
\end{table}

%% file: Tables/main_results_science.tex
\begin{table}[!htpb] \centering 
  \caption{Labeled Datasets and Scientific Citations} 
  \label{tab:main_results_science} 
  \footnotesize
  \setlength{\tabcolsep}{9pt}
  \renewcommand{\arraystretch}{1.2}
  \begin{adjustbox}{max width=\linewidth}
\begin{threeparttable}

\begingroup
\centering

\begin{tabular}{lcccccc}
   \tabularnewline \midrule \midrule
    & \multicolumn{6}{c}{Scientific Citations}\\
    \cline{2-7} 
 & \multicolumn{1}{c}{Full} & \multicolumn{1}{c}{2010-2014} & \multicolumn{1}{c}{2015-2022} & \multicolumn{1}{c}{Full} & \multicolumn{1}{c}{2010-2014} & \multicolumn{1}{c}{2015-2022} \\ 
   Model:                       & (1)           & (2)           & (3)           & (4)           & (5)           & (6)\\  
   \midrule 
   CIFAR-10 (only)               & -0.026        & 0.497$^{**}$  & -0.051        & 0.108         & 0.701$^{***}$ & 0.082\\   
                                 & (0.090)       & (0.172)       & (0.101)       & (0.095)       & (0.186)       & (0.108)\\   
   CIFAR-10 (others)             & -0.294        & 1.035$^{*}$   & -0.346$^{*}$  & -0.155        & 1.374$^{*}$   & -0.207\\   
                                 & (0.191)       & (0.489)       & (0.156)       & (0.216)       & (0.584)       & (0.179)\\   
   ImageNet                      &               &               &               & 0.386$^{***}$ & 0.594$^{*}$   & 0.384$^{***}$\\   
                                 &               &               &               & (0.092)       & (0.284)       & (0.092)\\   
   log(Nb. Authors)              & 0.351$^{***}$ & -0.130        & 0.438$^{***}$ & 0.341$^{***}$ & -0.127        & 0.427$^{***}$\\   
                                 & (0.072)       & (0.178)       & (0.076)       & (0.072)       & (0.178)       & (0.076)\\   
   log(Nb. References)           & 1.202$^{***}$ & 1.367$^{***}$ & 1.180$^{***}$ & 1.180$^{***}$ & 1.324$^{***}$ & 1.157$^{***}$\\   
                                 & (0.135)       & (0.265)       & (0.137)       & (0.138)       & (0.257)       & (0.140)\\   
   International Collab.         & 0.324$^{***}$ & 0.335$\textsuperscript{\textdagger}$   & 0.322$^{***}$ & 0.321$^{***}$ & 0.294         & 0.322$^{***}$\\   
                                 & (0.041)       & (0.173)       & (0.038)       & (0.042)       & (0.179)       & (0.039)\\   
   Share Company Affil.          & 1.271$^{***}$ & 1.268$^{**}$  & 1.284$^{***}$ & 1.224$^{***}$ & 1.079$^{*}$   & 1.240$^{***}$\\   
                                 & (0.151)       & (0.471)       & (0.155)       & (0.145)       & (0.454)       & (0.151)\\   
   Nb. Datasets                  & 0.036         & 0.337$^{*}$   & 0.041         & 0.029         & 0.320$\textsuperscript{\textdagger}$   & 0.034\\   
                                 & (0.060)       & (0.139)       & (0.047)       & (0.061)       & (0.165)       & (0.046)\\   
   Nb. Tasks                     & 0.008$^{***}$ & 0.007$^{**}$  & 0.007$^{***}$ & 0.004$^{***}$ & 0.003         & 0.004$^{**}$\\   
                                 & (0.001)       & (0.003)       & (0.001)       & (0.001)       & (0.003)       & (0.001)\\   
   Nb. Modalities                & 0.189$^{***}$ & 0.213         & 0.194$^{***}$ & 0.290$^{***}$ & 0.279         & 0.298$^{***}$\\   
                                 & (0.045)       & (0.187)       & (0.047)       & (0.047)       & (0.192)       & (0.048)\\   
   \midrule 
   Pub. Venue Type Fixed Effect  & YES           & YES           & YES           & YES           & YES           & YES\\  
   Subject Area Fixed Effect     & YES           & YES           & YES           & YES           & YES           & YES\\  
   Publication Year Fixed Effect & YES           & YES           & YES           & YES           & YES           & YES\\  
   \midrule 
   Observations                  & 28,393        & 1,734         & 26,659        & 28,393        & 1,734         & 26,659\\  
   Dependent variable mean       & 16.365        & 39.354        & 14.870        & 16.365        & 39.354        & 14.870\\  
   Pseudo R$^2$                  & 0.41097       & 0.27955       & 0.42151       & 0.41527       & 0.28836       & 0.42589\\  
   \midrule \midrule

\end{tabular}

\par\endgroup

\begin{tablenotes} 
\item \footnotesize{\textit{Notes}: This table reports estimates of regressions of the models described in equations \ref{eqn:main_cifar} and \ref{eqn:main_cifar_imagenet}. The dependent variable is the total number scientific citations received by a paper. The response variables are indicator variables that are equal to one if a paper mentions only CIFAR-10, CIFAR-10 among other datasets or ImageNet in the title, abstract or keywords. Columns (1) reports our baseline results of the estimates stemming from a Poisson regression.  Column (2) and (3) reports estimates of the same equation in a subset of the sample comprised of papers published from 2010 to 2014 and those published from 2015 to 2022, respectively. Columns (4 - 5) report estimates when adding a dataset indicator variable also for papers using ImageNet. Exponentiating the coefficients and differencing from one yields numbers interpretable as elasticities. All the specifications include publication venue type, publication year and scientific field fixed effects. Standard errors are clustered at the journal/conference level. Significance levels: \textdagger p<0.1; * p<0.05; ** p<0.01; *** p<0.001.}
\end{tablenotes} 
\end{threeparttable}
\end{adjustbox}
\end{table}

%% file: Tables/interviews_list.tex
\begin{table}[ht] \centering 
\caption{List of interviews}
\label{tab:interviews_list}
    \renewcommand{\arraystretch}{1.1} 
\begin{adjustbox}{max width=\textwidth}
\small
\begin{tabularx}{\textwidth}{XXXXXX}

\midrule\midrule
\textbf{Interview number} & \textbf{Interviewee} & \textbf{Affiliation} & \textbf{Position} & \textbf{Interviewer} & \textbf{Date} \\
\midrule
1 & Bruno Casella & University of Turin & PhD student & Daniel Souza & 12/07/2022 \\ 
2 & Rob Fergus & NYU/ DeepMind & Professor/ Researcher Scientist & Daniel Souza/ Aldo Geuna/ Jeff Rodriguez & 21/07/2022 \\ 
3 & Gianluca Mittone & University of   Turin & PhD student & Daniel Souza & 26/07/2022 \\ 
4 & Yann LeCun & NYU/Meta AI & Professor/VP \& Chief AI Scientist & Daniel Souza/ Aldo Geuna & 28/07/2022 \\ 
5 & Yoshua Bengio & Université de   Montréal & Full Professor & Daniel Souza/ Aldo Geuna/ Jeff Rodriguez & 17/10/2022 \\ 
6 & Rachel Parker & CIFAR & Sr Director,   Research & Daniel Souza/ Aldo Geuna/ Jeff Rodriguez & 18/11/2022 \\ 
7 & Melvin Silverman & CIFAR & Former VP of Research & Daniel Souza/ Aldo Geuna/ Jeff Rodriguez & 08/12/2022 \\ 
\midrule\midrule

\end{tabularx}

\end{adjustbox}
\end{table}

%% file: Tables/response_robustness.tex
\definecolor{mydarkgreen}{RGB}{0, 100, 0}
\begin{table}\centering
\caption{Summary Statistics for Response Analysis}
\label{tab:response_robustness_statistics} \footnotesize\setlength{\tabcolsep}{0.8pt}
\begin{adjustbox}{max width=\linewidth}
\begin{threeparttable}
\begin{tabular}{@{\extracolsep{5pt}}lcccccccc} 

\hline
\toprule
& \\ 
\multicolumn{9}{c}{\Large\textbf{\textcolor{orange}{Descriptive Statistics for Total Population of Papers}}}\\[-1.8ex]
& \\ 
\hline \\[-1.8ex] 
& \\ 

{\large{Statistic}} & \multicolumn{1}{c}{\large{Year}} & \multicolumn{1}{c}{\large{Number of Authors}} & \multicolumn{1}{c}{\large{Citation Count}} & \multicolumn{1}{c}{\large{Type Publications}} & \multicolumn{1}{c}{\large{Int. Collaboration}} & \multicolumn{1}{c}{\large{OLDs}} & \multicolumn{1}{c}{\large{ImageNet}} \ & \multicolumn{1}{c}{\large{Company Affil.}} \\ 

\hline \\[-1.8ex] 
\hline \\[-1.8ex] 

N & 6060 & 6056 & 6060 & 6060 & 6060 & 6013 & 6060 & 5874 \\ 
Ndist & 14 & 22 & 266 & 2 & 2 & 8 & 2 & 2 \\ 
Mean & 2020.09 & 4.06 & 41.44 & 0.37 & 0.23 & 2.19 & 0.25 & 0.11 \\ 
St. Dev. & 1.78 & 1.95 & 1184.43 & 0.48 & 0.42 & 1.03 & 0.43 & 0.31 \\ 
Min & 2010 & 1 & 0 & 0 & 0 & 1 & 0 & 0 \\ 
Pctl(25) & 2019 & 3 & 0 & 0 & 0 & 1 & 0 & 0 \\ 
Pctl(75) & 2021 & 5 & 8 & 1 & 0 & 3 & 0 & 0 \\ 
Max & 2023 & 36 & 90038 & 1 & 1 & 8 & 1 & 1 \\ 
& \\ 

\multicolumn{9}{l}{\large\textit{Kolmogorov-Smirnov test for Total Population * Corresponding Email}} \\ \hline
        D & 0.11061 & 0.051373 & 0.0635 & 0.19911 & 0.023343 & 0.028154 & 0.00090898 & 0.0033341 \\ 
        P-value & \textless  2.2e-16 & 4.661e-05 & 1.666e-07 & \textless 2.2e-16 & 0.2207 & 0.08453 & 1 & 1 \\ 
     
\toprule
\multicolumn{9}{c}{\Large\textbf{\textcolor{blue}{Descriptive Statistics for Papers with Corresponding Email Addresses}}}\\[-1.8ex]
& \\ 
\hline \\[-1.8ex] 

N & 3033 & 3033 & 3033 & 3033 & 3033 & 2987 & 3033 & 2935 \\ 
Ndist & 14 & 21 & 131 & 2 & 2 & 8 & 2 & 2 \\ 
Mean & 2020.57 & 4.26 & 10.84 & 0.57 & 0.26 & 2.26 & 0.25 & 0.1 \\ 
St. Dev. & 1.59 & 2.12 & 54.59 & 0.5 & 0.44 & 1.05 & 0.43 & 0.31 \\ 
Min & 2010 & 1 & 0 & 0 & 0 & 1 & 0 & 0 \\ 
Pctl(25) & 2020 & 3 & 0 & 0 & 0 & 2 & 0 & 0 \\ 
Pctl(75) & 2022 & 5 & 6 & 1 & 1 & 3 & 0 & 0 \\ 
Max & 2023 & 36 & 1508 & 1 & 1 & 8 & 1 & 1 \\ 
& \\ 
\multicolumn{9}{l}{\large\textit{Kolmogorov-Smirnov test for Corresponding Email * Responded to Survey}} \\ \hline
D & 0.15676 & 0.053131 & 0.056518 & 0.047216 & 0.021125 & 0.033868 & 0.045209 & 0.0074897 \\ 
P-value & 3.681e-06 & 0.4337 & 0.3569 & 0.5866 & 0.9998 & 0.9282 & 0.6419 & 1 \\ 

\toprule
\multicolumn{9}{c}{\Large\textbf{\textcolor{mydarkgreen}{Descriptive Statistics for Papers Whose Authors Responded to Survey}}}\\[-1.8ex]
& \\ 
\hline \\[-1.8ex] 

N & 295 & 295 & 295 & 295 & 295 & 283 & 295 & 280 \\ 
Ndist & 8 & 12 & 43 & 2 & 2 & 7 & 2 & 2 \\ 
Mean & 2020.99 & 3.97 & 9.86 & 0.62 & 0.28 & 2.21 & 0.2 & 0.1 \\ 
St. Dev. & 1.42 & 1.86 & 51.95 & 0.49 & 0.45 & 1.08 & 0.4 & 0.3 \\ 
Min & 2016 & 1 & 0 & 0 & 0 & 1 & 0 & 0 \\ 
Pctl(25) & 2020 & 3 & 0 & 0 & 0 & 1 & 0 & 0 \\ 
Pctl(75) & 2022 & 5 & 6 & 1 & 1 & 3 & 0 & 0 \\ 
Max & 2023 & 12 & 816 & 1 & 1 & 7 & 1 & 1 \\ 

& \\
\multicolumn{9}{l}{\large\textit{Kolmogorov-Smirnov test for Total Population * Responded to Survey}} \\ \hline
        D & 0.25667 & 0.029133 & 0.11405 & 0.24632 & 0.044468 & 0.0085604 & 0.0443 & 0.010824 \\ 
        P-value &  \textless  2.2e-16 & 0.9708 & 0.001327 & 2.998e-15 &  0.6342 & 1 & 0.6389 & 1 \\ \hline

\hline \\[-1.8ex] 

\bottomrule
\end{tabular} 
\begin{tablenotes} 
\item \footnotesize{\textit{Notes}: The tables above present descriptive statistics of variables for the population of papers analyzed in our study. These variables include publication year, number of authors, citation count, journal publications (using journal publications from aggregation type as a benchmark of comparison), international collaboration, number of datasets used, use of the Imagenet dataset and affiliation of authors. The table shows the number of observations (N), the number of unique values (Ndist), the mean, standard deviation, minimum, maximum, and 25th and 75th percentiles for each variable. Table 1 provides statistics for the entire population, while Tables 2 and 3 present statistics for subsets of papers based on whether they had corresponding email addresses and whether their authors responded to our survey. At the bottom of each table, we report the results of the Kolmogorov-Smirnov test, which assesses the distributional differences between the variables in different tables. Specifically, we report the results of the KS test between Table 1 and Table 3, Table 1 and Table 2, and Table 2 and Table 3, to identify any significant differences in the distribution of variables between the tables. These tables offer valuable insights into the characteristics of the papers in our sample and provide a foundation for further analysis.}
\end{tablenotes}
\end{threeparttable}
\end{adjustbox}
\end{table}

%% file: Tables/task_list.tex
\begin{table}[!htpb]    \centering
  \caption{Tasks performed using CIFAR-10 and ImageNet} 
  \label{tab:task_list}
      \tiny

\begin{threeparttable}
\begin{tabular}{ll}
\toprule
CIFAR-10 & ImageNet \\
\midrule
Image Classification & Image Classification \\
Image Generation & Image Generation \\
Semi-Supervised Image Classification & Semi-Supervised Image Classification \\
Image Clustering & Image Clustering \\
Long-tail Learning & Long-tail Learning \\
Neural Architecture Search & Neural Architecture Search \\
Density Estimation & Density Estimation \\
Binarization & Binarization \\
Stochastic Optimization & Stochastic Optimization \\
Quantization & Quantization \\
Small Data Image Classification & Small Data Image Classification \\
Image Compression & Image Compression \\
Conditional Image Generation & Conditional Image Generation \\
Adversarial Defense & Adversarial Defense \\
Object Recognition & Object Recognition \\
Unsupervised Image Classification & Unsupervised Image Classification \\
Adversarial Robustness & Adversarial Robustness \\
Network Pruning & Network Pruning \\
Classification with Binary Weight Network & Classification with Binary Weight Network \\
Data Augmentation & Data Augmentation \\
Robust classification & Robust classification \\
Classification with Binary Neural Network & Classification with Binary Neural Network \\
Open-World Semi-Supervised Learning & Open-World Semi-Supervised Learning \\
Neural Network Compression & Neural Network Compression \\
Anomaly Detection & Biologically-plausible Training \\
Graph Classification & CW Attack Detection \\
Image Retrieval & Classification \\
Out-of-Distribution Detection & Classification Consistency \\
Learning with noisy labels & Color Image Denoising \\
Image Classification with Label Noise & Continual Learning \\
Semi-Supervised Image Classification (Cold Start) & Contrastive Learning \\
Personalized Federated Learning & Data Free Quantization \\
Unsupervised Anomaly Detection with Specified Settings -- 30\% anomaly & Domain Generalization \\
Unsupervised Anomaly Detection with Specified Settings -- 20\% anomaly & Few-Shot Image Classification \\
Unsupervised Anomaly Detection with Specified Settings -- 1\% anomaly & Few-Shot Learning \\
Unsupervised Anomaly Detection with Specified Settings -- 0.1\% anomaly & Generalized Zero-Shot Learning \\
Unsupervised Anomaly Detection with Specified Settings -- 10\% anomaly & Image Classification with Differential Privacy \\
Adversarial Attack & Image Colorization \\
Sequential Image Classification & Image Compressed Sensing \\
Model Poisoning & Image Deblurring \\
Sparse Learning and binarization & Image Inpainting \\
Novel Class Discovery & Image Recognition \\
Hard-label Attack & Image Super-Resolution \\
Clean-label Backdoor Attack (0.05\%) & Incremental Learning \\
Nature-Inspired Optimization Algorithm & JPEG Decompression \\
Long-tail Learning on CIFAR-10-LT ($\rho$=100) & Knowledge Distillation \\
 & Linear-Probe Classification \\
 & Model Compression \\
 & Object Detection \\
 & Parameter Prediction \\
 & Partial Domain Adaptation \\
 & Prompt Engineering \\
 & Self-Supervised Image Classification \\
 & Sparse Learning \\
 & Unconditional Image Generation \\
 & Unsupervised Domain Adaptation \\
 & Variational Inference \\
 & Video Matting \\
 & Video Visual Relation Detection \\
 & Weakly Supervised Object Detection \\
 & Weakly-Supervised Object Localization \\
 & Zero-Shot Learning \\
 & Zero-Shot Object Detection \\
 & Zero-Shot Transfer Image Classification \\
\bottomrule
\end{tabular}

\begin{tablenotes} 
\item \footnotesize{\textit{Notes}: This table lists all the tasks associated with CIFAR-10 and ImageNet, the two most commonly used labeled datasets in Papers With Code. Data collected on July 17, 2023, and compiled by the authors.}
\end{tablenotes} 
\end{threeparttable}
 
\end{table}

%% file: Tables/datasets_descriptives.tex
\begin{table}[!ht]
    \centering
    \caption{Open Labeled Datasets Characteristics} 
    \label{tab:datasets_descriptives}
    \setlength{\tabcolsep}{7pt} 
    \renewcommand{\arraystretch}{1.5} 
    \begin{adjustbox}{max width=\textwidth}
        \begin{threeparttable}
            \small
            \begin{tabular}{lp{4cm}p{3cm}p{1.5cm}p{1.5cm}p{1.5cm}}
            \hline
                Dataset & Full Name & Created by & Introduced Year & Categories & Instances \\ \hline
                ImageNet & ImageNet Large Scale Visual Recognition Challenge & Princeton University & 2009 & 21,841 & 14,197,122 \\ 
                MNIST & Modified National Institute of Standards and Technology & AT\&T Bell Laboratories & 1998 & 10 & 70,000 \\ 
                COCO & Common Objects in Context & Microsoft & 2014 & 80 & 330,000 \\ 
                CIFAR-10 & Canadian Institute for Advanced Research 10 & University of Toronto & 2009 & 10 & 60,000 \\ 
                PASCAL VOC & Pattern Analysis, Statistical Modelling and Computational Learning - Visual Object Classes Challenge & University of Oxford & 2005 & 20 & 27,450 \\ 
                CIFAR-100 & Canadian Institute for Advanced Research 100 & University of Toronto & 2009 & 100 & 60,000 \\ 
                CUB-200-2011 & Caltech-UCSD Birds-200-2011 & California Institute of Technology & 2011 & 200 & 11,788 \\ 
                BSD & Berkeley Segmentation Dataset & Berkeley Vision and Learning Center & 2003 & 1 & 500 \\ 
                SVHN & Street View House Numbers & Stanford University & 2011 & 10 & 604,388 \\ 
                CelebA & Celebrities Attributes Dataset & Chinese University of Hong Kong & 2014 & 10,177 & 202,599 \\ 
                FRGC & Facial Recognition Grand Challenge & National Institute of Standards and Technology & 2006 & 1 & 50,000 \\ 
                Extended Yale B & Extended Yale Face Database B & Yale University & 2001 & 38 & 2,414 \\ 
                Fashion-MNIST & Dataset for benchmarking machine learning algorithms & Zalando Research & 2017 & 10 & 70,000 \\ 
                Flickr30k & Flickr 30k Dataset & University of Illinois & 2014 & 1 & 31,783 \\ 
                Cityscapes dataset & Dataset for urban scene understanding and autonomous driving & Daimler AG and University of Tübingen & 2016 & 30 & 5,000 \\ \hline
            \end{tabular}
            \begin{tablenotes}
            \item \footnotesize{\textit{Notes}: This table provides information on 15 datasets from our sample, including their names, supporting institutions, introduction years, number of categories, and instance counts. Elaborated by the authors.}

            \end{tablenotes}
        \end{threeparttable}
    \end{adjustbox}
\end{table}

%% file: Tables/robustness_negbin.tex
\begin{table}[!htbp] \centering 
  \caption{Robustness Check: Negative Binomial} 
  \label{tab:robustness_negbin} 
  \footnotesize
  \setlength{\tabcolsep}{9pt}
  \renewcommand{\arraystretch}{1.2}
  \begin{adjustbox}{max width=\linewidth}
\begin{threeparttable}

\begingroup
\centering

\begin{tabular}{lcccccc}
   \tabularnewline \midrule \midrule
    & \multicolumn{3}{c}{Patent Citations} & \multicolumn{3}{c}{Scientific Citations}\\
    \cline{2-7} 
 & \multicolumn{1}{c}{Full} & \multicolumn{1}{c}{2010-2014} & \multicolumn{1}{c}{2015-2022} & \multicolumn{1}{c}{Full} & \multicolumn{1}{c}{2010-2014} & \multicolumn{1}{c}{2015-2022} \\ 
   Model:                  & (1)           & (2)           & (3)           & (4)           & (5)           & (6)\\  
   \midrule 
   CIFAR-10 (only)         & 0.357$^{*}$   & 1.491$^{*}$   & 0.292$\textsuperscript{\textdagger}$   & 0.013         & 0.606$\textsuperscript{\textdagger}$   & 0.013\\   
                           & (0.153)       & (0.596)       & (0.161)       & (0.069)       & (0.352)       & (0.074)\\   
   CIFAR-10 (others)       & -0.048        & 0.459         & -0.039        & -0.168$\textsuperscript{\textdagger}$  & 1.353$^{*}$   & -0.175$^{*}$\\   
                           & (0.142)       & (0.550)       & (0.138)       & (0.097)       & (0.672)       & (0.084)\\   
   ImageNet                & 0.168$^{*}$   & 0.598$\textsuperscript{\textdagger}$   & 0.084         & 0.426$^{***}$ & 0.403         & 0.440$^{***}$\\   
                           & (0.083)       & (0.316)       & (0.089)       & (0.052)       & (0.313)       & (0.054)\\   
   log(Nb. Authors)        & 0.536$^{***}$ & 0.031         & 0.639$^{***}$ & 0.344$^{***}$ & -0.205$^{*}$  & 0.395$^{***}$\\   
                           & (0.079)       & (0.167)       & (0.084)       & (0.040)       & (0.100)       & (0.041)\\   
   log(Nb. References)     & 0.603$^{***}$ & 1.601$^{***}$ & 0.502$^{***}$ & 0.947$^{***}$ & 1.023$^{***}$ & 0.958$^{***}$\\   
                           & (0.095)       & (0.235)       & (0.099)       & (0.110)       & (0.128)       & (0.114)\\   
   International Collab.   & 0.046         & 0.333$\textsuperscript{\textdagger}$   & 0.002         & 0.416$^{***}$ & 0.288$^{**}$  & 0.416$^{***}$\\   
                           & (0.068)       & (0.192)       & (0.079)       & (0.047)       & (0.101)       & (0.047)\\   
   Share Company Affil.    & 0.920$^{***}$ & 2.329$^{***}$ & 0.783$^{***}$ & 1.226$^{***}$ & 0.931$^{*}$   & 1.224$^{***}$\\   
                           & (0.156)       & (0.565)       & (0.158)       & (0.148)       & (0.454)       & (0.143)\\   
   Nb. Datasets            & 0.031         & -0.109        & 0.047         & -0.003        & 0.102         & 0.009\\   
                           & (0.067)       & (0.149)       & (0.068)       & (0.038)       & (0.199)       & (0.036)\\   
   Nb. Tasks               & 0.006$^{***}$ & 0.005         & 0.006$^{***}$ & 0.003$^{***}$ & 0.005$^{*}$   & 0.002$^{***}$\\   
                           & (0.001)       & (0.004)       & (0.001)       & (0.001)       & (0.003)       & (0.001)\\   
   Nb. Modalities          & 0.159$^{*}$   & -0.207        & 0.198$^{**}$  & 0.218$^{***}$ & 0.258$^{*}$   & 0.226$^{***}$\\   
                           & (0.069)       & (0.318)       & (0.070)       & (0.040)       & (0.124)       & (0.042)\\   
   \midrule 
   Observations            & 28,393        & 1,734         & 26,659        & 28,393        & 1,734         & 26,659\\  
   Dependent variable mean & 0.15676       & 0.51096       & 0.13373       & 16.365        & 39.354        & 14.870\\  
   Pseudo R$^2$            & 0.15171       & 0.10799       & 0.15478       & 0.10241       & 0.04174       & 0.10559\\  
   Over-dispersion         & 0.17058       & 0.18649       & 0.18141       & 0.50494       & 0.56484       & 0.50624\\  
   \midrule \midrule

\end{tabular}
\par\endgroup

\begin{tablenotes} 
\item \footnotesize{\textit{Notes}: This table reports estimates of regressions of the model described in equation \ref{eqn:main_cifar_imagenet}. The dependent variable for columns (1)-(3) is the total number of patent families citing the focal papers, while for columns (4)-(6) it is the total number of scientific citations received by the focal papers. The response variables are indicator variables that are equal to one if a paper mentions only CIFAR-10, CIFAR-10 among other datasets or ImageNet in the title, abstract or keywords. Columns (1) and (4) reports our baseline results of the estimates stemming from a Poisson regression.  Column (2)-(3) and (5)-(6) reports estimates of the same equation in a subset of the sample comprised of papers published from 2010 to 2014 and those published from 2015 to 2022, respectively. Exponentiating the coefficients and differencing from one yields numbers interpretable as elasticities. All the specifications include publication venue type, publication year and scientific field fixed effects. Standard errors are clustered at the journal/conference level. Significance levels: \textdagger p<0.1; * p<0.05; ** p<0.01; *** p<0.001.}
\end{tablenotes} 
\end{threeparttable}
\end{adjustbox}
\end{table}

%% file: Tables/robustness_restricted_patent.tex
\begin{table}[!htbp] \centering 
  \caption{Robustness Check: Restricted Sample - Patent Citations} 
  \label{tab:robustness_restricted_sample_patents} 
  \footnotesize
  \setlength{\tabcolsep}{9pt}
  \renewcommand{\arraystretch}{1.2}
  \begin{adjustbox}{max width=\linewidth}
\begin{threeparttable}

\begingroup
\centering
\begin{tabular}{lcccccc}
   \tabularnewline \midrule \midrule
    & \multicolumn{6}{c}{Patents Citations}\\
    \cline{2-7} 
 & \multicolumn{1}{c}{Full} & \multicolumn{1}{c}{2010-2014} & \multicolumn{1}{c}{2015-2022} & \multicolumn{1}{c}{Full} & \multicolumn{1}{c}{2010-2014} & \multicolumn{1}{c}{2015-2022} \\ 
   Model:                       & (1)           & (2)           & (3)           & (4)           & (5)           & (6)\\  
   \midrule 
   CIFAR-10 (only)               & 0.769$^{***}$ & 1.305$^{**}$  & 0.621$^{**}$  & 0.876$^{***}$ & 1.883$^{***}$ & 0.668$^{**}$\\   
                                 & (0.181)       & (0.421)       & (0.209)       & (0.199)       & (0.500)       & (0.215)\\   
   CIFAR-10 (others)             & 0.049         & 0.184         & 0.073         & 0.189         & 1.029$\textsuperscript{\textdagger}$   & 0.138\\   
                                 & (0.216)       & (0.368)       & (0.192)       & (0.227)       & (0.543)       & (0.197)\\   
   ImageNet                      &               &               &               & 0.338$^{*}$   & 1.170$^{**}$  & 0.167\\   
                                 &               &               &               & (0.140)       & (0.411)       & (0.122)\\   
   log(Nb. Authors)              & 0.500$^{***}$ & 0.437$^{*}$   & 0.608$^{***}$ & 0.492$^{***}$ & 0.421$^{*}$   & 0.605$^{***}$\\   
                                 & (0.101)       & (0.186)       & (0.118)       & (0.101)       & (0.200)       & (0.118)\\   
   log(Nb. References)           & 0.509$^{**}$  & 1.536$^{***}$ & 0.347$^{*}$   & 0.453$^{**}$  & 1.267$^{***}$ & 0.321$\textsuperscript{\textdagger}$\\   
                                 & (0.173)       & (0.310)       & (0.173)       & (0.158)       & (0.199)       & (0.167)\\   
   International Collab.         & -0.015        & 0.050         & -0.032        & -0.030        & -0.111        & -0.037\\   
                                 & (0.102)       & (0.419)       & (0.105)       & (0.103)       & (0.460)       & (0.104)\\   
   Share Company Affil.          & 1.425$^{***}$ & 3.533$^{***}$ & 1.111$^{***}$ & 1.360$^{***}$ & 3.017$^{***}$ & 1.082$^{***}$\\   
                                 & (0.244)       & (0.288)       & (0.184)       & (0.235)       & (0.308)       & (0.181)\\   
   Nb. Datasets                  & -0.170        & -0.124        & -0.077        & -0.170        & -0.297        & -0.074\\   
                                 & (0.106)       & (0.193)       & (0.099)       & (0.106)       & (0.205)       & (0.099)\\   
   Nb. Tasks                     & 0.021$^{***}$ & 0.036$^{***}$ & 0.015$^{***}$ & 0.018$^{***}$ & 0.030$^{***}$ & 0.014$^{***}$\\   
                                 & (0.003)       & (0.007)       & (0.004)       & (0.003)       & (0.009)       & (0.004)\\   
   Nb. Modalities                & -0.080        &               & 0.190         & -0.038        &               & 0.205\\   
                                 & (0.179)       &               & (0.164)       & (0.173)       &               & (0.164)\\   
   \midrule 
   Pub. Venue Type Fixed Effect  & YES           & YES           & YES           & YES           & YES           & YES\\  
   Subject Area Fixed Effect     & YES           & YES           & YES           & YES           & YES           & YES\\  
   Publication Year Fixed Effect & YES           & YES           & YES           & YES           & YES           & YES\\  
   \midrule 
   Observations                  & 15,407        & 799           & 14,504        & 15,407        & 799           & 14,504\\  
   Dependent variable mean       & 0.17005       & 0.61452       & 0.14679       & 0.17005       & 0.61452       & 0.14679\\  
   Pseudo R$^2$                  & 0.29363       & 0.36343       & 0.28420       & 0.29624       & 0.39371       & 0.28484\\  
   \midrule \midrule

\end{tabular}
\par\endgroup

\begin{tablenotes} 
\item \footnotesize{\textit{Notes}: This table reports estimates of regressions of the models described in equations \ref{eqn:main_cifar} and \ref{eqn:main_cifar_imagenet} using a sample that includes only conference proceedings and datasets with at least 100 papers indexed by Papers With Code and 5 or more (10\%) tasks overlapping with CIFAR-10. The dependent variable is the total number of patent families that cited the focal paper. The response variables are indicator variables that are equal to one if a paper mentions only CIFAR-10, CIFAR-10 among other datasets or ImageNet in the title, abstract or keywords. Columns (1) reports our baseline results of the estimates stemming from a Poisson regression.  Column (2) and (3) reports estimates of the same equation in a subset of the sample comprised of papers published from 2010 to 2014 and those published from 2015 to 2022, respectively. Columns (4 - 5) report estimates when adding a dataset indicator variable also for papers using ImageNet. Exponentiating the coefficients and differencing from one yields numbers interpretable as elasticities. All the specifications include publication venue type, publication year and scientific field fixed effects. Standard errors are clustered at the journal/conference level. Significance levels: \textdagger p<0.1; * p<0.05; ** p<0.01; *** p<0.001.}
\end{tablenotes} 
\end{threeparttable}
\end{adjustbox}
\end{table}

%% file: Tables/robustness_restricted_science.tex
\begin{table}[!htbp] \centering 
  \caption{Robustness Check: Restricted Sample - Scientific Citations} 
  \label{tab:robustness_restricted_sample_science} 
  \footnotesize
  \setlength{\tabcolsep}{9pt}
  \renewcommand{\arraystretch}{1.2}
  \begin{adjustbox}{max width=\linewidth}
\begin{threeparttable}

\begingroup
\centering
\begin{tabular}{lcccccc}
   \tabularnewline \midrule \midrule
    & \multicolumn{6}{c}{Scientific Citations}\\
    \cline{2-7} 
 & \multicolumn{1}{c}{Full} & \multicolumn{1}{c}{2010-2014} & \multicolumn{1}{c}{2015-2022} & \multicolumn{1}{c}{Full} & \multicolumn{1}{c}{2010-2014} & \multicolumn{1}{c}{2015-2022} \\  
   Model:                       & (1)           & (2)           & (3)           & (4)           & (5)           & (6)\\   
   \midrule 
   CIFAR-10 (only)               & 0.217$\textsuperscript{\textdagger}$   & 0.773$^{*}$   & 0.157         & 0.418$^{***}$ & 0.997$^{***}$ & 0.368$^{**}$\\   
                                 & (0.128)       & (0.305)       & (0.148)       & (0.119)       & (0.298)       & (0.134)\\   
   CIFAR-10 (others)             & -0.313        & 0.735         & -0.357$\textsuperscript{\textdagger}$  & -0.099        & 1.182$\textsuperscript{\textdagger}$   & -0.142\\   
                                 & (0.231)       & (0.454)       & (0.185)       & (0.236)       & (0.603)       & (0.185)\\   
   ImageNet                      &               &               &               & 0.552$^{***}$ & 0.624$\textsuperscript{\textdagger}$   & 0.577$^{***}$\\   
                                 &               &               &               & (0.073)       & (0.340)       & (0.074)\\   
   log(Nb. Authors)              & 0.336$^{***}$ & 0.193         & 0.372$^{***}$ & 0.319$^{***}$ & 0.197         & 0.353$^{***}$\\   
                                 & (0.087)       & (0.216)       & (0.093)       & (0.087)       & (0.217)       & (0.093)\\   
   log(Nb. References)           & 1.069$^{***}$ & 0.994$^{*}$   & 1.080$^{***}$ & 0.992$^{***}$ & 0.918$^{*}$   & 1.001$^{***}$\\   
                                 & (0.166)       & (0.408)       & (0.165)       & (0.157)       & (0.378)       & (0.158)\\   
   International Collab.         & 0.221$^{***}$ & 0.096         & 0.244$^{***}$ & 0.199$^{***}$ & 0.009         & 0.229$^{***}$\\   
                                 & (0.058)       & (0.278)       & (0.051)       & (0.058)       & (0.283)       & (0.052)\\   
   Share Company Affil.          & 1.271$^{***}$ & 2.000$^{***}$ & 1.268$^{***}$ & 1.179$^{***}$ & 1.754$^{***}$ & 1.182$^{***}$\\   
                                 & (0.128)       & (0.471)       & (0.117)       & (0.117)       & (0.471)       & (0.108)\\   
   Nb. Datasets                  & -0.010        & 0.411         & 0.017         & 0.007         & 0.318         & 0.040\\   
                                 & (0.126)       & (0.277)       & (0.110)       & (0.131)       & (0.318)       & (0.113)\\   
   Nb. Tasks                     & 0.015$^{***}$ & 0.021$^{***}$ & 0.013$^{**}$  & 0.009$\textsuperscript{\textdagger}$   & 0.016$^{*}$   & 0.007\\   
                                 & (0.004)       & (0.005)       & (0.005)       & (0.005)       & (0.007)       & (0.005)\\   
   Nb. Modalities                & 0.235         &               & 0.310$\textsuperscript{\textdagger}$   & 0.346$^{*}$   &               & 0.425$^{**}$\\   
                                 & (0.178)       &               & (0.165)       & (0.168)       &               & (0.158)\\   
   \midrule 
   Pub. Venue Type Fixed Effect  & YES           & YES           & YES           & YES           & YES           & YES\\  
   Subject Area Fixed Effect     & YES           & YES           & YES           & YES           & YES           & YES\\  
   Publication Year Fixed Effect & YES           & YES           & YES           & YES           & YES           & YES\\  
   \midrule 
   Observations                  & 15,600        & 895           & 14,703        & 15,600        & 895           & 14,703\\  
   Dependent variable mean       & 16.558        & 41.165        & 15.062        & 16.558        & 41.165        & 15.062\\  
   Pseudo R$^2$                  & 0.41044       & 0.30952       & 0.42224       & 0.42279       & 0.32472       & 0.43591\\  
   \midrule \midrule
\end{tabular}
\par\endgroup

\begin{tablenotes} 
\item \footnotesize{\textit{Notes}: This table reports estimates of regressions of the models described in equations \ref{eqn:main_cifar} and \ref{eqn:main_cifar_imagenet} using a sample that includes only conference proceedings and datasets with at least 100 papers indexed by Papers With Code and 5 or more (10\%) tasks overlapping with CIFAR-10. The dependent variable is the total number scientific citations received by a paper. The response variables are indicator variables that are equal to one if a paper mentions only CIFAR-10, CIFAR-10 among other datasets or ImageNet in the title, abstract or keywords. Columns (1) reports our baseline results of the estimates stemming from a Poisson regression.  Column (2) and (3) reports estimates of the same equation in a subset of the sample comprised of papers published from 2010 to 2014 and those published from 2015 to 2022, respectively. Columns (4 - 5) report estimates when adding a dataset indicator variable also for papers using ImageNet. Exponentiating the coefficients and differencing from one yields numbers interpretable as elasticities. All the specifications include publication venue type, publication year and scientific field fixed effects. Standard errors are clustered at the journal/conference level. Significance levels: \textdagger p<0.1; * p<0.05; ** p<0.01; *** p<0.001.}
\end{tablenotes} 
\end{threeparttable}
\end{adjustbox}
\end{table}

%% file: Tables/robustness_enlarged_patent.tex
\begin{table}[!htbp] \centering 
  \caption{Robustness Check: Enlarged Sample - Patent Citations} 
  \label{tab:robustness_enlarged_sample_patent} 
  \footnotesize
  \setlength{\tabcolsep}{9pt}
  \renewcommand{\arraystretch}{1.2}
  \begin{adjustbox}{max width=\linewidth}
\begin{threeparttable}

\begingroup
\centering
\begin{tabular}{lcccccc}
   \tabularnewline \midrule \midrule
    & \multicolumn{6}{c}{Patents Citations}\\
    \cline{2-7} 
 & \multicolumn{1}{c}{Full} & \multicolumn{1}{c}{2010-2014} & \multicolumn{1}{c}{2015-2022} & \multicolumn{1}{c}{Full} & \multicolumn{1}{c}{2010-2014} & \multicolumn{1}{c}{2015-2022} \\ 
   Model:                       & (1)           & (2)           & (3)           & (4)           & (5)           & (6)\\ 
   \midrule 
   CIFAR-10 (only)               & 0.412$^{*}$   & 0.692$^{**}$  & 0.359$\textsuperscript{\textdagger}$   & 0.481$^{**}$  & 1.089$^{***}$ & 0.379$\textsuperscript{\textdagger}$\\   
                                 & (0.169)       & (0.214)       & (0.199)       & (0.179)       & (0.214)       & (0.206)\\   
   CIFAR-10 (others)             & -0.052        & 0.349         & -0.012        & 0.034         & 0.952         & 0.014\\   
                                 & (0.187)       & (0.426)       & (0.171)       & (0.202)       & (0.624)       & (0.181)\\   
   ImageNet                      &               &               &               & 0.225$^{*}$   & 0.958$^{**}$  & 0.072\\   
                                 &               &               &               & (0.104)       & (0.353)       & (0.104)\\   
   log(Nb. Authors)              & 0.514$^{***}$ & -0.130        & 0.723$^{***}$ & 0.508$^{***}$ & -0.118        & 0.721$^{***}$\\   
                                 & (0.105)       & (0.157)       & (0.113)       & (0.105)       & (0.154)       & (0.114)\\   
   log(Nb. References)           & 0.639$^{***}$ & 1.745$^{***}$ & 0.434$^{**}$  & 0.620$^{***}$ & 1.620$^{***}$ & 0.428$^{**}$\\   
                                 & (0.152)       & (0.157)       & (0.150)       & (0.148)       & (0.146)       & (0.149)\\   
   International Collab.         & 0.097         & 0.191         & 0.061         & 0.094         & 0.113         & 0.061\\   
                                 & (0.079)       & (0.239)       & (0.092)       & (0.079)       & (0.272)       & (0.092)\\   
   Share Company Affil.          & 1.155$^{***}$ & 2.509$^{***}$ & 0.971$^{***}$ & 1.125$^{***}$ & 2.188$^{***}$ & 0.962$^{***}$\\   
                                 & (0.198)       & (0.417)       & (0.147)       & (0.194)       & (0.430)       & (0.147)\\   
   Nb. Datasets                  & 0.019         & 0.160         & 0.051         & 0.012         & 0.180         & 0.049\\   
                                 & (0.091)       & (0.121)       & (0.081)       & (0.090)       & (0.135)       & (0.081)\\   
   Nb. Tasks                     & 0.008$^{***}$ & 0.010$^{**}$  & 0.006$^{***}$ & 0.006$^{***}$ & 0.002         & 0.006$^{**}$\\   
                                 & (0.002)       & (0.004)       & (0.002)       & (0.002)       & (0.004)       & (0.002)\\   
   Nb. Modalities                & 0.089         & -0.646$\textsuperscript{\textdagger}$  & 0.181$^{*}$   & 0.142$\textsuperscript{\textdagger}$   & -0.507        & 0.197$^{*}$\\   
                                 & (0.070)       & (0.381)       & (0.077)       & (0.078)       & (0.396)       & (0.085)\\   
   \midrule 
   Pub. Venue Type Fixed Effect  & YES           & YES           & YES           & YES           & YES           & YES\\  
   Subject Area Fixed Effect     & YES           & YES           & YES           & YES           & YES           & YES\\  
   Publication Year Fixed Effect & YES           & YES           & YES           & YES           & YES           & YES\\  
   \midrule 
   Observations                  & 28,433        & 1,658         & 26,705        & 28,433        & 1,658         & 26,705\\  
   Dependent variable mean       & 0.15855       & 0.54403       & 0.13503       & 0.15855       & 0.54403       & 0.13503\\  
   Pseudo R$^2$                  & 0.26884       & 0.25634       & 0.26651       & 0.26965       & 0.27031       & 0.26660\\  
   \midrule \midrule

\end{tabular}
\par\endgroup

\begin{tablenotes} 
\item \footnotesize{\textit{Notes}: This table reports estimates of regressions of the models described in equations \ref{eqn:main_cifar} and \ref{eqn:main_cifar_imagenet} using an enlarged sample that encompasses all kinds of publication outlets. The dependent variable is the total number of patent families that cited the focal paper. The response variables are indicator variables that are equal to one if a paper mentions only CIFAR-10, CIFAR-10 among other datasets or ImageNet in the title, abstract or keywords. Columns (1) reports our baseline results of the estimates stemming from a Poisson regression.  Column (2) and (3) reports estimates of the same equation in a subset of the sample comprised of papers published from 2010 to 2014 and those published from 2015 to 2022, respectively. Columns (4 - 5) report estimates when adding a dataset indicator variable also for papers using ImageNet. Exponentiating the coefficients and differencing from one yields numbers interpretable as elasticities. All the specifications include publication venue type, publication year and scientific field fixed effects. Standard errors are clustered at the journal/conference level. Significance levels: \textdagger p<0.1; * p<0.05; ** p<0.01; *** p<0.001.}
\end{tablenotes} 
\end{threeparttable}
\end{adjustbox}
\end{table}

%% file: Tables/robustness_enlarged_science.tex
\begin{table}[!htbp] \centering 
  \caption{Robustness Check: Enlarged Sample - Scientific Citations} 
  \label{tab:robustness_enlarged_sample_science} 
  \footnotesize
  \setlength{\tabcolsep}{9pt}
  \renewcommand{\arraystretch}{1.2}
  \begin{adjustbox}{max width=\linewidth}
\begin{threeparttable}

\begingroup
\centering
\begin{tabular}{lcccccc}
   \tabularnewline \midrule \midrule
    & \multicolumn{6}{c}{Scientific Citations}\\
    \cline{2-7} 
 & \multicolumn{1}{c}{Full} & \multicolumn{1}{c}{2010-2014} & \multicolumn{1}{c}{2015-2022} & \multicolumn{1}{c}{Full} & \multicolumn{1}{c}{2010-2014} & \multicolumn{1}{c}{2015-2022} \\ 
   Model:                       & (1)           & (2)           & (3)           & (4)           & (5)           & (6)\\ 
   \midrule 
   CIFAR-10 (only)               & 0.192         & 1.250$^{**}$  & 0.082         & 0.363$^{*}$   & 1.416$^{***}$ & 0.251$\textsuperscript{\textdagger}$\\   
                                 & (0.145)       & (0.397)       & (0.133)       & (0.150)       & (0.388)       & (0.136)\\   
   CIFAR-10 (others)             & -0.102        & 0.313         & -0.107        & 0.045         & 0.550         & 0.038\\   
                                 & (0.167)       & (0.425)       & (0.166)       & (0.188)       & (0.474)       & (0.183)\\   
   ImageNet                      &               &               &               & 0.437$^{***}$ & 0.530$^{*}$   & 0.429$^{***}$\\   
                                 &               &               &               & (0.100)       & (0.208)       & (0.109)\\   
   log(Nb. Authors)              & 0.419$^{***}$ & 0.188         & 0.468$^{***}$ & 0.401$^{***}$ & 0.181         & 0.452$^{***}$\\   
                                 & (0.057)       & (0.166)       & (0.062)       & (0.057)       & (0.160)       & (0.062)\\   
   log(Nb. References)           & 1.186$^{***}$ & 1.337$^{***}$ & 1.178$^{***}$ & 1.164$^{***}$ & 1.290$^{***}$ & 1.157$^{***}$\\   
                                 & (0.101)       & (0.180)       & (0.098)       & (0.102)       & (0.178)       & (0.100)\\   
   International Collab.         & 0.263$^{***}$ & 0.367$^{*}$   & 0.254$^{***}$ & 0.256$^{***}$ & 0.332$\textsuperscript{\textdagger}$   & 0.250$^{***}$\\   
                                 & (0.044)       & (0.166)       & (0.049)       & (0.046)       & (0.170)       & (0.052)\\   
   Share Company Affil.          & 1.313$^{***}$ & 1.950$^{**}$  & 1.297$^{***}$ & 1.254$^{***}$ & 1.711$^{**}$  & 1.244$^{***}$\\   
                                 & (0.137)       & (0.595)       & (0.146)       & (0.130)       & (0.607)       & (0.139)\\   
   Nb. Datasets                  & 0.049         & 0.476$^{**}$  & 0.037         & 0.048         & 0.442$^{**}$  & 0.038\\   
                                 & (0.048)       & (0.148)       & (0.040)       & (0.047)       & (0.160)       & (0.039)\\   
   Nb. Tasks                     & 0.007$^{***}$ & 0.009$^{***}$ & 0.006$^{***}$ & 0.003$^{**}$  & 0.005         & 0.002$\textsuperscript{\textdagger}$\\   
                                 & (0.001)       & (0.003)       & (0.001)       & (0.001)       & (0.003)       & (0.001)\\   
   Nb. Modalities                & 0.136$^{*}$   & -0.022        & 0.154$^{**}$  & 0.254$^{***}$ & 0.039         & 0.272$^{***}$\\   
                                 & (0.059)       & (0.182)       & (0.059)       & (0.059)       & (0.179)       & (0.058)\\   
   \midrule 
   Pub. Venue Type Fixed Effect  & YES           & YES           & YES           & YES           & YES           & YES\\  
   Subject Area Fixed Effect     & YES           & YES           & YES           & YES           & YES           & YES\\  
   Publication Year Fixed Effect & YES           & YES           & YES           & YES           & YES           & YES\\  
   \midrule 
   Observations                  & 33,693        & 2,062         & 31,630        & 33,693        & 2,062         & 31,630\\  
   Dependent variable mean       & 18.511        & 45.833        & 16.731        & 18.511        & 45.833        & 16.731\\  
   Pseudo R$^2$                  & 0.43621       & 0.33513       & 0.44532       & 0.44171       & 0.34236       & 0.45083\\  
   \midrule \midrule

\end{tabular}
\par\endgroup

\begin{tablenotes} 
\item \footnotesize{\textit{Notes}: This table reports estimates of regressions of the models described in equations \ref{eqn:main_cifar} and \ref{eqn:main_cifar_imagenet} using an enlarged sample that encompasses all types of publication outlets and papers missing patent citation information. The dependent variable is the total number of scientific publications that cited the focal paper. The response variables are indicator variables that are equal to one if a paper mentions only CIFAR-10, CIFAR-10 among other datasets or ImageNet in the title, abstract or keywords. Columns (1) reports our baseline results of the estimates stemming from a Poisson regression.  Column (2) and (3) reports estimates of the same equation in a subset of the sample comprised of papers published from 2010 to 2014 and those published from 2015 to 2022, respectively. Columns (4 - 5) report estimates when adding a dataset indicator variable also for papers using ImageNet. Exponentiating the coefficients and differencing from one yields numbers interpretable as elasticities. All the specifications include publication venue type, publication year and scientific field fixed effects. Standard errors are clustered at the journal/conference level. Significance levels: \textdagger p<0.1; * p<0.05; ** p<0.01; *** p<0.001.}
\end{tablenotes} 

\end{threeparttable}
\end{adjustbox}
\end{table}

%% file: Tables/robustness_indicator_patents.tex
\begin{table}[!htbp] \centering 
  \caption{Robustness Check: Alternative Datasets Indicator Variables - Patent Citations} 
  \label{tab:robustness_alternative_indicator_patent} 
  \footnotesize
  \setlength{\tabcolsep}{9pt}
  \renewcommand{\arraystretch}{1.2}
  \begin{adjustbox}{max width=\linewidth}
\begin{threeparttable}

\begingroup
\centering
\begin{tabular}{lcccccc}
   \tabularnewline \midrule \midrule
    & \multicolumn{6}{c}{Patents Citations}\\
    \cline{2-7} 
 & \multicolumn{1}{c}{Full} & \multicolumn{1}{c}{2010-2014} & \multicolumn{1}{c}{2015-2022} & \multicolumn{1}{c}{Full} & \multicolumn{1}{c}{2010-2014} & \multicolumn{1}{c}{2015-2022} \\ 
   Model:                       & (1)           & (2)           & (3)           & (4)           & (5)           & (6)\\  
   \midrule 
   CIFAR-10                      & 0.126         & 0.491$^{*}$   & 0.131         & 0.207         & 1.011$^{**}$  & 0.156\\   
                                 & (0.143)       & (0.226)       & (0.146)       & (0.158)       & (0.336)       & (0.158)\\   
   ImageNet                      &               &               &               & 0.228$^{*}$   & 0.963$^{**}$  & 0.073\\   
                                 &               &               &               & (0.103)       & (0.342)       & (0.105)\\   
   log(Nb. Authors)              & 0.483$^{***}$ & -0.105        & 0.686$^{***}$ & 0.477$^{***}$ & -0.092        & 0.684$^{***}$\\   
                                 & (0.099)       & (0.156)       & (0.109)       & (0.100)       & (0.152)       & (0.109)\\   
   log(Nb. References)           & 0.674$^{***}$ & 1.756$^{***}$ & 0.472$^{**}$  & 0.655$^{***}$ & 1.623$^{***}$ & 0.465$^{**}$\\   
                                 & (0.149)       & (0.159)       & (0.147)       & (0.145)       & (0.151)       & (0.146)\\   
   International Collab.         & 0.084         & 0.189         & 0.046         & 0.081         & 0.111         & 0.046\\   
                                 & (0.079)       & (0.236)       & (0.093)       & (0.079)       & (0.268)       & (0.093)\\   
   Share Company Affil.          & 1.154$^{***}$ & 2.516$^{***}$ & 0.971$^{***}$ & 1.123$^{***}$ & 2.196$^{***}$ & 0.961$^{***}$\\   
                                 & (0.197)       & (0.418)       & (0.144)       & (0.193)       & (0.426)       & (0.145)\\   
   Nb. Datasets                  & -0.046        & 0.132         & -0.005        & -0.051        & 0.170         & -0.006\\   
                                 & (0.082)       & (0.098)       & (0.076)       & (0.082)       & (0.108)       & (0.076)\\   
   Nb. Tasks                     & 0.008$^{***}$ & 0.010$^{*}$   & 0.007$^{***}$ & 0.006$^{***}$ & 0.002         & 0.006$^{**}$\\   
                                 & (0.002)       & (0.004)       & (0.002)       & (0.002)       & (0.004)       & (0.002)\\   
   Nb. Modalities                & 0.102         & -0.600        & 0.191$^{*}$   & 0.155$^{*}$   & -0.478        & 0.208$^{*}$\\   
                                 & (0.070)       & (0.392)       & (0.076)       & (0.078)       & (0.406)       & (0.084)\\   
   \midrule 
   Pub. Venue Type Fixed Effect  & YES           & YES           & YES           & YES           & YES           & YES\\  
   Subject Area Fixed Effect     & YES           & YES           & YES           & YES           & YES           & YES\\  
   Publication Year Fixed Effect & YES           & YES           & YES           & YES           & YES           & YES\\  
   \midrule 
   Observations                  & 27,905        & 1,620         & 26,220        & 27,905        & 1,620         & 26,220\\  
   Dependent variable mean       & 0.15951       & 0.54691       & 0.13596       & 0.15951       & 0.54691       & 0.13596\\  
   Pseudo R$^2$                  & 0.26509       & 0.25357       & 0.26186       & 0.26593       & 0.26791       & 0.26195\\  
   \midrule \midrule

\end{tabular}
\par\endgroup

\begin{tablenotes} 
\item \footnotesize{\textit{Notes}: This table reports estimates of regressions of the models described in equations \ref{eqn:main_cifar} and \ref{eqn:main_cifar_imagenet}. The dependent variable is the total number of patent families that cited the focal paper. The response variables are indicator variables that are equal to one if a paper mentions CIFAR-10 or ImageNet in the title, abstract or keywords. Columns (1) reports our baseline results of the estimates stemming from a Poisson regression.  Column (2) and (3) reports estimates of the same equation in a subset of the sample comprised of papers published from 2010 to 2014 and those published from 2015 to 2022, respectively. Columns (4 - 5) report estimates when adding a dataset indicator variable also for papers using ImageNet. Exponentiating the coefficients and differencing from one yields numbers interpretable as elasticities. All the specifications include publication venue type, publication year and scientific field fixed effects. Standard errors are clustered at the journal/conference level. Significance levels: \textdagger p<0.1; * p<0.05; ** p<0.01; *** p<0.001.}
\end{tablenotes} 
\end{threeparttable}
\end{adjustbox}
\end{table}

%% file: Tables/robustness_indicator_science.tex
\begin{table}[!htbp] \centering 
  \caption{Robustness Check: Alternative Datasets Indicator Variables - Scientific Citations} 
  \label{tab:robustness_alternative_indicator_science} 
  \footnotesize
  \setlength{\tabcolsep}{9pt}
  \renewcommand{\arraystretch}{1.2}
  \begin{adjustbox}{max width=\linewidth}
\begin{threeparttable}

\begingroup
\centering
\begin{tabular}{lcccccc}
   \tabularnewline \midrule \midrule
    & \multicolumn{6}{c}{Scientific Citations}\\
    \cline{2-7} 
 & \multicolumn{1}{c}{Full} & \multicolumn{1}{c}{2010-2014} & \multicolumn{1}{c}{2015-2022} & \multicolumn{1}{c}{Full} & \multicolumn{1}{c}{2010-2014} & \multicolumn{1}{c}{2015-2022} \\  
   Model:                       & (1)           & (2)           & (3)           & (4)           & (5)           & (6)\\   
   \midrule 
   CIFAR-10                      & -0.212        & 0.891$^{*}$   & -0.255$\textsuperscript{\textdagger}$  & -0.074        & 1.177$^{*}$   & -0.117\\   
                                 & (0.153)       & (0.392)       & (0.133)       & (0.174)       & (0.468)       & (0.154)\\   
   ImageNet                      &               &               &               & 0.386$^{***}$ & 0.575$^{*}$   & 0.384$^{***}$\\   
                                 &               &               &               & (0.092)       & (0.274)       & (0.093)\\   
   log(Nb. Authors)              & 0.352$^{***}$ & -0.125        & 0.440$^{***}$ & 0.342$^{***}$ & -0.122        & 0.429$^{***}$\\   
                                 & (0.073)       & (0.181)       & (0.077)       & (0.072)       & (0.181)       & (0.076)\\   
   log(Nb. References)           & 1.202$^{***}$ & 1.362$^{***}$ & 1.180$^{***}$ & 1.180$^{***}$ & 1.320$^{***}$ & 1.157$^{***}$\\   
                                 & (0.135)       & (0.261)       & (0.136)       & (0.138)       & (0.253)       & (0.140)\\   
   International Collab.         & 0.322$^{***}$ & 0.339$^{*}$   & 0.320$^{***}$ & 0.319$^{***}$ & 0.301$\textsuperscript{\textdagger}$   & 0.320$^{***}$\\   
                                 & (0.040)       & (0.172)       & (0.038)       & (0.042)       & (0.178)       & (0.039)\\   
   Share Company Affil.          & 1.275$^{***}$ & 1.262$^{**}$  & 1.289$^{***}$ & 1.227$^{***}$ & 1.082$^{*}$   & 1.244$^{***}$\\   
                                 & (0.150)       & (0.480)       & (0.154)       & (0.144)       & (0.464)       & (0.150)\\   
   Nb. Datasets                  & 0.012         & 0.377$^{**}$  & 0.015         & 0.005         & 0.371$^{**}$  & 0.008\\   
                                 & (0.056)       & (0.119)       & (0.047)       & (0.056)       & (0.140)       & (0.047)\\   
   Nb. Tasks                     & 0.007$^{***}$ & 0.007$^{**}$  & 0.007$^{***}$ & 0.004$^{***}$ & 0.003         & 0.004$^{**}$\\   
                                 & (0.001)       & (0.003)       & (0.001)       & (0.001)       & (0.003)       & (0.001)\\   
   Nb. Modalities                & 0.192$^{***}$ & 0.216         & 0.199$^{***}$ & 0.294$^{***}$ & 0.282         & 0.303$^{***}$\\   
                                 & (0.045)       & (0.189)       & (0.047)       & (0.047)       & (0.194)       & (0.048)\\   
   \midrule 
   Pub. Venue Type Fixed Effect  & YES           & YES           & YES           & YES           & YES           & YES\\  
   Subject Area Fixed Effect     & YES           & YES           & YES           & YES           & YES           & YES\\  
   Publication Year Fixed Effect & YES           & YES           & YES           & YES           & YES           & YES\\  
   \midrule 
   Observations                  & 28,393        & 1,734         & 26,659        & 28,393        & 1,734         & 26,659\\  
   Dependent variable mean       & 16.365        & 39.354        & 14.870        & 16.365        & 39.354        & 14.870\\  
   Pseudo R$^2$                  & 0.41067       & 0.27869       & 0.42113       & 0.41498       & 0.28699       & 0.42552\\  
   \midrule \midrule

\end{tabular}
\par\endgroup

\begin{tablenotes} 
\item \footnotesize{\textit{Notes}: This table reports estimates of regressions of the models described in equations \ref{eqn:main_cifar} and \ref{eqn:main_cifar_imagenet}. The dependent variable is the total number scientific citations received by a paper. The response variables are indicator variables that are equal to one if a paper mentions CIFAR-10 or ImageNet in the title, abstract or keywords. Columns (1) reports our baseline results of the estimates stemming from a Poisson regression.  Column (2) and (3) reports estimates of the same equation in a subset of the sample comprised of papers published from 2010 to 2014 and those published from 2015 to 2022, respectively. Columns (4 - 5) report estimates when adding a dataset indicator variable also for papers using ImageNet. Exponentiating the coefficients and differencing from one yields numbers interpretable as elasticities. All the specifications include publication venue type, publication year and scientific field fixed effects. Standard errors are clustered at the journal/conference level. Significance levels: \textdagger p<0.1; * p<0.05; ** p<0.01; *** p<0.001.}
\end{tablenotes} 
\end{threeparttable}
\end{adjustbox}
\end{table}

%% file: Tables/robustness_3years_window_patents.tex
\begin{table}[!htbp] \centering 
  \caption{Robustness Check: Labeled Datasets and Patent Citations - 3-Years Window} 
  \label{tab:robustness_cit_3years_patent} 
  \footnotesize
  \setlength{\tabcolsep}{9pt}
  \renewcommand{\arraystretch}{1.2}
  \begin{adjustbox}{max width=\linewidth}
\begin{threeparttable}

\begingroup
\centering
\begin{tabular}{lcccccc}
   \tabularnewline \midrule \midrule
    & \multicolumn{6}{c}{Patents Citations - 3 Years Window}\\
    \cline{2-7} 
 & \multicolumn{1}{c}{Full} & \multicolumn{1}{c}{2010-2014} & \multicolumn{1}{c}{2015-2022} & \multicolumn{1}{c}{Full} & \multicolumn{1}{c}{2010-2014} & \multicolumn{1}{c}{2015-2022} \\  
   Model:                       & (1)           & (2)           & (3)           & (4)           & (5)           & (6)\\  
   \midrule 
   CIFAR-10 (only)               & 0.431$^{*}$   & 0.530$^{*}$   & 0.409$\textsuperscript{\textdagger}$   & 0.494$^{*}$   & 0.894$^{***}$ & 0.432$\textsuperscript{\textdagger}$\\   
                                 & (0.188)       & (0.232)       & (0.215)       & (0.196)       & (0.218)       & (0.224)\\   
   CIFAR-10 (others)             & -0.078        & -1.519$^{*}$  & 0.053         & 0.005         & -0.982        & 0.084\\   
                                 & (0.209)       & (0.652)       & (0.200)       & (0.222)       & (0.761)       & (0.210)\\   
   ImageNet                      &               &               &               & 0.216$^{*}$   & 0.867$^{**}$  & 0.081\\   
                                 &               &               &               & (0.107)       & (0.326)       & (0.119)\\   
   log(Nb. Authors)              & 0.530$^{***}$ & 0.082         & 0.684$^{***}$ & 0.523$^{***}$ & 0.091         & 0.681$^{***}$\\   
                                 & (0.103)       & (0.246)       & (0.121)       & (0.104)       & (0.243)       & (0.121)\\   
   log(Nb. References)           & 0.684$^{***}$ & 1.831$^{***}$ & 0.513$^{**}$  & 0.665$^{***}$ & 1.716$^{***}$ & 0.506$^{**}$\\   
                                 & (0.181)       & (0.188)       & (0.186)       & (0.177)       & (0.180)       & (0.185)\\   
   International Collab.         & 0.119         & 0.262         & 0.075         & 0.117         & 0.204         & 0.075\\   
                                 & (0.090)       & (0.234)       & (0.108)       & (0.090)       & (0.262)       & (0.108)\\   
   Share Company Affil.          & 1.294$^{***}$ & 2.944$^{***}$ & 1.101$^{***}$ & 1.264$^{***}$ & 2.658$^{***}$ & 1.091$^{***}$\\   
                                 & (0.191)       & (0.510)       & (0.171)       & (0.188)       & (0.535)       & (0.171)\\   
   Nb. Datasets                  & 0.034         & 0.368$^{**}$  & 0.027         & 0.029         & 0.389$^{**}$  & 0.025\\   
                                 & (0.105)       & (0.126)       & (0.103)       & (0.104)       & (0.133)       & (0.103)\\   
   Nb. Tasks                     & 0.008$^{***}$ & 0.010$^{***}$ & 0.007$^{***}$ & 0.006$^{**}$  & 0.002         & 0.006$^{**}$\\   
                                 & (0.002)       & (0.003)       & (0.002)       & (0.002)       & (0.004)       & (0.002)\\   
   Nb. Modalities                & 0.064         & -0.592        & 0.160$\textsuperscript{\textdagger}$   & 0.113         & -0.483        & 0.178$\textsuperscript{\textdagger}$\\   
                                 & (0.083)       & (0.393)       & (0.091)       & (0.089)       & (0.396)       & (0.098)\\   
   \midrule 
   Pub. Venue Type Fixed Effect  & YES           & YES           & YES           & YES           & YES           & YES\\  
   Subject Area Fixed Effect     & YES           & YES           & YES           & YES           & YES           & YES\\  
   Publication Year Fixed Effect & YES           & YES           & YES           & YES           & YES           & YES\\  
   \midrule 
   Observations                  & 10,114        & 1,579         & 8,429         & 10,114        & 1,579         & 8,429\\  
   Dependent variable mean       & 0.31016       & 0.32552       & 0.31119       & 0.31016       & 0.32552       & 0.31119\\  
   Pseudo R$^2$                  & 0.13808       & 0.26681       & 0.13424       & 0.13902       & 0.27749       & 0.13437\\  
   \midrule \midrule

\end{tabular}
\par\endgroup

\begin{tablenotes} 
\item \footnotesize{\textit{Notes}: This table reports estimates of regressions of the models described in equations \ref{eqn:main_cifar} and \ref{eqn:main_cifar_imagenet}. The dependent variable is the total number of patent citations received by a paper within 3 years of the publication year. The response variables are indicator variables that are equal to one if a paper mentions only CIFAR-10, CIFAR-10 among other datasets or ImageNet in the title, abstract or keywords. Columns (1) reports our baseline results of the estimates stemming from a Poisson regression.  Column (2) and (3) reports estimates of the same equation in a subset of the sample comprised of papers published from 2010 to 2014 and those published from 2015 to 2019, respectively. Columns (4 - 5) report estimates when adding a dataset indicator variable also for papers using ImageNet. Exponentiating the coefficients and differencing from one yields numbers interpretable as elasticities. All the specifications include publication venue type, publication year and scientific field fixed effects. Standard errors are clustered at the journal/conference level. Significance levels: \textdagger p<0.1; * p<0.05; ** p<0.01; *** p<0.001.}
\end{tablenotes} 
\end{threeparttable}
\end{adjustbox}
\end{table}

%% file: Tables/robustness_3years_window_science.tex
\begin{table}[!htbp] \centering 
  \caption{Robustness Check: Labeled Datasets and Scientific Citations - 3-Years Window} 
  \label{tab:robustness_cit_3years_science} 
  \footnotesize
  \setlength{\tabcolsep}{9pt}
  \renewcommand{\arraystretch}{1.2}
  \begin{adjustbox}{max width=\linewidth}
\begin{threeparttable}

\begingroup
\centering

\begin{tabular}{lcccccc}
   \tabularnewline \midrule \midrule
    & \multicolumn{6}{c}{Scientific Citations - 3 Years Window}\\
    \cline{2-7} 
 & \multicolumn{1}{c}{Full} & \multicolumn{1}{c}{2010-2014} & \multicolumn{1}{c}{2015-2022} & \multicolumn{1}{c}{Full} & \multicolumn{1}{c}{2010-2014} & \multicolumn{1}{c}{2015-2022} \\ 
   Model:                       & (1)           & (2)           & (3)           & (4)           & (5)           & (6)\\  
   \midrule 
   CIFAR-10 (only)               & 0.084         & 0.800$^{***}$ & 0.063         & 0.217$^{*}$   & 0.960$^{***}$ & 0.196$^{*}$\\   
                                 & (0.096)       & (0.235)       & (0.093)       & (0.088)       & (0.230)       & (0.090)\\   
   CIFAR-10 (others)             & -0.388$^{*}$  & -0.284        & -0.367$^{*}$  & -0.249        & -0.034        & -0.227\\   
                                 & (0.174)       & (0.452)       & (0.170)       & (0.211)       & (0.483)       & (0.205)\\   
   ImageNet                      &               &               &               & 0.393$^{***}$ & 0.473$^{**}$  & 0.391$^{***}$\\   
                                 &               &               &               & (0.102)       & (0.180)       & (0.102)\\   
   log(Nb. Authors)              & 0.465$^{***}$ & -0.039        & 0.517$^{***}$ & 0.452$^{***}$ & -0.038        & 0.503$^{***}$\\   
                                 & (0.097)       & (0.108)       & (0.103)       & (0.097)       & (0.107)       & (0.104)\\   
   log(Nb. References)           & 1.315$^{***}$ & 1.514$^{***}$ & 1.297$^{***}$ & 1.292$^{***}$ & 1.483$^{***}$ & 1.274$^{***}$\\   
                                 & (0.132)       & (0.184)       & (0.134)       & (0.136)       & (0.185)       & (0.138)\\   
   International Collab.         & 0.308$^{***}$ & 0.269$^{**}$  & 0.311$^{***}$ & 0.310$^{***}$ & 0.245$^{*}$   & 0.315$^{***}$\\   
                                 & (0.041)       & (0.098)       & (0.047)       & (0.039)       & (0.103)       & (0.045)\\   
   Share Company Affil.          & 1.200$^{***}$ & 1.365$^{***}$ & 1.202$^{***}$ & 1.149$^{***}$ & 1.233$^{***}$ & 1.152$^{***}$\\   
                                 & (0.199)       & (0.373)       & (0.200)       & (0.190)       & (0.350)       & (0.191)\\   
   Nb. Datasets                  & 0.089$^{*}$   & 0.533$^{***}$ & 0.072$\textsuperscript{\textdagger}$   & 0.083$\textsuperscript{\textdagger}$   & 0.532$^{***}$ & 0.067\\   
                                 & (0.044)       & (0.092)       & (0.043)       & (0.043)       & (0.099)       & (0.043)\\   
   Nb. Tasks                     & 0.006$^{***}$ & 0.002         & 0.006$^{***}$ & 0.002$^{*}$   & -0.002        & 0.003$^{*}$\\   
                                 & (0.001)       & (0.002)       & (0.001)       & (0.001)       & (0.002)       & (0.001)\\   
   Nb. Modalities                & 0.194$^{**}$  & 0.118         & 0.205$^{***}$ & 0.294$^{***}$ & 0.165         & 0.307$^{***}$\\   
                                 & (0.060)       & (0.119)       & (0.061)       & (0.059)       & (0.120)       & (0.060)\\   
   \midrule 
   Pub. Venue Type Fixed Effect  & YES           & YES           & YES           & YES           & YES           & YES\\  
   Subject Area Fixed Effect     & YES           & YES           & YES           & YES           & YES           & YES\\  
   Publication Year Fixed Effect & YES           & YES           & YES           & YES           & YES           & YES\\  
   \midrule 
   Observations                  & 10,346        & 1,734         & 8,612         & 10,346        & 1,734         & 8,612\\  
   Dependent variable mean       & 21.480        & 11.685        & 23.452        & 21.480        & 11.685        & 23.452\\  
   Pseudo R$^2$                  & 0.33006       & 0.32315       & 0.32350       & 0.33576       & 0.32905       & 0.32935\\  
   \midrule \midrule

\end{tabular}
\par\endgroup

\begin{tablenotes} 
\item \footnotesize{\textit{Notes}: This table reports estimates of regressions of the models described in equations \ref{eqn:main_cifar} and \ref{eqn:main_cifar_imagenet}. The dependent variable is the total number of scientific citations received by a paper within 3 years of the publication year. The response variables are indicator variables that are equal to one if a paper mentions only CIFAR-10, CIFAR-10 among other datasets or ImageNet in the title, abstract or keywords. Columns (1) reports our baseline results of the estimates stemming from a Poisson regression.  Column (2) and (3) reports estimates of the same equation in a subset of the sample comprised of papers published from 2010 to 2014 and those published from 2015 to 2019, respectively. Columns (4 - 5) report estimates when adding a dataset indicator variable also for papers using ImageNet. Exponentiating the coefficients and differencing from one yields numbers interpretable as elasticities. All the specifications include publication venue type, publication year and scientific field fixed effects. Standard errors are clustered at the journal/conference level. Significance levels: \textdagger p<0.1; * p<0.05; ** p<0.01; *** p<0.001.}
\end{tablenotes} 
\end{threeparttable}
\end{adjustbox}
\end{table}

%% file: references.bib
@article{rose_pybliometrics_2019,
	title = {pybliometrics: {Scriptable} bibliometrics using a {Python} interface to {Scopus}},
	volume = {10},
	issn = {2352-7110},
	shorttitle = {pybliometrics},
	url = {https://www.sciencedirect.com/science/article/pii/S2352711019300573},
	doi = {10.1016/j.softx.2019.100263},
	language = {en},
	urldate = {2023-05-14},
	journal = {SoftwareX},
	author = {Rose, Michael E. and Kitchin, John R.},
	month = jul,
	year = {2019},
	pages = {100263},
}

@misc{graham_fractional_2015,
	title = {Fractional {Max}-{Pooling}},
	urldate = {2023-05-06},
	publisher = {arXiv},
	author = {Graham, Benjamin},
	month = may,
	year = {2015},
	note = {arXiv:1412.6071 [cs]
version: 4},
}

@article{stokel-walker_what_2023,
	title = {What {ChatGPT} and generative {AI} mean for science},
	volume = {614},
	copyright = {2023 Springer Nature Limited},
	url = {https://www.nature.com/articles/d41586-023-00340-6},
	doi = {10.1038/d41586-023-00340-6},
	language = {en},
	number = {7947},
	urldate = {2023-03-16},
	journal = {Nature},
	author = {Stokel-Walker, Chris and Van Noorden, Richard},
	month = feb,
	year = {2023},
	note = {Bandiera\_abtest: a
Cg\_type: News Feature
Number: 7947
Publisher: Nature Publishing Group
Subject\_term: Publishing, Machine learning, Mathematics and computing},
	pages = {214--216},
}

@book{stokes_pasteurs_2011,
	title = {Pasteur's {Quadrant}: {Basic} {Science} and {Technological} {Innovation}},
	isbn = {978-0-8157-1907-6},
	shorttitle = {Pasteur's {Quadrant}},
	language = {en},
	publisher = {Brookings Institution Press},
	author = {Stokes, Donald E.},
	month = mar,
	year = {2011},
	note = {Google-Books-ID: TLKDbvJX86YC},
}

@article{krizhevskyImageNetClassificationDeep2017,
	title = {{ImageNet} classification with deep convolutional neural networks},
	volume = {60},
	issn = {0001-0782},
	url = {https://doi.org/10.1145/3065386},
	doi = {10.1145/3065386},
	number = {6},
	urldate = {2022-09-02},
	journal = {Commun. ACM},
	author = {Krizhevsky, Alex and Sutskever, Ilya and Hinton, Geoffrey E.},
	month = may,
	year = {2017},
	pages = {84--90},
}

@article{torralba80MillionTiny2008,
	title = {80 million tiny images: a large dataset for non-parametric object and scene recognition},
	language = {en},
	journal = {IEEE TRANSACTIONS ON PATTERN ANALYSIS AND MACHINE INTELLIGENCE},
	author = {Torralba, Antonio and Fergus, Rob and Freeman, William T},
	year = {2008},
	pages = {12},
}

@book{kuhnStructureScientificRevolutions1970,
	address = {Chicago},
	edition = {2d ed.},
	series = {International encyclopedia of unified science. {Foundations} of the unity of science, v. 2, no. 2},
	title = {The structure of scientific revolutions},
	isbn = {978-0-226-45803-8},
	language = {en},
	publisher = {University of Chicago Press},
	author = {Kuhn, Thomas S.},
	collaborator = {Ralph Ellison Collection (Library of Congress)},
	year = {1970},
}

@incollection{raimbaultEmergenceTechnoscientificFields2021,
	address = {Cham},
	series = {Sociology of the {Sciences} {Yearbook}},
	title = {The {Emergence} of {Technoscientific} {Fields} and the {New} {Political} {Sociology} of {Science}},
	language = {en},
	urldate = {2022-09-18},
	booktitle = {Community and {Identity} in {Contemporary} {Technosciences}},
	publisher = {Springer International Publishing},
	author = {Raimbault, Benjamin and Joly, Pierre-Benoît},
	editor = {Kastenhofer, Karen and Molyneux-Hodgson, Susan},
	year = {2021},
	pages = {85--106},
}

@book{kastenhoferCommunityIdentityContemporary2021,
	address = {Cham},
	series = {Sociology of the {Sciences} {Yearbook}},
	title = {Community and {Identity} in {Contemporary} {Technosciences}},
	volume = {31},
	url = {https://link.springer.com/10.1007/978-3-030-61728-8},
	language = {en},
	urldate = {2022-09-19},
	publisher = {Springer International Publishing},
	editor = {Kastenhofer, Karen and Molyneux-Hodgson, Susan},
	year = {2021},
	doi = {10.1007/978-3-030-61728-8},
}

@article{bensaude-vincentMattersInterestObjects2011,
	title = {Matters of {Interest}: {The} {Objects} of {Research} in {Science} and {Technoscience}},
	volume = {42},
	issn = {1572-8587},
	shorttitle = {Matters of {Interest}},
	url = {https://doi.org/10.1007/s10838-011-9172-y},
	doi = {10.1007/s10838-011-9172-y},
	language = {en},
	number = {2},
	urldate = {2022-09-19},
	journal = {J Gen Philos Sci},
	author = {Bensaude-Vincent, Bernadette and Loeve, Sacha and Nordmann, Alfred and Schwarz, Astrid},
	month = nov,
	year = {2011},
	pages = {365--383},
}

@article{schmidhuberDeepLearningNeural2015,
	title = {Deep learning in neural networks: {An} overview},
	volume = {61},
	issn = {0893-6080},
	shorttitle = {Deep learning in neural networks},
	url = {https://www.sciencedirect.com/science/article/pii/S0893608014002135},
	doi = {10.1016/j.neunet.2014.09.003},
	language = {en},
	urldate = {2022-10-16},
	journal = {Neural Networks},
	author = {Schmidhuber, Jürgen},
	month = jan,
	year = {2015},
	pages = {85--117},
}

@article{frickelGeneralTheoryScientific2005,
	title = {A {General} {Theory} of {Scientific}/{Intellectual} {Movements}},
	volume = {70},
	issn = {0003-1224},
	url = {https://doi.org/10.1177/000312240507000202},
	doi = {10.1177/000312240507000202},
	language = {en},
	number = {2},
	urldate = {2022-12-15},
	journal = {Am Sociol Rev},
	author = {Frickel, Scott and Gross, Neil},
	month = apr,
	year = {2005},
	note = {Publisher: SAGE Publications Inc},
	pages = {204--232},
}

@article{bengioDeepLearningAI2021,
	title = {Deep learning for {AI}},
	volume = {64},
	issn = {0001-0782},
	url = {https://doi.org/10.1145/3448250},
	doi = {10.1145/3448250},
	number = {7},
	urldate = {2022-12-17},
	journal = {Commun. ACM},
	author = {Bengio, Yoshua and Lecun, Yann and Hinton, Geoffrey},
	month = jun,
	year = {2021},
	pages = {58--65},
}

@article{lecunDeepLearning2015,
	title = {Deep learning},
	volume = {521},
	issn = {0028-0836, 1476-4687},
	url = {http://www.nature.com/articles/nature14539},
	doi = {10.1038/nature14539},
	language = {en},
	number = {7553},
	urldate = {2022-12-17},
	journal = {Nature},
	author = {LeCun, Yann and Bengio, Yoshua and Hinton, Geoffrey},
	month = may,
	year = {2015},
	pages = {436--444},
}

@article{waldropWhatAreLimits2019,
	title = {What are the limits of deep learning?},
	volume = {116},
	url = {https://www.pnas.org/doi/abs/10.1073/pnas.1821594116},
	doi = {10.1073/pnas.1821594116},
	number = {4},
	urldate = {2022-12-18},
	journal = {Proceedings of the National Academy of Sciences},
	author = {Waldrop, M. Mitchell},
	month = jan,
	year = {2019},
	note = {Publisher: Proceedings of the National Academy of Sciences},
	pages = {1074--1077},
}

@misc{goldman10YearsLater2022,
	title = {10 years later, deep learning ‘revolution’ rages on, say {AI} pioneers {Hinton}, {LeCun} and {Li}},
	url = {https://venturebeat.com/ai/10-years-on-ai-pioneers-hinton-lecun-li-say-deep-learning-revolution-will-continue/},
	language = {en-US},
	urldate = {2022-12-19},
	journal = {VentureBeat},
	author = {Goldman, Sharon},
	month = sep,
	year = {2022},
}

@incollection{weberRiseDatadrivenAI2021,
	title = {: {On} the rise of data-driven {AI} and its epistemontological foundations},
	isbn = {978-0-429-19853-3},
	booktitle = {The {Routledge} {Social} {Science} {Handbook} of {AI}},
	publisher = {Routledge},
	author = {Weber, Jutta and Prietl, Bianca},
	year = {2021},
	note = {Num Pages: 16},
}

@article{mertonPrioritiesScientificDiscovery1957,
	title = {Priorities in {Scientific} {Discovery}: {A} {Chapter} in the {Sociology} of {Science}},
	volume = {22},
	issn = {0003-1224},
	shorttitle = {Priorities in {Scientific} {Discovery}},
	url = {https://www.jstor.org/stable/2089193},
	doi = {10.2307/2089193},
	number = {6},
	urldate = {2022-12-20},
	journal = {American Sociological Review},
	author = {Merton, Robert K.},
	year = {1957},
	note = {Publisher: [American Sociological Association, Sage Publications, Inc.]},
	pages = {635--659},
}

@article{klingerDeepLearningDeep2021,
	title = {Deep learning, deep change? {Mapping} the evolution and geography of a general purpose technology},
	volume = {126},
	issn = {1588-2861},
	shorttitle = {Deep learning, deep change?},
	url = {https://doi.org/10.1007/s11192-021-03936-9},
	doi = {10.1007/s11192-021-03936-9},
	language = {en},
	number = {7},
	urldate = {2022-12-23},
	journal = {Scientometrics},
	author = {Klinger, Joel and Mateos-Garcia, Juan and Stathoulopoulos, Konstantinos},
	month = jul,
	year = {2021},
	pages = {5589--5621},
}

@article{jumperHighlyAccurateProtein2021,
	title = {Highly accurate protein structure prediction with {AlphaFold}},
	volume = {596},
	copyright = {2021 The Author(s)},
	issn = {1476-4687},
	url = {https://www.nature.com/articles/s41586-021-03819-2},
	doi = {10.1038/s41586-021-03819-2},
	language = {en},
	number = {7873},
	urldate = {2022-12-23},
	journal = {Nature},
	author = {Jumper, John and Evans, Richard and Pritzel, Alexander and Green, Tim and Figurnov, Michael and Ronneberger, Olaf and Tunyasuvunakool, Kathryn and Bates, Russ and Žídek, Augustin and Potapenko, Anna and Bridgland, Alex and Meyer, Clemens and Kohl, Simon A. A. and Ballard, Andrew J. and Cowie, Andrew and Romera-Paredes, Bernardino and Nikolov, Stanislav and Jain, Rishub and Adler, Jonas and Back, Trevor and Petersen, Stig and Reiman, David and Clancy, Ellen and Zielinski, Michal and Steinegger, Martin and Pacholska, Michalina and Berghammer, Tamas and Bodenstein, Sebastian and Silver, David and Vinyals, Oriol and Senior, Andrew W. and Kavukcuoglu, Koray and Kohli, Pushmeet and Hassabis, Demis},
	month = aug,
	year = {2021},
	note = {Number: 7873
Publisher: Nature Publishing Group},
	pages = {583--589},
}

@article{craftsArtificialIntelligenceGeneralpurpose2021,
	title = {Artificial intelligence as a general-purpose technology: an historical perspective},
	volume = {37},
	issn = {0266-903X},
	shorttitle = {Artificial intelligence as a general-purpose technology},
	url = {https://doi.org/10.1093/oxrep/grab012},
	doi = {10.1093/oxrep/grab012},
	number = {3},
	urldate = {2022-12-23},
	journal = {Oxford Review of Economic Policy},
	author = {Crafts, Nicholas},
	month = sep,
	year = {2021},
	pages = {521--536},
}

@article{callawayItWillChange2020,
	title = {‘{It} will change everything’: {DeepMind}’s {AI} makes gigantic leap in solving protein structures},
	volume = {588},
	copyright = {2021 Nature},
	shorttitle = {‘{It} will change everything’},
	url = {https://www.nature.com/articles/d41586-020-03348-4},
	doi = {10.1038/d41586-020-03348-4},
	language = {en},
	number = {7837},
	urldate = {2022-12-23},
	journal = {Nature},
	author = {Callaway, Ewen},
	month = nov,
	year = {2020},
	note = {Bandiera\_abtest: a
Cg\_type: News
Number: 7837
Publisher: Nature Publishing Group
Subject\_term: Computational biology and bioinformatics, Structural biology, Drug discovery},
	pages = {203--204},
}

@article{bianchiniArtificialIntelligenceScience2022,
	title = {Artificial intelligence in science: {An} emerging general method of invention},
	volume = {51},
	issn = {0048-7333},
	shorttitle = {Artificial intelligence in science},
	url = {https://www.sciencedirect.com/science/article/pii/S0048733322001275},
	doi = {10.1016/j.respol.2022.104604},
	language = {en},
	number = {10},
	urldate = {2022-12-23},
	journal = {Research Policy},
	author = {Bianchini, Stefano and Müller, Moritz and Pelletier, Pierre},
	month = dec,
	year = {2022},
	pages = {104604},
}

@article{chahDeepRabbitHole2019,
	title = {Down the deep rabbit hole: {Untangling} deep learning from machine learning and artificial intelligence},
	copyright = {Copyright (c) 2019 First Monday},
	issn = {1396-0466},
	shorttitle = {Down the deep rabbit hole},
	url = {https://firstmonday.org/ojs/index.php/fm/article/view/8237},
	doi = {10.5210/fm.v24i2.8237},
	language = {en},
	urldate = {2023-01-13},
	journal = {First Monday},
	author = {Chah, Niel},
	month = feb,
	year = {2019},
}

@article{kerstingMachineLearningArtificial2018,
	title = {Machine {Learning} and {Artificial} {Intelligence}: {Two} {Fellow} {Travelers} on the {Quest} for {Intelligent} {Behavior} in {Machines}},
	volume = {1},
	issn = {2624-909X},
	shorttitle = {Machine {Learning} and {Artificial} {Intelligence}},
	url = {https://www.frontiersin.org/articles/10.3389/fdata.2018.00006},
	urldate = {2023-01-13},
	journal = {Frontiers in Big Data},
	author = {Kersting, Kristian},
	year = {2018},
}

@misc{farrowTuringAwardHonours2019,
	title = {Turing {Award} honours {CIFAR}’s ‘pioneers of {AI}’},
	url = {https://cifar.ca/cifarnews/2019/03/27/turing-award-honours-cifar-s-pioneers-of-ai/},
	language = {en-US},
	urldate = {2023-01-13},
	journal = {CIFAR},
	author = {Farrow, Jon},
	month = mar,
	year = {2019},
}

@article{krizhevsky2009learning,
	title = {Learning multiple layers of features from tiny images},
	author = {Krizhevsky, Alex},
	year = {2009},
	note = {Publisher: Toronto, ON, Canada},
}

@misc{brownellHowArtificialIntelligence2016a,
	title = {How the artificial intelligence revolution was born in a {Vancouver} hotel},
	url = {https://financialpost.com/technology/how-the-artificial-intelligence-revolution-was-born-in-a-vancouver-hotel},
	language = {en-CA},
	urldate = {2023-01-13},
	journal = {Financial Post},
	author = {Brownell, Claire},
	year = {2016},
}

@article{hintonFastLearningAlgorithm2006,
	title = {A {Fast} {Learning} {Algorithm} for {Deep} {Belief} {Nets}},
	volume = {18},
	issn = {0899-7667},
	url = {https://doi.org/10.1162/neco.2006.18.7.1527},
	doi = {10.1162/neco.2006.18.7.1527},
	number = {7},
	urldate = {2023-01-14},
	journal = {Neural Computation},
	author = {Hinton, Geoffrey E. and Osindero, Simon and Teh, Yee-Whye},
	month = jul,
	year = {2006},
	pages = {1527--1554},
}

@article{pavlikCollaboratingChatGPTConsidering2023,
	title = {Collaborating {With} {ChatGPT}: {Considering} the {Implications} of {Generative} {Artificial} {Intelligence} for {Journalism} and {Media} {Education}},
	issn = {1077-6958},
	shorttitle = {Collaborating {With} {ChatGPT}},
	url = {https://doi.org/10.1177/10776958221149577},
	doi = {10.1177/10776958221149577},
	language = {en},
	urldate = {2023-01-14},
	journal = {Journalism \& Mass Communication Educator},
	author = {Pavlik, John V.},
	month = jan,
	year = {2023},
	note = {Publisher: SAGE Publications Inc},
	pages = {10776958221149577},
}

@misc{jeblickChatGPTMakesMedicine2022,
	title = {{ChatGPT} {Makes} {Medicine} {Easy} to {Swallow}: {An} {Exploratory} {Case} {Study} on {Simplified} {Radiology} {Reports}},
	shorttitle = {{ChatGPT} {Makes} {Medicine} {Easy} to {Swallow}},
	urldate = {2023-01-14},
	publisher = {arXiv},
	author = {Jeblick, Katharina and Schachtner, Balthasar and Dexl, Jakob and Mittermeier, Andreas and Stüber, Anna Theresa and Topalis, Johanna and Weber, Tobias and Wesp, Philipp and Sabel, Bastian and Ricke, Jens and Ingrisch, Michael},
	month = dec,
	year = {2022},
	note = {arXiv:2212.14882 [cs]},
}

@book{mitchellMachineLearning1997,
	title = {Machine {Learning}},
	isbn = {978-0-07-115467-3},
	language = {en},
	publisher = {McGraw-Hill},
	author = {Mitchell, Tom M.},
	year = {1997},
	note = {Google-Books-ID: EoYBngEACAAJ},
}

@misc{europeancommissionAIOpenData2018,
	title = {{AI} and {Open} {Data}: a crucial combination},
	url = {https://data.europa.eu/en/publications/datastories/ai-and-open-data-crucial-combination},
	urldate = {2023-04-01},
	journal = {Data stories},
	author = {{European Commission}},
	month = jul,
	year = {2018},
}

@article{davidDigitalTechnologyBoomerang2005,
	title = {The {Digital} {Technology} {Boomerang}: {New} {Intellectual} {Property} {Rights} {Threaten} {Global} “{Open} {Science}”},
	shorttitle = {The {Digital} {Technology} {Boomerang}},
	url = {https://ideas.repec.org//p/wpa/wuwpdc/0502012.html},
	language = {en},
	urldate = {2023-04-26},
	journal = {Development and Comp Systems},
	author = {David, Paul A.},
	month = feb,
	year = {2005},
	note = {Number: 0502012
Publisher: University Library of Munich, Germany},
}

@book{mertonSociologyScienceTheoretical1973,
	title = {The {Sociology} of {Science}: {Theoretical} and {Empirical} {Investigations}},
	isbn = {978-0-226-52092-6},
	shorttitle = {The {Sociology} of {Science}},
	language = {en},
	publisher = {University of Chicago Press},
	author = {Merton, Robert K.},
	year = {1973},
	note = {Google-Books-ID: zPvcHuUMEMwC},
}

@incollection{davidEconomicLogicOpen2003,
	title = {The economic logic of open science and the balance between private property rights and the public domain in scientific data and information: {A} primer},
	isbn = {978-0-309-08850-3},
	language = {en},
	booktitle = {The {Role} of {Scientific} and {Technical} {Data} and {Information} in the {Public} {Domain}: {Proceedings} of a {Symposium}},
	publisher = {National Academies Press},
	author = {David, Paul A.},
	editor = {Esanu, Julie M and Uhlir, Paul F.},
	month = sep,
	year = {2003},
	note = {Google-Books-ID: OULvaEq8YFoC},
}

@article{parthaNewEconomicsScience1994,
	series = {Special {Issue} in {Honor} of {Nathan} {Rosenberg}},
	title = {Toward a new economics of science},
	volume = {23},
	issn = {0048-7333},
	url = {https://www.sciencedirect.com/science/article/pii/0048733394010021},
	doi = {10.1016/0048-7333(94)01002-1},
	language = {en},
	number = {5},
	urldate = {2023-04-26},
	journal = {Research Policy},
	author = {Partha, Dasgupta and David, Paul A.},
	month = sep,
	year = {1994},
	pages = {487--521},
}

@article{davidCanOpenScience2004,
	title = {Can "{Open} {Science}" be {Protected} from the {Evolving} {Regime} of {IPR} {Protections}?},
	volume = {160},
	issn = {0932-4569},
	url = {https://www.jstor.org/stable/40752435},
	number = {1},
	urldate = {2023-04-26},
	journal = {Journal of Institutional and Theoretical Economics (JITE) / Zeitschrift für die gesamte Staatswissenschaft},
	author = {David, Paul A.},
	year = {2004},
	note = {Publisher: Mohr Siebeck GmbH \& Co. KG},
	pages = {9--34},
}

@article{oecdMakingOpenScience2015,
	address = {Paris},
	title = {Making {Open} {Science} a {Reality}},
	url = {https://www.oecd-ilibrary.org/science-and-technology/making-open-science-a-reality_5jrs2f963zs1-en},
	language = {en},
	urldate = {2023-04-26},
	institution = {OECD},
	author = {OECD},
	month = oct,
	year = {2015},
	doi = {10.1787/5jrs2f963zs1-en},
}

@inproceedings{deanDeepLearningRevolution2020,
  title = {1.1 {{The Deep Learning Revolution}} and {{Its Implications}} for {{Computer Architecture}} and {{Chip Design}}},
  booktitle = {2020 {{IEEE International Solid- State Circuits Conference}} - ({{ISSCC}})},
  author = {Dean, Jeffrey},
  year = {2020},
  month = feb,
  pages = {8--14},
  publisher = {IEEE},
  address = {San Francisco, CA, USA},
  doi = {10.1109/ISSCC19947.2020.9063049},
  urldate = {2024-05-12},
  copyright = {https://ieeexplore.ieee.org/Xplorehelp/downloads/license-information/IEEE.html},
  isbn = {978-1-72813-205-1},
  langid = {english}
}

@article{martensImpactDataAccess2018,
  title = {The Impact of Data Access Regimes on Artificial Intelligence and Machine Learning},
  author = {Martens, Bertin},
  year = {2018},
  journal = {European Commission, Joint Research Centre (JRC)},
  volume = {2018},
  number = {09},
  langid = {english}
}

@misc{li_andreeto_ranzato_perona_2022, title={Caltech 101}, abstractNote={Pictures of objects belonging to 101 categories. About 40 to 800 images per category. Most categories have about 50 images. Collected in September 2003 by Fei-Fei Li, Marco Andreetto, and Marc'Aurelio Ranzato. The size of each image is roughly 300 x 200 pixels. We have carefully clicked outlines of each object in these pictures, these are included under the 'Annotations.tar'. There is also a MATLAB script to view the annotations, 'show_annotations.m'.}, publisher={CaltechDATA}, author={Li, Fei-Fei and Andreeto, Marco and Ranzato, Marc'Aurelio and Perona, Pietro}, year={2022}, month= apr}

@inproceedings{Sevilla2022,
  title = {Compute Trends Across Three Eras of Machine Learning},
  url = {http://dx.doi.org/10.1109/IJCNN55064.2022.9891914},
  DOI = {10.1109/ijcnn55064.2022.9891914},
  booktitle = {2022 International Joint Conference on Neural Networks (IJCNN)},
  publisher = {IEEE},
  author = {Sevilla,  Jaime and Heim,  Lennart and Ho,  Anson and Besiroglu,  Tamay and Hobbhahn,  Marius and Villalobos,  Pablo},
  year = {2022},
  month = jul 
}

@misc{KochPeterson2024,
  author = {Koch,  Bernard J. and Peterson,  David},
  title = {From Protoscience to Epistemic Monoculture: How Benchmarking Set the Stage for the Deep Learning Revolution},
  publisher = {arXiv},
  year = {2024},
  copyright = {Creative Commons Attribution Non Commercial No Derivatives 4.0 International}
}

@inbook{Goltsev2004,
  title = {A Process of Differentiation in the Assembly Neural Network},
  ISBN = {9783540304999},
  ISSN = {1611-3349},
  booktitle = {Lecture Notes in Computer Science},
  publisher = {Springer Berlin Heidelberg},
  author = {Goltsev,  Alexander and Kussul,  Ernst and Baidyk,  Tatyana},
  year = {2004},
  pages = {452–457}
}

@article{MartnezPlumed2021,
  title = {Futures of artificial intelligence through technology readiness levels},
  volume = {58},
  ISSN = {0736-5853},
  url = {http://dx.doi.org/10.1016/j.tele.2020.101525},
  DOI = {10.1016/j.tele.2020.101525},
  journal = {Telematics and Informatics},
  publisher = {Elsevier BV},
  author = {Martínez-Plumed,  Fernando and Gómez,  Emilia and Hernández-Orallo,  José},
  year = {2021},
  month = may,
  pages = {101525}
}

@article{fadfc20e-20f4-3bc9-8e9f-d067328c5712,
 ISSN = {00028282},
 URL = {http://www.jstor.org/stable/116885},
 author = {Paul A. David},
 journal = {The American Economic Review},
 number = {2},
 pages = {15--21},
 publisher = {American Economic Association},
 title = {Common Agency Contracting and the Emergence of "Open Science" Institutions},
 urldate = {2024-07-19},
 volume = {88},
 year = {1998}
}

@article{fortnow_viewpointtime_2009,
	title = {{ViewpointTime} for computer science to grow up},
	volume = {52},
	issn = {0001-0782, 1557-7317},
	url = {https://dl.acm.org/doi/10.1145/1536616.1536631},
	doi = {10.1145/1536616.1536631},
	language = {en},
	number = {8},
	urldate = {2024-07-20},
	journal = {Communications of the ACM},
	author = {Fortnow, Lance},
	month = aug,
	year = {2009},
	pages = {33--35},
}

@article{meyer_viewpointresearch_2009,
	title = {{ViewpointResearch} evaluation for computer science},
	volume = {52},
	issn = {0001-0782, 1557-7317},
	url = {https://dl.acm.org/doi/10.1145/1498765.1498780},
	doi = {10.1145/1498765.1498780},
	language = {en},
	number = {4},
	urldate = {2024-07-20},
	journal = {Communications of the ACM},
	author = {Meyer, Bertrand and Choppy, Christine and Staunstrup, Jørgen and Van Leeuwen, Jan},
	month = apr,
	year = {2009},
	pages = {31--34},
}

@article{franceschet_role_2010,
	title = {The role of conference publications in {CS}},
	volume = {53},
	issn = {0001-0782, 1557-7317},
	url = {https://dl.acm.org/doi/10.1145/1859204.1859234},
	doi = {10.1145/1859204.1859234},
	language = {en},
	number = {12},
	urldate = {2024-07-20},
	journal = {Communications of the ACM},
	author = {Franceschet, Massimo},
	month = dec,
	year = {2010},
	pages = {129--132},
}

@article{jordan_94_2024,
	title = {94\% on {CIFAR}-10 in 3.29 {Seconds} on a {Single} {GPU}},
	copyright = {Creative Commons Attribution 4.0 International},
	url = {https://arxiv.org/abs/2404.00498},
	doi = {10.48550/ARXIV.2404.00498},
	urldate = {2024-07-21},
	author = {Jordan, Keller},
	year = {2024},
}

@article{marx_reliance_2022,
	title = {Reliance on science by inventors: {Hybrid} extraction of in‐text patent‐to‐article citations},
	volume = {31},
	issn = {1058-6407, 1530-9134},
	shorttitle = {Reliance on science by inventors},
	url = {https://onlinelibrary.wiley.com/doi/10.1111/jems.12455},
	doi = {10.1111/jems.12455},
	language = {en},
	number = {2},
	urldate = {2024-07-21},
	journal = {Journal of Economics \& Management Strategy},
	author = {Marx, Matt and Fuegi, Aaron},
	month = apr,
	year = {2022},
	pages = {369--392},
}

@article{marx_reliance_2020,
	title = {Reliance on science: {Worldwide} front‐page patent citations to scientific articles},
	volume = {41},
	issn = {0143-2095, 1097-0266},
	shorttitle = {Reliance on science},
	url = {https://onlinelibrary.wiley.com/doi/10.1002/smj.3145},
	doi = {10.1002/smj.3145},
	language = {en},
	number = {9},
	urldate = {2024-07-21},
	journal = {Strategic Management Journal},
	author = {Marx, Matt and Fuegi, Aaron},
	month = sep,
	year = {2020},
	pages = {1572--1594},
}

@mastersthesis{shermatov_national_2024,
	address = {Paris, France},
	title = {National {Computing} {Capacities} for {Frontier} {AI}/{DL} {Research}},
	school = {Sorbonne University Association \& University of Turin},
	author = {Shermatov, Fazliddin},
	year = {2024},
}

@article{fukushima_neocognitron_1980,
	title = {Neocognitron: {A} self-organizing neural network model for a mechanism of pattern recognition unaffected by shift in position},
	volume = {36},
	copyright = {http://www.springer.com/tdm},
	issn = {0340-1200, 1432-0770},
	shorttitle = {Neocognitron},
	url = {http://link.springer.com/10.1007/BF00344251},
	doi = {10.1007/BF00344251},
	language = {en},
	number = {4},
	urldate = {2024-07-23},
	journal = {Biological Cybernetics},
	author = {Fukushima, Kunihiko},
	month = apr,
	year = {1980},
	pages = {193--202},
}
